\documentclass[times, review, 10pt]{elsarticle}
\usepackage{setspace}
\usepackage{geometry}
\usepackage{verbatim} 
\usepackage{amsmath,amsthm,amssymb}
\usepackage{enumitem}
\usepackage{algorithmic}	
\usepackage{algorithm}
\usepackage{url} 
\usepackage[hidelinks]{hyperref}
\usepackage{xcolor}
\usepackage{colortbl}
\usepackage{subfigure}
\usepackage{tikz}
\usepackage{svg}
\usepackage{multirow}
\usepackage{booktabs}
\usepackage{afterpage}
\usepackage{tablefootnote}
\usepackage{makecell}
\usepackage{todonotes}
\usepackage{soul}

\def\x{\bar{x}}

\theoremstyle{definition} 
\newtheorem{definition}{Definition}

\newtheorem{proposition}{Proposition}
\newtheorem{theorem}{Theorem}

\newtheorem{lemma}{Lemma}

\usepackage{etoolbox}

\journal{Pattern Recognition}

\begin{document}

\begin{frontmatter}



\title{Integrated Subset Selection and Bandwidth Estimation Algorithm for Geographically Weighted Regression}

\author[label1]{Hyunwoo Lee}
\ead{hyunwoolee@vt.edu}
\affiliation[label1]{organization={Grado Department of Industrial and Systems Engineering},
            addressline={Virginia Tech},
            city={Blacksburg},
            postcode={24060},
            state={VA},
            country={USA}}

\author[label2]{Young Woong Park\corref{cor1}}
\ead{ywpark@iastate.edu}
\cortext[cor1]{Corresponding author}
\affiliation[label2]{organization={Ivy College of Business},
            addressline={Iowa State University},
            city={Ames},
            postcode={50011},
            state={IA},
            country={USA}}


\begin{abstract}
This study proposes a mathematical programming-based algorithm for the integrated selection of variable subsets and bandwidth estimation in geographically weighted regression, a local regression method that allows the kernel bandwidth and regression coefficients to vary across study areas. Unlike standard approaches in the literature, in which bandwidth and regression parameters are estimated separately for each focal point on the basis of different criteria, our model uses a single objective function for the integrated estimation of regression and bandwidth parameters across all focal points, based on the regression likelihood function and variance modeling. The proposed model further integrates a procedure to select a single subset of independent variables for all focal points, whereas existing approaches may return heterogeneous subsets across focal points. We then propose an alternative direction method to solve the nonconvex mathematical model and show that it converges to a partial minimum. The computational experiment indicates that the proposed algorithm provides competitive explanatory power with stable spatially varying patterns, with the ability to select the best subset and account for additional constraints.
\end{abstract}




\end{frontmatter}


\section{Introduction}
When observations (data points) are located over multiple geographic regions with varying characteristics, the spatial characteristics could be an important factor in regression analysis. However, typical multiple linear regression neglects spatial characteristics. Geographically weighted regression (GWR) is a local regression model that can construct spatially varying relationships between variables \citep{brunsdon1996geographically,brunsdon1998geographically,charlton2009geographically}. GWR allows local variations in the estimated coefficients by building a local model for each focal point (calibration location). The geographic range and the extent of weights of each local model are chosen for each focal point. GWR has been used as an exploratory tool to analyze the effects of the independent variables over the study region. Datasets that are well-suited for GWR often consist of socio-economic and socio-demographic information. Examples include the Georgia dataset \citep{fotheringham2003geographically}, the school dataset \citep{fotheringham2001spatial}, the Irish famine dataset \citep{fotheringham2013demographic}, the obesity dataset \citep{oshan2020targeting}, and the Ohio art culture dataset, first introduced in this paper. For practical applications, one can apply GWR to any dataset combining socio-economic, socio-demographic, and geographic location information. For a comprehensive overview and implementation guidelines, readers are referred to the recent review by Comber et al. \cite{comber2023route}.

In GWR, the weight function, also referred to as the weight scheme, defines how each focal point is influenced by observations, with closer observations having larger weights. The weight function's kernel bandwidth (parameter) then determines the observation weights. The differently weighted observations are used in the coefficient estimation of the local regression model for each focal point. Among various weight function forms studied \citep{fotheringham2003geographically, wheeler2009geographically}, an exponential function is typically used to reflect the distance decay effect between the focal point and observations. To estimate the kernel bandwidth parameters, cross-validation (CV) \citep{brunsdon1996geographically} and corrected Akaike information criterion (AICc) \citep{fotheringham2003geographically} have been extensively used, while another approach parametrizes the kernel bandwidth in the estimation process \citep{paez2002general}. The three approaches are distinguished in how they handle the unequal variances of the observations, which leads to different estimation processes.

The standard GWR estimation process consists of two steps: (1) the coefficient estimation step minimizing the weighted sum of squared errors (WSSE) and (2) the bandwidth estimation step minimizing the CV score or AICc values. The first step builds regression models given observation weights. However, variance can still be heterogeneous because the error variance is not directly modeled. In the second step, search algorithms such as golden section search and binary search help search for the bandwidth parameter $\gamma$ with lower CV scores or AICc values. Paez et al. \citep{paez2002general} proposed a different approach from the standard GWR estimation process to deal with the unequal error variance issues. For each focal point, they directly model the error variance and propose a parametrization of the estimation of the bandwidth parameter that allows an objective function to include the regression coefficient and bandwidth parameter. Adopting the ideas of parametrization of Paez et al. \citep{paez2002general}, we propose directly modeling the error variance to achieve a single objective function integrating the regression coefficient and bandwidth parameter estimation steps. However, our approach is distinguished from Paez et al. \citep{paez2002general} and the standard GWR estimation procedure in three respects. First, we propose an integrated GWR framework for both global and local bandwidth settings, whereas the existing approaches work for one bandwidth setting: a global setting for the standard GWR estimation procedure and a local setting for Paez et al. \citep{paez2002general} and Comber et al. \citep{comber2018hyper}. Second, we integrate a subset selection procedure into the estimation framework. Third, our approach simultaneously builds all local models, whereas the existing approaches consider one local model at a time \citep{paez2002general, comber2018hyper, wheeler2009simultaneous}. This simultaneous model estimation allows us to choose consistent subsets over all local models.

The regression subset selection problem, also referred to as variable selection, is a procedure in which a subset of $p$ independent (explanatory) variables is selected to reduce model complexity and enhance model interpretability while maintaining explanatory power. Stepwise regression, least absolute shrinkage and selection operator (lasso) \citep{tibshirani1996regression}, and ridge regression \citep{hoerl1970ridge} are the most popular algorithms for the subset selection task. Meta-heuristic algorithms are also employed for the subset selection task, for example, genetic algorithms \cite{gokberk2007learning}, and tabu search \citep{zhang2002optimal}. In addition to these subset selection algorithms, mathematical programming-based models have recently received increasing attention given improved computing and solver powers, exactness, and flexibility. Using mathematical programs, prior research has examined various objective functions, including mean square error \citep{miyashiro2015subset}, the sum of squared errors (SSE) \citep{bertsimas2016best, bertsimas2016or}, the sum of absolute errors \citep{konno2009choosing}, AIC \citep{miyashiro2015mixed}, Bayesian information criterion \citep{gomez2021mixed}, $L_0$ penalized quantile regression \citep{dai2023variable}, mean absolute error, and minimal redundancy maximal relevance \citep{park2020}. Furthermore, a line of research incorporates regression assumptions, diagnostics, and multicollinearity issues into the mathematical programming models \citep{bertsimas2016or, carrizosa2020integer, chung2020mathematical}. Recently, efficient algorithms for solving the regression subset selection problem have been developed using penalized methods and alternating optimization techniques \citep{hazimeh2020fast,moreira2022alternating}.

In the context of GWR subset selection, most of the literature has focused on local subset selection, with limited discussion on global subset selection methods. Moreover, despite recent advances in solving subset selection problems through mathematical programming, its application in GWR remains largely unexplored, highlighting its potential for future research. Current research has used ridge regression \citep{wheeler2007diagnostic, barcena2014alleviating}, lasso regressions \citep{wheeler2009simultaneous}, elastic net \citep{li2018geographically, comber2018geographically}, and forward selection approach \citep{comber2018hyper} to select a subset for each focal point. However, these local subset selection approaches may return heterogeneous subsets across focal points, which leads to inconsistent local models and a lack of interpretability across all models. Considering the goal of GWR, capturing spatially varying patterns across focal points, finding an optimal single global subset is critical for interpretability and coherence in the result. To select a global subset over all focal points, Fotheringham et al. \citep{fotheringham2013demographic} propose a forward selection (FS) approach, in which variables with the lowest AIC are successively added to the global subset. This approach is later implemented by Gollini et al. \citep{gollini2015gwmodel} in their R packages. However, the FS approach can be stuck at local optimal and does not guarantee optimality \citep{park2020}. To build optimal and consistent local models, we present a mathematical programming model that returns an optimal subset given bandwidth parameters.
 
In this study, we propose a mathematical programming model to simultaneously estimate regression coefficients, bandwidth parameters, and best variable subset. However, the proposed mathematical program becomes non-convex, which is difficult to solve. Therefore, we propose an alternative direction method (ADM) \citep{gorski2007biconvex, geissler2017penalty} to solve the problem and show that it converges to a partial minimum\footnote{A partial minimum is a solution that minimizes the objective function with respect to a subset of variables, while the remaining variables are held fixed. Refer to Definition \ref{definition1} for a formal specification.}. The ADM is a special case of alternating optimization tasks \citep{bazaraa2006nonlinear} that leverages the block structure of subproblem. It has been successfully applied to various optimization, including bilevel optimization problems \citep{kleinert2021computing}, convex quadratic programs \citep{sun2010modified}, mixed integer programs \citep{geissler2017penalty, gottlich2021penalty}, unsupervised feature selection \citep{zuo2025unsupervised}, elastic net support vector machines \citep{liang2024linearized}, and best subset selection in linear regression \citep{moreira2022alternating}. ADM is particularly well-suited where the parameter set can be partitioned into two blocks - a property exhibited by our proposed mathematical model. This feature enables us to apply ADM effectively in our study.

Our study contributes to the literature in three ways.
First, we propose an integrated procedure based on a mathematical programming model that estimates both regression coefficients and bandwidth parameters with a single objective function. This differs from the popular approaches in the literature, which estimate regression coefficients and bandwidth parameters using inconsistent objective functions. Second, the proposed mathematical programming model incorporates a subset selection procedure to select global and consistent subsets for all local models. Third, to tackle the difficult-to-solve non-convex mathematical programming model, we propose an ADM with guaranteed convergence to a partial minimum. Fourth, we present one of the most comprehensive experiments in the GWR literature, featuring multiple performance measures evaluated across diverse datasets.

The structure of the paper is as follows. In Section \ref{section2}, we overview GWR and derive a likelihood function for all focal points. In Section \ref{section3}, we propose mathematical models for integrated subset selection and bandwidth estimation and describe the ADM for GWR with the convergence result. Section \ref{section4} presents the computational experiments conducted on various datasets. Finally, Section \ref{section5} concludes our study by discussing key findings, limitations, and potential future research directions.

\section{GWR Likelihood} \label{section2}
This section first presents a brief overview of the basic GWR models and standard estimation procedure in Section \ref{section2.1}. Then, we propose a new likelihood function by combining likelihood functions of all focal points in Section \ref{section2.2}. We use the new likelihood function to develop the mathematical models and algorithms in a subsequent section.

\subsection{GWR Overview}\label{section2.1}
Let $\boldsymbol{X}$ be an $n \times m$ matrix of the independent variable (IV) with $n$ observations and $m$ features and $\boldsymbol{Y}$ be an $n \times 1$ vector of the dependent variable (DV). The multiple linear regression (MLR) can be written as
$$
\textstyle y_i = \beta_{1} + \sum_{j = 2}^{m} \beta_{j} x_{ij} + \epsilon_i, i \in I, \\ 
$$
where $\boldsymbol{\beta}$ is an $m \times 1$ vector of coefficient parameters and $\boldsymbol{\epsilon}$ is an $n \times 1$ vector of random error terms. Ordinary least square is the most common estimator, and the following closed-form expression gives the estimated regression coefficients: $\textstyle \hat{\boldsymbol{\beta}} = [\boldsymbol{X}^T \boldsymbol{X}]^{-1}\boldsymbol{X}^T Y$.

By contrast, GWR builds a local regression model and estimates coefficients for each focal point $o$, which we write as
$$
\textstyle y_{i}= \beta_{o1} + \sum_{j = 2}^{m} \beta_{oj} x_{oj} + \epsilon_{oi}, i \in I,
$$
where $\boldsymbol{\beta}_o$ is an $m \times 1$ vector of coefficient parameters and $\boldsymbol{\epsilon}_o$ is an $n \times 1$ vector of random error terms. To estimate $\hat{\boldsymbol{\beta}}_o$, we use all other observations $i$ and their locational relationship to focal point $o$. The estimated coefficient $\boldsymbol{\beta_o}$ can only be used to predict $y_o$. Thus, $n$ local regression models need to be built to make predictions for all focal points. The standard GWR estimation procedure uses weighted least squares to assign different weights to observations based on distance, and the coefficients can be estimated by 
$\textstyle \hat{\boldsymbol{\beta}}_{o} = [\boldsymbol{X}^T \boldsymbol{W}_o \boldsymbol{X}]^{-1}\boldsymbol{X}^T \boldsymbol{W}_o \boldsymbol{Y},$ where $\boldsymbol{W}_o = diag([W_{o1}, …, W_{on}])$ is an $n \times n$ diagonal matrix and $W_{oi}$ is the weight given to observation $i$ for focal model $o$. It is important to note that the matrix $\boldsymbol{W}_o$ is used solely during the estimation process. Once estimation is complete, the resulting coefficients $\hat{\boldsymbol{\beta}}_{o}$s are utilized for evaluating the regression model. However, this approach assumes that the random errors are normally distributed with constant variance \citep{wheeler2005multicollinearity, leung2000statistical}, which may not be realistic. The literature has used two approaches to account for unequal variance. The standard GWR estimation procedures adopt the assumption of normally distributed random errors while imposing weights only on the error terms. By contrast, Paez et al. \citep{paez2002general} directly model an error variance that varies for each observation. Without loss of generality, we assume that when an intercept is included, $x_{i1} = 1, \forall i \in I$. We provide a detailed description of the two approaches next.

\vspace{\baselineskip}
\noindent
\textbf{Approach 1:} $\boldsymbol{\epsilon_o} \sim N(o,\sigma^2)$ \\
The most common approach in the literature deals with unequal variance by imposing weights on the error terms. For each focal point $o$, the error term, likelihood, and regression likelihood (log-likelihood) are defined as
\begin{equation*}
    \begin{split}
    & \textstyle \epsilon_{oi} = y_i - \sum_{j=1}^{m} x_{ij} \beta_{oj}, \epsilon_{oi} \sim N(o,\sigma^2),\\
    & \textstyle f(\epsilon_{oi}|\theta) = \frac{1}{\sigma \sqrt{2 \pi}} e^{-\frac{\epsilon_{oi}^2}{2\sigma^2}} = \frac{1}{\sigma \sqrt{2 \pi}}  e^{{-\frac{1}{2\sigma^2} W_{oi}(y_i - \sum_{j=1}^{m} x_{ij} \beta_{oj})}^2}, \text{ and} \\
    & \textstyle L_o = \sum_{i=1}^{n} \log f(\epsilon_{oi}|\theta) = -\frac{1}{2} \sum_{i=1}^{n} \log \sigma^2 - \frac{n}{2} \log 2 \pi - \frac{1}{2\sigma^2} \sum_{i=1}^{n} W_{oi}{(y_i - \sum_{j=1}^{m} x_{ij} \beta_{oj})}^2. \\
    \end{split}
\end{equation*}
The regression coefficients are estimated by maximizing $L_o$. Because $\sigma$ is assumed to be known, the first two terms of $L_o$ remain constant, and $L_o$ can be maximized by minimizing $\sum_{i=1}^n W_{oi}{(y_i - \sum_{j=1}^{m} x_{ij} \beta_{oj})}^2$, assuming $W_{oi}$s are given.

The weights $W_{oi}$s are defined by weight function and the bandwidth parameter $\gamma$. Popular choices for the weight function include the exponential kernel function $W_{oi} = exp(-\frac{d_{oi}}{\gamma})$, Gaussian kernel function $W_{oi} = \exp(-\frac{1}{2}(\frac{d_{oi}^2}{\gamma}))$, and bi-square kernel function $\textstyle W_{oi} = 1 - \frac{d_{oi}^2}{\gamma}$. Among them, exponential functions are the most common as they can represent the distance decay effect well in GWR. Throughout this study, we use the exponential form of $\textstyle W_{oi} = \exp(-\gamma_o d_{oi}^2)$, following Paez et al. \citep{paez2002general}, because it provides a simple functional form for $L_0$.

For the bandwidth parameter estimation, a CV score or AICc is commonly used. First, CV is an iterative process that searches for the bandwidth parameter by minimizing the prediction error of all focal points over all possible $\gamma$, which can be written as
$$
    \operatorname*{argmin}_\gamma \sum_{o \in O} (y_o - \hat{y}_o(\gamma))^2,
$$
where $\hat{y}_o(\gamma)$ is the predicted value for focal point $o$ when $\gamma$ is used. Second, the AICc aims to minimize the estimation error of the DV considering the trade-off between the goodness of fit of the model and model complexity. AICc for GWR is defined as
\begin{equation*}
    AIC_c = 2n\log(\hat{\sigma}) + n\log(2\pi) + n (\frac{n+trace(\boldsymbol{H})}{n-2-trace(\boldsymbol{H})}),
\end{equation*}
where $\boldsymbol{H}$ is the hat matrix for GWR that maps the vector of DV values to the vector of predicted values as $\boldsymbol{HY} = \boldsymbol{\hat{Y}}$ and the trace of a matrix is the sum of the matrix diagonal elements. Each row of the hat matrix is calculated as $\boldsymbol{H}_o = \boldsymbol{X}_o( \boldsymbol{X}^T \boldsymbol{W}_o \boldsymbol{X})^{-1}\boldsymbol{X}^T\boldsymbol{X}_o$, where $\boldsymbol{X}_o$ is the $o$-th row vector of $\boldsymbol{X}$. To find the best bandwidth of the objective function, both the CV and AICc approaches need a search algorithm such as a golden section search or binary search.

\vspace{\baselineskip}
\noindent
\textbf{Approach 2:} $\boldsymbol{\epsilon}_o \sim N(o,\boldsymbol{\Omega_o})$ \\
Instead of imposing weights on errors, Paez et al. \citep{paez2002general} proposed an approach to directly model the variance. The covariance matrix $\boldsymbol{\Omega}_o$ has the diagonal element of $\omega_{oi} = \sigma_{oi}^2 $. The error term, likelihood, and regression likelihood (log-likelihood) are defined as
\begin{align}
    & \textstyle \epsilon_{oi} = y_i - \sum_{j=1}^{m} x_{ij} \beta_{oj}, \epsilon_{oi} \sim N(o,\sigma_{oi}^2) \nonumber, \\
    & \textstyle f(\epsilon_{oi}|\theta) = \frac{1}{\sigma_{oi} \sqrt{2 \pi}} e^{-\frac{\epsilon_{oi}^2}{2\sigma_{oi}^2}} = \frac{1}{\sigma_{oi} \sqrt{2 \pi}}  e^{{-\frac{(y_i - \sum_{j=1}^{m} x_{ij} \beta_{oj})^2}{2\sigma_{oi}^2}}} \nonumber, \text{ and} \\
    & \textstyle L_o = \sum_{i=1}^{n} \log f(\epsilon_{oi}|\theta) = -\frac{1}{2} \sum_{i=1}^{n} \log \sigma_{oi}^2 - \frac{n}{2} \log 2 \pi -         \frac{1}{2} \sum_{i=1}^{n} \frac{{(y_i - \sum_{j=1}^{m} x_{ij} \beta_{oj})}^2}{\sigma_{oi}^2}.  \label{ll_derive0}  
\end{align}
Instead of giving the weights $W_{oi}$ to each likelihood, by modeling $\frac{1}{\sigma_{oi}^2}$, the distance decay effects can be achieved. Paez et al. \citep{paez2002general} model the variance $\sigma_{oi}^2$ as $\sigma_{oi}^2 = \sigma^2 \exp{(\gamma_o d_{oi}^2)}$, which consists of a common variance and weight function. In our model, we directly model the variance $\sigma_{oi}^2$ to $\sigma_{oi}^2 = \exp{(\gamma_o d_{oi}^2)}$. We note that this adjusted derivation has several merits. The likelihood estimator $\sigma_o^2$ of Paez et al. \citep{paez2002general} includes the logarithm of the WSSE, which tends to cause issues in balancing the WSSE and the term for $\gamma_o$ and returns negative values for $\gamma_o$. This setting unintentionally imposes smaller weights for nearer observations, which does not adhere to the principle of GWR. In contrast, our derivation works better for balancing the terms. Furthermore, $\frac{1}{\sigma_{oi}^2}$ can be directly linked to the weight function $\exp{(-\gamma_o d_{oi}^2)}$, providing a clear form.

\subsection{Likelihood Derivation for All Focal Points}\label{section2.2}

When we plug $\sigma_{oi}^2 = \exp{(\gamma_o d_{oi}^2)}$ into \eqref{ll_derive0}, the last term of the log-likelihood becomes the WSSE and the first term is no longer a constant. The regression likelihood becomes a single objective function containing bandwidth $\gamma_o$ and coefficient $\beta_{oj}$s. Therefore, both estimation procedures can be achieved by maximizing the log-likelihood function. The following optimization problem, for focal point $o$, maximizes \eqref{ll_derive0}:
\begin{equation}
    \textstyle \text{min} \quad -L_o = \gamma_o\frac{1}{2} \sum_{i=1}^{n} d_{oi}^2 + \frac{1}{2} \sum_{i=1}^n \frac{{(y_i - \sum_{j \in J} x_{ij} \beta_{oj})}^2}{\exp(\gamma_o d_{oi}^2)}. \label{ll_derive0.5}
\end{equation}
Note that when $\gamma_o$ increases, the first term increases and the second term decreases. The objective function then tries to balance the two terms. To maximize $L_o$ in \eqref{ll_derive0}, Paez et al. \citep{paez2002general} proposed a solution by taking derivatives of \eqref{ll_derive0} with respect to each parameter $\theta = \{ \boldsymbol{\beta}_o, \sigma^2, \gamma\}$, repeating the procedure for each focal point. Our approach to solve the problem in  \eqref{ll_derive0.5} includes two distinct points. First, instead of solving the problem for each focal point $o$, we solve a single problem for all focal points to control the global model consistency. Second, we integrate a subset selection procedure to reduce the model complexity and improve interpretability. Because the set of focal points does not need to be equal to the set of observations, we set the number of focal points to $c$. Then, we propose a single objective function for all focal points, which allows us to solve all local GWR models simultaneously. By assuming that each focal point has its own bandwidth parameter $\gamma_o$, we derive the following maximum-likelihood function:
\begin{equation*}
    \textstyle \operatorname*{argmax}_\theta \prod_{o=1}^{c} \prod_{i=1}^{n} f(\epsilon_{oi}|\theta) = \operatorname*{argmax} \sum_{o=1}^{c} L_o.
\end{equation*}
Then, $\sum_{o=1}^c -L_o$ is the objective function to be minimized with parameter set $\theta = \{ \boldsymbol{\beta_o}, \gamma_o | o = \{1,2, ..., c\}\}$. Thus, we can formulate the following optimization problem for the GWR model for all focal points:
\begin{equation}
    \textstyle Minimize \quad \sum_{o=1}^c -L_o = \frac{1}{2} \sum_{o=1}^c \gamma_o \sum_{i=1}^{n} d_{oi}^2 + \frac{1}{2} \sum_{o=1}^c \sum_{i=1}^n \frac{{(y_i - \sum_{j \in J} x_{ij} \beta_{oj})}^2}{\exp(\gamma_o d_{oi}^2)}.\label{ll_derive1}
\end{equation}
We assume that individual $\gamma_o$ exists for each focal point. When the kernel bandwidth is the same across all the regions, we use the notation $\gamma$ and refer to it as global bandwidth.

\section{Mathematical Models and Algorithms for GWR}\label{section3}

In this section, we develop an integrated procedure for subset selection, regression coefficient, and bandwidth estimation based on mathematical programming and an ADM algorithm. In Section \ref{section3.1}, using the likelihood function proposed in Section \ref{section2.2}, we propose a mathematical programming model that incorporates the coefficient estimation and bandwidth estimation while selecting a subset of a fixed number of explanatory variables. In Section \ref{section3.2}, to solve the non-convex mathematical programming model, we propose the ADM algorithm and show that the algorithm converges to a partial minimum.

\subsection{Mathematical Models for Integrated Subset Selection and Bandwidth Estimation} \label{section3.1}
Throughout this study, we use the following notations: \
\begin{itemize}[noitemsep]
\item[] $c$: number of focal points; \vspace{-0.1cm}
\item[] $n$: number of observation locations (neighborhood points); \vspace{-0.1cm}
\item[] $m$: number of IVs;  \vspace{-0.1cm}
\item[] $p$: number of selected IVs; \vspace{-0.1cm}
\item[] $O=\{1, ..., c\}$: index set of focal points; \vspace{-0.1cm}
\item[] $I=\{1, ..., n\}$: index set of observation locations; \vspace{-0.1cm}
\item[] $J=\{1, ..., m\}$: index set of IVs; \vspace{-0.1cm}
\item[] $HC_{\rho} = \{ (j,k) | j \in J, k \in J, j \neq k, \rho_{jk} \geq \rho \} $: index set of pairs of IVs with correlation $\rho_{jk}$ at least $\rho$; \vspace{-0.1cm}
\item[] $M_{\beta}, M_{\gamma}$: Big-M values for coefficients $\beta$ and bandwidth parameter $\gamma$;
\item[] $\boldsymbol{X}$: data matrix corresponding to all IVs, where $\boldsymbol{X} = [x_{ij}] \in \mathbb{R}^{n \times m}$; and \vspace{-0.1cm}
\item[] $\boldsymbol{D}= [d_{oi}] \in \mathbb{R}^{c \times n}$: distance matrix between observation locations and focal points.
\end{itemize}
For all mathematical models derived, we use the following decision variables:
\begin{itemize}[noitemsep]    
\item[] $\beta_{oj}$: coefficient of IV $j \in J$ at focal point $o \in O$; \vspace{-0.1cm}
\item[]$e_{oi}$: fitting error of observation $i \in I$;  \vspace{-0.1cm}
\item[]$\gamma_{o}$: bandwidth parameter at focal point $o \in O$; and \vspace{-0.1cm}
\item[]$z_{j}$: 1 if IV $j \in J$ is selected; 0 otherwise.
\end{itemize}
Our proposed mathematical programming model incorporating subset selection is shown as follows.
\begin{subequations}
\label{mae_derive1}
\begin{align}
\min \quad & \textstyle \sum_{o=1}^c \gamma_o \sum_{i=1}^n d_{oi}^2 + \sum_{o=1}^c \sum_{i=1}^n \frac{{e_{oi}}^2}{\exp(\gamma_o d_{oi}^2)} \label{mae_derive1_a} \\[-1pt]
s.t. \quad & \textstyle e_{oi} = y_i - \sum_{j=1}^{m} x_{ij} \beta_{oj}, & o \in O, \quad i \in I, \label{mae_derive1_b} \\[-1pt]
& \textstyle -M_{\beta} z_j \leq \beta_{oj} \leq M_{\beta} z_j, & o \in O, \quad j \in J, \label{mae_derive1_c}\\[-1pt]
& \textstyle \sum_{j=1}^m z_j = p, & j \in J, \label{mae_derive1_d}\\[-1pt]
& \textstyle z_j + z_k \leq 1, & (j,k) \in HC_{\rho}, \label{mae_derive1_e}\\[-1pt]
& \textstyle z_j \in \{0, 1\}, & j \in J,  \label{mae_derive1_f}\\[-1pt]
& \textstyle 0 \leq \gamma_o \leq M_{\gamma}, & o \in O \label{mae_derive1_g}
\end{align}
\end{subequations}
The proposed model in \eqref{mae_derive1} determines the regression coefficients and bandwidth parameters by minimizing the sum of the negative log-likelihood function. Note that the scaling factor of $1/2$ in \eqref{ll_derive1} is omitted, as it does not affect the optimal solution. Constraint \eqref{mae_derive1_b} defines the error terms and Constraint \eqref{mae_derive1_c} indicates that the coefficient for IV $j$ must be 0 if not selected ($z_j = 0$). Note that the regression coefficients $\beta_{oj}$ for all focal points in $O$ are subject to the same $z_j$ value. Therefore, Constraint \eqref{mae_derive1_c} forces the model to select an identical subset across all focal point models. Constraint \eqref{mae_derive1_d} ensures that exactly $p$ variables must be selected. Constraint \eqref{mae_derive1_e} forces the model to select at most one of the highly correlated pair IVs with threshold $\rho$. We set $\rho = 0.9$ for the experiments conducted herein. Constraint \eqref{mae_derive1_f} defines the binary decision variable, and Constraint \eqref{mae_derive1_g} imposes upper and lower bounds on the bandwidth parameter $\gamma_o$. In Constraint \eqref{mae_derive1_g}, the upper bound $M_{\gamma}$ for $\gamma_o$ is a sufficiently large value. We later show that a valid value for $M_{\gamma}$ always exists.

Note that the problem in \eqref{mae_derive1} allows local bandwidth setting using $\gamma_o$s, where each focal point has its own bandwidth function. By replacing $\gamma_o$s with $\gamma$, we can also impose the global bandwidth setting, where all focal points share a common bandwidth function. However, regardless of the global or local setting, the problem in \eqref{mae_derive1} is a non-convex program and is difficult to solve in its current form. To tackle this problem, we partition the decision variable of \eqref{mae_derive1} into two sets, 
\begin{equation} \label{Beta_set}
B(p) := \textstyle \Bigl\{\beta_{oj}: -Mz_j \leq \beta_{oj} \leq Mz_j, \ z_j \in \{0,1\}, \forall j \in J, \ \sum_{j=1}^m z_j = p, \ z_j + z_k \leq 1, \forall (j,k) \in HC_{\rho} \Bigr\}
\end{equation}
and 
\begin{equation} \label{Gamma_set}
\Gamma := \Bigl\{\gamma_{o}: 0 \leq \gamma_{o} \leq M_{\gamma}, \forall o \in O \Bigr\},
\end{equation}
and alternately fix one of the sets and solve for the other set.

The first step, which we refer to as the $B(p)$ step, reduces \eqref{mae_derive1} using fixed $\Gamma$. When all $\gamma_o$s in $\Gamma$ are fixed to the appropriate values, the first term of \eqref{mae_derive1_a} becomes a constant, and Constraint \eqref{mae_derive1_g} can be dropped. Let $\bar{\gamma}_o$ be the fixed values. Then, we can obtain the following mixed integer quadratic program (MIQP):
\begin{equation}\label{mae_derive2}
\displaystyle \min \biggl\{\sum_{o=1}^c \bar{\gamma}_o \sum_{i=1}^n d_{oi}^2 + \sum_{o=1}^c \sum_{i=1}^n \frac{{(y_i - \sum_{j=1}^{m} x_{ij} \beta_{oj})}^2}{\exp(\bar{\gamma}_o d_{oi}^2)} \ | \ \beta_{oj} \in B(p)   \biggr\}. 
\end{equation}
The MIQP model in \eqref{mae_derive2} is equivalent to minimizing WSSE and is solvable by most of the numerical solvers. We denote \eqref{mae_derive2} by $MP_{\beta}$ because it estimates the regression coefficients $\beta$. 

The second step, which we refer to as the $\Gamma$ step, reduces \eqref{mae_derive1} using fixed $B(p)$. When the regression coefficients in $B(p)$ are fixed to the appropriate values in \eqref{mae_derive1}, $e_{oi}$ is immediately calculated in \eqref{mae_derive1_b} and substituted into \eqref{mae_derive1_a}. Additionally, $z_{j}$ values are fixed properly to satisfy Constraints \eqref{mae_derive1_c}-\eqref{mae_derive1_f}. As such, we can drop Constraints \eqref{mae_derive1_c}-\eqref{mae_derive1_f}. Let $\bar{\beta}_{oj} \mbox{ and } \bar{z}_j$ be the fixed values. Then, we can obtain the following optimization model:
\begin{equation}\label{mae_derive3}
\displaystyle \min \biggl\{ \sum_{o=1}^c \gamma_o \sum_{i=1}^n d_{oi}^2 + \sum_{o=1}^c \sum_{i=1}^n \frac{ {(y_i - \sum_{j=1}^{m} x_{ij} \bar{\beta}_{oj})}^2} {\exp(\gamma_o d_{oi}^2)} \ | \  \gamma_o \in \Gamma \biggr\}.
\end{equation}

Note that the model in \eqref{mae_derive3} is a multivariate (or univariate) constrained optimization problem with respect to $\gamma_o$s (or $\gamma$). We refer to \eqref{mae_derive3} as $MP_{\gamma}$ because it estimates the bandwidth parameters in $\Gamma$. Note that $\gamma_o$ in $\Gamma$ is bounded by $M_{\gamma}$. This bound is crucial for the convergence proof presented in \ref{section3.2}. In the following lemma, we demonstrate that imposing this bound does not affect the solution of the problem.

\begin{lemma} \label{lemma1}
Assume $d_{oi} > o$ and $e_{oi}>0$ for some $o \in O$ and $i \in I$. Let \(\Gamma' := \{\gamma_{o} : \gamma_o \geq 0, \forall o \in O\}\). Then, there exists a constant $M_{\gamma}$ such that solving \eqref{mae_derive3} with $\Gamma$ and solving \eqref{mae_derive3} with $\Gamma'$ yield the same optimal solution.
\end{lemma}
\begin{proof}
Consider the problem in \eqref{mae_derive3} with \(\Gamma'\), which has no upper bound. Let \(\{\gamma_o', o \in O\}\) be an optimal solution for this problem with an objective function value of \(\xi'\). Because the objective function of \eqref{mae_derive3} becomes unbounded as $\gamma_o \rightarrow \infty$ for any $o \in O$, $\gamma_o'$ must be finite for all $o \in O$. To define $\Gamma$ for \eqref{mae_derive3}, set $M_{\gamma} = \max \{\gamma_o' : o \in O\} $. Let $\{\gamma_o^*, o \in O\}$ be an optimal solution for \eqref{mae_derive3} with an objective function value $\xi^*$. First, we have $\xi^* \leq \xi'$ because $\{\gamma'_o, o \in O\}$ is a feasible solution for \eqref{mae_derive3}.
Second, we have $\xi^* \geq \xi'$ because $\Gamma \subseteq \Gamma^{'}$. Combining these two inequalities, we conclude that $\xi^* = \xi'$, demonstrating that the optimal objective function values are identical when \eqref{mae_derive3} is solved with $\Gamma$ and $\Gamma'$. Furthermore, we show the strict convexity of the objective function in \eqref{mae_derive3} with respect to $\gamma_o$s. By taking the second derivative with respect to $\gamma_o$s, the objective function becomes $\sum_{o=1}^c  d_{oi}^4 \sum_{i=1}^n \frac{ {e_{oi}}^2} {\exp(\gamma_o d_{oi}^2)} $, where $e_{oi} = y_i - \sum_{j=1}^{m} x_{ij} \beta_{oj}$. This expression is greater than zero whenever $d_{oi} > o$ and $e_{oi}>0$ for some $o \in O$ and $i \in I$, which are typically met in practice. Hence, the strict convexity of the objective function in \eqref{mae_derive3}, along with the convexity of the sets $\Gamma$ and $\Gamma'$, guarantees a unique optimal solution.
\end{proof}

Note that a valid value for the big-M values for $\beta$ in Constraint \eqref{mae_derive1_c} can be obtained mathematically \citep{park2020}.

\subsection{Algorithms for Integrated Subset Selection and Bandwidth Estimation} \label{section3.2}

In this section, we review the ADM and propose an ADM algorithm for the problem in \eqref{mae_derive1} based on $MP_\beta$ \eqref{mae_derive2} and $MP_\gamma$ \eqref{mae_derive3} with convergence results. Consider an optimization problem in which the decision variables can be partitioned into two disjoint subsets. If this problem is convex with respect to one decision variable subset given fixed values for the other subset, it is referred to as \textit{biconvex optimization problem}. Gorski et al. \citep{gorski2007biconvex} study the biconvex optimization problem and propose solving it by alternately fixing one subset and solving for the other subset. This method, named the \textit{alternate convex search} algorithm, can solve not only biconvex problems but also a broader class of optimization problems. Later, Geissler et al. \citep{geissler2017penalty} reformulate Gorski et al.’s theorem in a more concise manner, maintaining the same range of solvable problems. In their version, the partitioned subproblems are required to have nonempty and compact sets instead of convex sets. This reformulation improved accessibility to the underlying principles while preserving the algorithm's generality.

Consider the general problem:
\begin{equation}\label{mae_PADM1}
\min_{u \in U, v \in V} f(u, v) \quad \text{s.t.} \quad g(u, v) = 0, \quad h(u, v) \geq 0, \quad u \in U, \quad v \in V,
\end{equation}
where the objective funtion $f: \mathbb{R}^{n_{u}+n_{v}} \rightarrow \mathbb{R}$ and the constraint functions $g: \mathbb{R}^{n_{u}+n_{v}} \rightarrow \mathbb{R}^{q}, h: \mathbb{R}^{n_{u}+n_{v}} \rightarrow \mathbb{R}^{s}$ are continuous and the sets $U$ and $V$ are nonempty and compact.

ADMs are iterative procedures that solve the problem in \eqref{mae_PADM1} by alternatingly solving two simpler subproblems. Given an iterate $\left(u^{t}, v^{t}\right)$, the problem in \eqref{mae_PADM1} with $v$ fixed to $v^{t}$ is solved into the direction of $u$, yielding a new solution $u^{t+1}$. Subsequently, $u$ is fixed to $u^{t+1}$, and the problem in \eqref{mae_PADM1} is solved into the direction of $v$, resulting a new solution $v^{t+1}$. Algorithm \ref{algo:ADM} formally states these procedures.

\begin{algorithm}[h!]
\caption{Standard ADM (Geissler et al. \cite{geissler2017penalty}) }  
\label{algo:ADM}                           
\begin{algorithmic}[1]  
\vspace{0.1cm}
\STATE Choose initial values $\left(u^{0}, v^{0}\right) \in U \times V$. \label{algo:ADM_1}   \FOR{$t=0,1, \ldots$} \label{algo:ADM_2}
\STATE Compute $u^{t+1} \in \underset{u}{\arg \min }\left\{f\left(u, v^{t}\right): g\left(u, v^{t}\right)=0, h\left(u, v^{t}\right) \geq 0, u \in U\right\}$. \label{algo:ADM_3}
\STATE Compute $v^{t+1} \in \underset{v}{\arg \min }\left\{f\left(u^{t+1}, v\right): g\left(u^{t+1}, v\right)=0, h\left(u^{t+1}, v\right) \geq 0, v \in V\right\}$. \label{algo:ADM_4}
\STATE Set $t \leftarrow t+1$. \label{algo:ADM_5}
\ENDFOR \label{algo:ADM_6}
\end{algorithmic}
\end{algorithm}

Consider the feasible set of \eqref{mae_PADM1} defined as 
$$
\Omega=\{(u, v) \in U \times V: g(u, v)=0, h(u, v) \geq 0\} \subseteq U \times V.
$$
To discuss the general convergence results of Algorithm \ref{algo:ADM}, we introduce the following definition.

\begin{definition} \label{definition1}
Let $\left(u^{*}, v^{*}\right) \in \Omega$ be a feasible point of the model \eqref{mae_PADM1}. Then, $\left(u^{*}, v^{*}\right)$ is called a partial minimum of the model \eqref{mae_PADM1} whenever it satisfies 
$$
\begin{array}{ll}
f\left(u^{*}, v^{*}\right) \leq f\left(u, v^{*}\right) & \text { for all }\left(u, v^{*}\right) \in \Omega \\
f\left(u^{*}, v^{*}\right) \leq f\left(u^{*}, v\right) & \text { for all }\left(u^{*}, v\right) \in \Omega .
\end{array}
$$
\end{definition}
The general convergence result of Algorithm \ref{algo:ADM} is stated in Theorems 4.5 and 4.9 of Gorski et al. \cite{gorski2007biconvex} and rephrased by Geissler et al. \citep{geissler2017penalty}. We present the theorem by Geissler et al. \citep{geissler2017penalty}.

\begin{theorem}\label{theorem1}[Thoerem 2, Geissler et al. \citep{geissler2017penalty}]
Let $\left\{\left(u^{t}, v^{t}\right)\right\}_{t=0}^{\infty}$ be a sequence generated by Algorithm \ref{algo:ADM} with $\left(u^{t+1}, v^{t+1}\right) \in \Theta\left(u^{t}, v^{t}\right)$, where $\Theta\left(u^{t}, v^{t}\right)=\left\{\left(u^{*}, v^{*}\right):\right. f\left(u^{*}, v^{t}\right) \leq f\left(u, v^{t}\right), \forall u \in U,  \left.f\left(u^{*}, v^{*}\right) \leq f\left(u^{*}, v\right), \forall v \in V\right\}.
$ Suppose that the solution of the first optimization problem (in Line \eqref{algo:ADM_3}) is always unique. Then, every convergent subsequence of $\left\{\left(u^{t}, v^{t}\right)\right\}_{t=0}^{\infty}$ converges to a partial minimum. In addition, if $k$ and $k^{\prime}$ are two limit points of such subsequences, then $f(w)=f\left(w^{\prime}\right)$ holds.
\end{theorem}

We adopt the ADM to solve our proposed model \eqref{mae_derive1}, verifying that it can be interpreted as a specific instance of \eqref{mae_PADM1}. First, the objective function \eqref{mae_derive1_a} is a real-valued function with respect to \(B(p)\) and \(\Gamma\). There are no common constraints corresponding to \eqref{mae_PADM1} because all constraints are disjoint and partitioned into \(B(p)\) and \(\Gamma\). The remaining requirement is that the sets \(B(p)\) and \(\Gamma\), corresponding to \(U\) and \(V\), must be nonempty and compact sets. We show this with the following proposition.

\begin{lemma} \label{lemma2}
The sets \(B(p)\) and \(\Gamma\) are nonempty and compact.
\end{lemma}
\begin{proof}
\(B(p)\) is nonempty because \(\beta_{oj} = 0, \forall j \in J, \forall o \in O\) is always a feasible solution and \(\Gamma\) is nonempty because \(\gamma_o = 0, \forall o \in O\) is also a feasible solution. The set \(B(p)\) is a mixed-binary set where the binary variables \(z_j\) define intersections of closed and bounded sets of \(\beta_{oj}\). Since the possible choices of \(z_j\) are finite due to the parameter $p$, \(B(p)\) consists of a finite number of such intersections. Similarly, \(\Gamma\) is also defined as an intersection of closed and bounded sets of \(\gamma_o\).
\end{proof}

In Algorithm \ref{algo_myalgo}, we propose the ADM algorithm for \eqref{mae_derive1}, referred to as the integrated GWR (IGWR) algorithm for the rest of the paper. The two subproblems, $MP_{\beta}$ \eqref{mae_derive2} and $MP_{\gamma}$ \eqref{mae_derive3}, are solved in an alternating manner in Algorithm \ref{algo_myalgo}. We note that all the parameters for all focal points, $B^*$ and $\Gamma^*$, are updated altogether in Algorithm \ref{algo_myalgo}. 

Before discussing the convergence of Algorithm \ref{algo_myalgo}, we define a tie-breaking rule for Line \ref{algo_myalgo_3} of Algorithm \ref{algo_myalgo}. If multiple optimal solutions exist, $MP_{\beta}$ returns $B(p)^*$ corresponding to the solution with the smallest sum of indexes, ensuring a unique solution for $MP_{\beta}$. However, $MP_{\beta}$ typically has a unique optimal solution because the differences in WSSE exceed numerical precision, such as $10^{-6}$, making the likelihood of multiple optimal solutions negligible.

We now establish the convergence of Algorithm \ref{algo_myalgo} to a partial minimum of the solution with the following proposition.

\begin{algorithm}[h!]
\caption{IGWR}  
\label{algo_myalgo}                           
\begin{algorithmic}[1]  
\vspace{0.1cm}
\REQUIRE $\gamma_{init}$, $p$ (subset cardinality), $\theta$ (threshold) \\ 
\ENSURE $\Gamma^*$ (bandwidth parameter set), $B(p)^*$ (coefficient set), $Obj^* $ (current best obj value)\\
\STATE $t \gets 0, \Gamma^0 \gets \gamma_{init}, Obj_{0} \gets 0 $, Relative Gap $\gets \infty$ \label{algo_myalgo_1}
\WHILE{Relative Gap $> \theta$} \label{algo_myalgo_2}
\STATE \quad $(B(p)^{t+1}, Obj_{t+1}) \gets$ Solve $MP_{\beta}$ with respect to $B(p)$ while fixing $\Gamma^t$ \label{algo_myalgo_3}
\STATE \quad $(\Gamma^{t+1}, Obj_{t+1}) \gets$ Solve $MP_{\gamma}$ with respect to $\Gamma$ while fixing $B(p)^{t+1}$  \label{algo_myalgo_4}
\STATE \quad RelativeGap $\gets |Obj_{t+1}-Obj_{t}| / (Obj_{t})$ \label{algo_myalgo_5}
\STATE \quad $ t \gets t+1 $ \label{algo_myalgo_6}
\ENDWHILE
\STATE $\Gamma^{*} \gets \Gamma^{t}, B(p)^{*} \gets B(p)^{t}, Obj_{*} \gets Obj_{t}$
\end{algorithmic}
\end{algorithm}

\begin{proposition}\label{proposition2}
Algorithm \ref{algo_myalgo} converges to a partial minimum solution of the model \eqref{mae_derive1}. This convergence and the properties of the solution hold irrespective of whether \(\Gamma\) or \(\Gamma'\) is used in Line \ref{algo_myalgo_4} of the algorithm.
\end{proposition}

\begin{proof}
By Lemma \ref{lemma2}, \(B(p)\) and \(\Gamma\) in IGWR are nonempty and compact sets, corresponding to \(U\) and \(V\) in the ADM respectively. Line \ref{algo_myalgo_3} of Algorithm \ref{algo_myalgo} produces a unique optimal solution due to the tie-breaking rule while Line \ref{algo_myalgo_4} of Algorithm \ref{algo_myalgo} ensures a unique optimal solution by the strict convexity of the objective function \eqref{mae_derive3} as shown in the proof of Lemma \ref{lemma1}. Given these unique optimal solutions for the two subproblems, Theorem \ref{theorem1} guarantees that Algorithm \ref{algo_myalgo} converges to a partial minimum. Furthermore, by Lemma \ref{lemma1}, replacing \(\Gamma\) with \(\Gamma'\) in Line \ref{algo_myalgo_4} does not affect the convergence outcome or the uniqueness of the optimal solution. Thus, the solution \((B(p)^*, \Gamma^*)\) remains invariant under this substitution.
\end{proof}

As demonstrated in Proposition \ref{proposition2}, the same partial minimum solution \((B(p)^*, \Gamma^*)\) is achieved irrespective of whether the optimization in Line \ref{algo_myalgo_4} of Algorithm \ref{algo_myalgo} is performed with \(\Gamma\) or \(\Gamma'\). In practice,  Line \ref{algo_myalgo_4} of Algorithm \ref{algo_myalgo} is solved using \(\Gamma'\) without the necessity to explicitly derive \(M_{\gamma}\) values. However, proving the equivalence of solving the problem with both sets \(\Gamma\) and \(\Gamma'\) is crucial for establishing the convergence results of the ADM because the partitioned sets in the ADM framework must both be nonempty and compact.

Our IGWR algorithm is a mathematical programming-based approach that requires more computation time compared to regularization methods or heuristic approaches. Specifically, Line \ref{algo_myalgo_3} of Algorithm \ref{algo_myalgo} is computationally intensive, while Line \ref{algo_myalgo_4} can be solved relatively quickly. To enhance the overall efficiency of the IGWR algorithm, we employ a warm-start strategy for solving Line \ref{algo_myalgo_3}, leveraging the $z_j$ values from the previous iteration. This approach improves the algorithm's traceability and computational performance. In practice, the computational time remains typically under 10 minutes, making it highly tractable for practical applications.

We next discuss two important issues in running IGWR in Algorithm \ref{algo_myalgo}. First, we need initial values $\gamma_{init}$ to solve $MP_\beta$ in Line \ref{algo_myalgo_3}. Because different initial values of $\gamma_{init}$ can lead to different partial minima, we need good initial values for $\gamma_{init}$. To determine the initial value for bandwidth parameter $\gamma_{init}$, we consider the intercept-only model. In other words, the regression coefficients in $B$ except for the intercept are fixed to zero, and $MP_\gamma$ is solved to determine $\gamma_{init}$. Second, our model can have scaling issues. Note that the objective function \eqref{mae_derive1_a} contains three factors: bandwidth $\gamma_o$, fitting error $e_{oi}$, and distance value $d_{oi}$. When the distance value $d_{oi}$ is relatively larger than the other two factors, the second term in \eqref{mae_derive1_a} tends to be zero because the denominator becomes larger than the numerator. This makes the first term in \eqref{mae_derive1_a} dominant, and $\gamma_o$ tends to turn to zero. The distance value $d_{oi}$ should be rescaled to have small values to avoid this scaling issue. Therefore, we scale the distance by $d_{oi} = 
d_{oi} / \max_{i,o} \{d_{oi}\}$, which transforms the distances to the range between 0 and 1. 

\subsection{Comparison of Models}\label{section3.3}
In this section, we compare the characteristics of the proposed model with four benchmarks in the literature: (1) basic GWR (BGWR) (2) FS approach based on BGWR \citep{fotheringham2013demographic, gollini2015gwmodel}, (3) multiscale GWR (MGWR) \citep{fotheringham2017multiscale, oshan2019mgwr}, and (4) geographically weighted Lasso (GWL) \citep{wheeler2009simultaneous}. The FS approach iteratively adds IVs when there is an improvement, where each iteration solves the standard GWR problem. The MGWR is designed to apply different bandwidths for each DV and IV pair. The GWL method is a combination of GWR and Lasso to select IVs and reduce multicollinearity.

Table \ref{tab:table1} presents the summary of the comparison. The rows represent the characteristics, which include the goal of the model, the objective functions for the regression coefficient and bandwidth parameter estimations, algorithms for the regression coefficient and bandwidth parameter estimations, the subset selection model, and bandwidth type. According to the table, the proposed model has three distinct characteristics compared with the benchmarks. First, it integrates the coefficient and bandwidth estimation steps using a single objective function, while the other approaches use inconsistent objective functions in the two steps. Second, our model is capable of selecting a consistent subset for all focal points, while other regularization-based methods for GWR \citep{ wheeler2009simultaneous, wheeler2007diagnostic, li2018geographically} or a hyper-local method \citep{comber2018hyper} return a local and inconsistent subset for each focal point. The FS approach determines the global subset, but it is a heuristic and does not guarantee optimality. Third, our model can work with both global and local bandwidth settings, while most approaches in the literature only allow the global bandwidth setting except for MGWR. One limitation is that our model is restricted to using an exponential kernel function, which is necessary for deriving our formulation. Nonetheless, the proposed model offers a flexible and consistent approach in terms of the objective function, subset selection, and bandwidth type.

\begin{table}[htbp]
  \centering
  \small
  \footnotesize
  \setlength{\tabcolsep}{3pt}
  \caption{Comparison of Proposed and Benchmark Models} \label{tab:table1}
  \scalebox{0.8}{
    \begin{tabular}{c|c|c|c|c|c}
    \toprule
          & Proposed model (IGWR) & Basic GWR (BGWR) & FS & MGWR  & GWL \\ \hline
    Goals & \makecell{Simultaneous coefficient and \\ bandwidth estimation with \\subset selection} & \makecell{Capturing spatially \\ varying pattern}  & Subset selection  & \makecell{Varying bandwidth for \\ each DV and IV pair} & \makecell{Reducing \\ multicollinearity}  \\ \hline
    \makecell{Objective\\ (coefficient)} & \makecell{Integrated maximum \\ likelihood function} & \multicolumn{4}{c}{Weighted sum of squares} \\ \hline
    \makecell{Objective\\ (bandwidth)} & \makecell{Integrated maximum \\ likelihood function} & \multicolumn{4}{c}{CV score or AICc value} \\ \hline
    \makecell{Algorithm \\ (coefficient)} & ADM $MP_{\beta}$ & \multicolumn{4}{c}{Weighted least squares} \\ \hline
    \makecell{Algorithm \\ (bandwidth)} & ADM $MP_{\gamma}$ & \multicolumn{4}{c}{Golden section search or binary search} \\ \hline 
    Subset selection & MIQP & N/A & Forward selection & None & Lasso \\ \hline 
    Selected subset & Homogeneous & N/A & Homogeneous & None & Heterogeneous \\  \hline
    Bandwidth type & \makecell{Global or local \\ (varies by focal point)} & N/A & Global & \makecell{Multiscale \\ (for each IV)} & Global \\ \hline
    Bandwidth function & \makecell{Exponential function} & Various functions & Various functions & Various functions & Various functions \\ \bottomrule
    \end{tabular}%
  }
\end{table}%
\vspace{-0.5cm}

\section{Computational Experiments}\label{section4}
In this section, we present computational experiments for the proposed algorithms and compare their performance with benchmark models. In Section \ref{section4.1}, we introduce three datasets—Georgia, Ohio, and the US Census—used in the experiments. We analyze the GWR outputs in detail using the Georgia and Ohio datasets, while the US Census dataset is primarily used to compare performance metrics with benchmark methods. Section \ref{section4.2} details the implementation of the IGWR algorithm and checks the effect of input parameters of the IGWR algorithm using the Georgia dataset, one of the most widely studied datasets. Section \ref{section4.3} compares our proposed model and benchmark models in terms of explanatory power and spatial patterns, focusing on the Georgia dataset. Section \ref{section4.4} exhibits the differences in subset selection results between the IGWR algorithm and the FS approach for the Ohio dataset. Lastly, we present comprehensive computational experiments on all models using the three datasets, which highlight the generalization of each model's characteristics.

For the computational experiments, we employ Intel(R) Core(TM) i5-8600 CPU @ 3.10GHz (6 CPUs) and 32GB RAM. The proposed model and the algorithm are implemented with Python, where Gurobi 11.0.0. is used to solve $MP_{\beta}$ and the \texttt{scikit-learn} package in Python \citep{pedregosa2011scikit} is used to solve $MP_{\gamma}$. We use four benchmark algorithms summarized in Table \ref{tab:table1}: BGWR, FS, MGWR, and GWL. For BGWR and FS, we use the \texttt{gwr.basic} function of the \texttt{GW model} package in R \citep{gollini2015gwmodel}. For MGWR, we utilize the \texttt{mgwr} package in Python \citep{oshan2019mgwr}. For GWL, we use the \texttt{gwrr} package in R \citep{wheeler2009simultaneous}. For FS, instead of using the \texttt{model.selection.gwr} function, we implement a revised FS approach. This adjustment is necessary because \texttt{model.selection.gwr} retains the bandwidth calibrated when $p=1$ without updating it, even though different IVs require different bandwidths. Our revised FS procedure addresses this issue. The codes and data are available in the online supplement. Lastly, it is important to note that GWR is typically used to capture spatial patterns rather than for prediction, and we adhere to this practice.

\subsection{Datasets}\label{section4.1}
 The first dataset, which we refer to as Georgia, includes 159 observations, six IVs, and a DV, with each observation representing a county. Georgia is one of the most popular datasets for GWR, originally introduced by Fotheringham et al. \cite{fotheringham2003geographically}. This dataset is widely used, as it is provided by default in the \texttt{GW model} and \texttt{mgwr} packages. The analysis aims to explain the percentage of the population with a bachelor's degree using county socioeconomic and demographic information. Table \ref{tab:table2} lists the variables.

\begin{table}[htbp]
  \centering
  \small
    \caption{Variables in Georgia Dataset}
    \label{tab:table2}
    \scalebox{0.8}{
    \begin{tabular}{c|c|l}
    \toprule
    Variable name & Class & Description \\ \midrule 
    $PctBach$ & DV & Percentage of the county population with a bachelor’s degree \\ \midrule 
    $TotPop90$ & \multirow{6}{*}{IV} & Population of the county in 1990 \\
    $PctRural$ & & Percentage of the county population defined as rural \\
    $PctEld$ & & Percentage of the county population aged 65 or over \\
    $PctFB$ & & Percentage of the county population born outside the United States\\
    $PctPov$ & & Percentage of the county population living below the poverty line \\
    $PctBlack$ & & Percentage of the county population that is black \\ \bottomrule
    \end{tabular}%
    }
\end{table}%

In this study, we present a new dataset for GWR with more IVs in a different application. The new dataset, which we refer to as Ohio, includes art and cultural organizations' expenditure, demographics, and socioeconomic information for each county in Ohio. The data came from the database of Southern Methodist University DataArts. The Ohio dataset includes 88 observations, 15 IVs, and a DV (See Table \ref{tab:table3} for the variables). The goal is to estimate the geographically varying effect of county demographic and socioeconomic variables on the county per-organization expenditure. Doing so will show which IVs are affecting the art organization's expenditure and help policy makers and funding agencies make decisions. 

\begin{figure}[h!]
\begin{center}
    \subfigure[DV: \textit{PctBach} in Georgia dataset]{%
        \includegraphics[height=3.5cm]{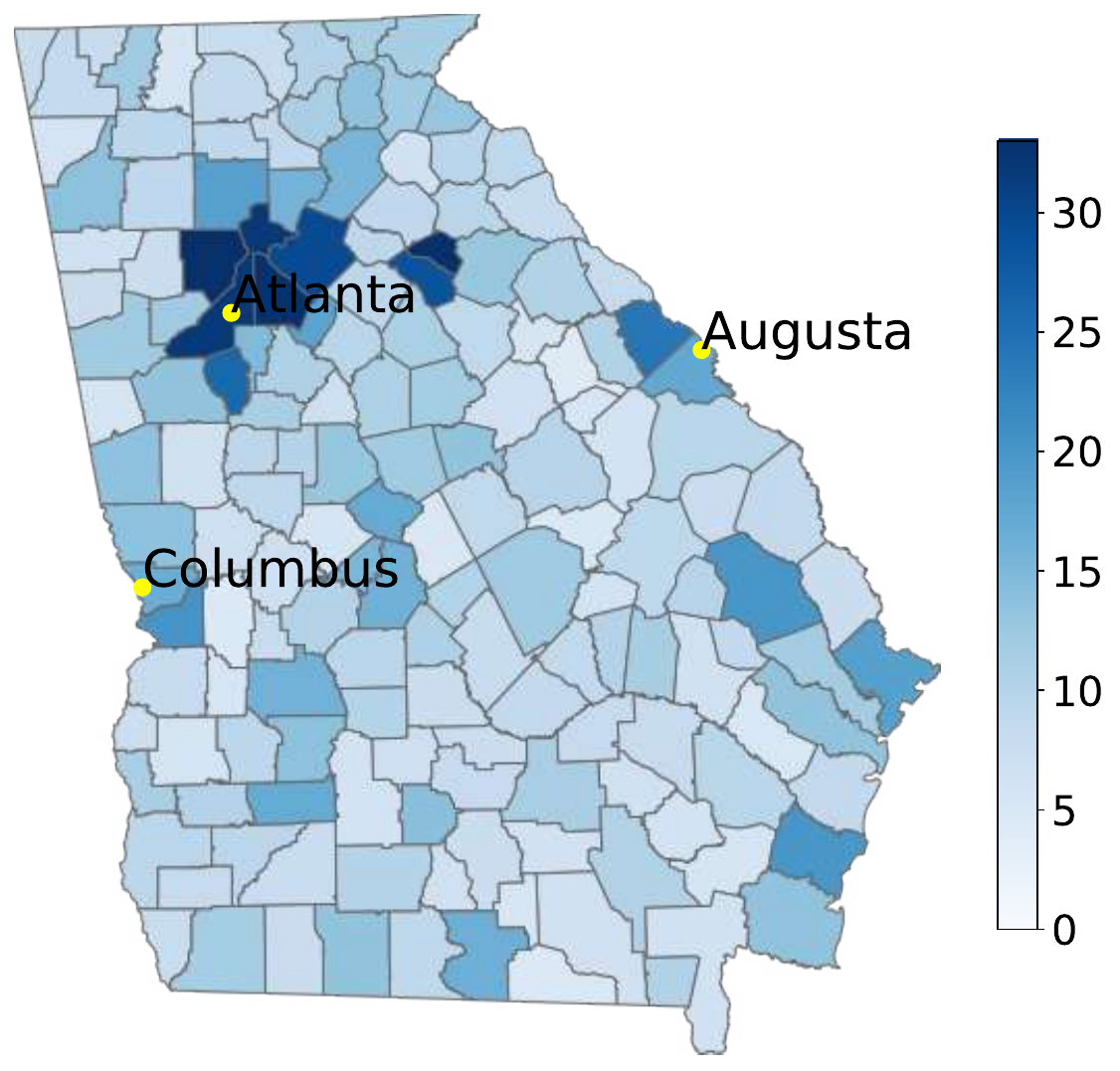} \label{heatmap_georgia}
        }\qquad \quad
    \subfigure[DV: \textit{cfexp\_per\_org} in Ohio dataset]{%
        \includegraphics[height=3.5cm]{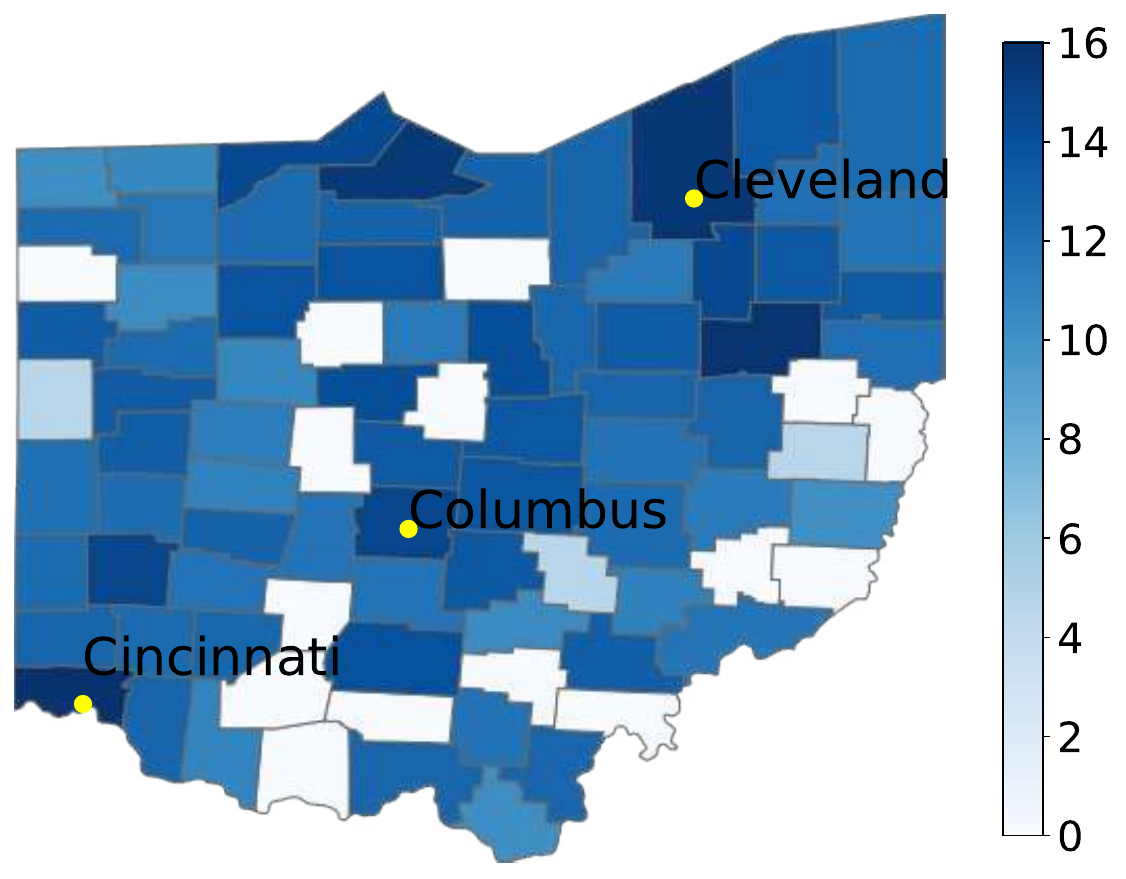} \label{heatmap_ohio}
    }
\end{center}
\vspace{-0.5cm}
    \caption{Heatmap of DV for Georgia and Ohio Datasets}
\label{Heatmap}
\end{figure}

\begin{figure}[h!]
\begin{center}
    \subfigure[Georgia dataset]{%
        \includegraphics[height=5.5cm]{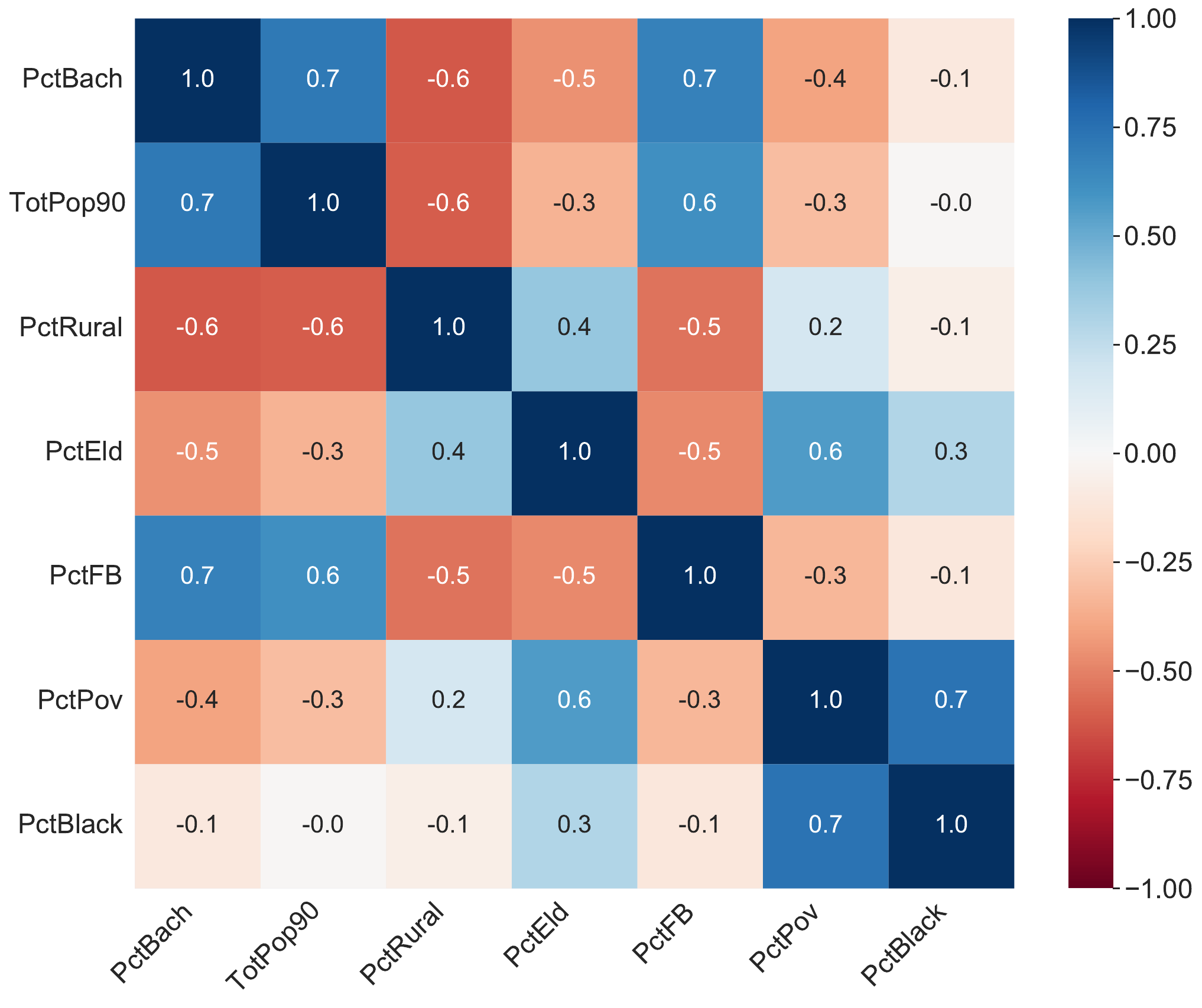} \label{heatmap_corr_georgia}
        }\qquad \quad
    \subfigure[Ohio dataset]{%
        \includegraphics[height=5.5cm]{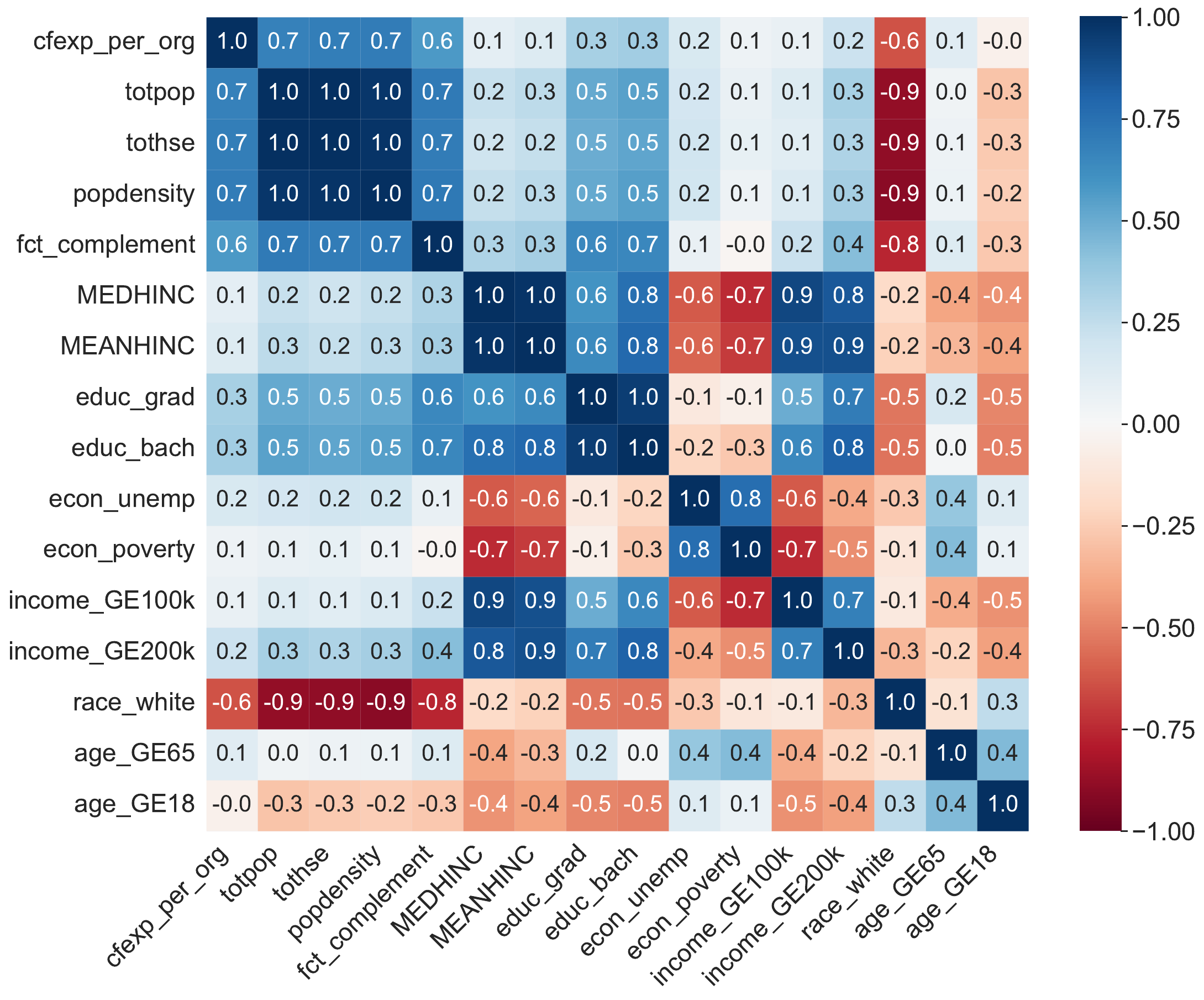} \label{heatmap_corr_ohio}
    }
\end{center}
\vspace{-0.5cm}
    \caption{Heatmap of Correlation Matrices}
\label{Heatmap_corr}
\end{figure}

Figure \ref{Heatmap} shows the heatmaps of the DV for the Georgia and Ohio datasets. For each map, yellow circles indicate the three largest cities in each state. Figure \ref{Heatmap_corr} shows the heatmaps of the correlation matrix for the two datasets. The correlation matrix of Ohio dataset in Figure \ref{heatmap_corr_ohio} shows many correlated IV pairs and indicates that subset selection is necessary. The Ohio dataset has more correlated IVs than the Georgia dataset, which implies that the subset selection procedure is required.

\begin{table}[h!]
  \centering
  \small
	\caption{Variables in Ohio Dataset}
	\label{tab:table3}
    \scalebox{0.8}{
    \begin{tabular}{c|c|l}
    \toprule
    Variable name & Class & Description \\ \midrule
    $cfexp\_per\_org$ & DV & Total expenditure per organization in the county [log-scaled] \\ \midrule 
    $totpop$ & \multirow{15}{*}{IV} & Total population of the county [log-scaled] \\
    $tothse$ & & Total household of the county [log-scaled] \\
    $popdensity$ & & Population density in the county [log-scaled] \\
    $fct\_complement$ & & Factor score for complement facilities (bar, restaurant, accommodation) in the county \\
    $medhinc$ & & Median household income (in thousands of dollars) in the county [log-scaled] \\
    $meanhinc$ & & Mean household income (in thousands of dollars) in the county [log-scaled] \\
    $educ\_grad$ & & Percentage of population with a graduate degree in the county \\
    $educ\_bach$ & & Percentage of population with bachelor's degree or above in the county \\
    $econ\_unemp$ & & Percentage of unemployed population in the county\\
    $econ\_poverty$ & & Poverty rate in percentage in the county \\
    $income\_GE100k$ & & Percentage of population with income $\geq$ 100k in the county \\
    $income\_GE200k$ & & Percentage of population with income $\geq$ 200k in the county \\
    $race\_white$ & & Percentage of the white population in the county \\
    $age\_GE65$ & & Percentage of population aged $\geq$ 65 in the county \\
    $age\_LE18$ & & Percentage of population aged $\leq$ 18 in the county \\ \bottomrule
    \end{tabular}%
    }
\end{table}%

The third dataset, the US Census County-level dataset, serves as additional instances to generalize the characterization of all models. This dataset, sourced from publicly available data at \url{https://www.ers.usda.gov/data-products/county-level-data-sets/}, consists of 11 columns (independent variables) and 3132 rows (counties). For this study, we filtered the data by state and selected the top 20 mainland states based on the number of counties. These states include: ``TX", ``GA", ``VA", ``KY", ``MO", ``KS", ``IL", ``NC", ``IA", ``TN", ``NE", ``IN", ``OH", ``MN", ``MI", ``MS", ``OK", ``AR", ``WI", and ``AL", represented by their abbreviations. Table \ref{tab:table_us_census} presents the DV and IVs for the US Census dataset.

\begin{table}[h!]
  \centering
  \small
	\caption{Variables in US Census Dataset}
	\label{tab:table_us_census}
    \scalebox{0.8}{
    \begin{tabular}{c|c|l}
    \toprule
    Variable name & Class & Description \\ \midrule
    $MEDINC$ & DV & Estimate of median household income, 2021 (divided by 1000) \\ \midrule 
    $RUC\_code$ & \multirow{11}{*}{IV} & Rural-Urban Continuum Code, 2013 \\
    $Pctpovall$ & &Estimated percent of people of all ages in poverty 2021  \\
    $Metro$ & & Metro nonmetro dummy 0=Nonmetro 1=Metro (Based on 2013 OMB Metropolitan Area delineation)  \\
    $Civilian\_labor\_force$ & & Civilian labor force annual average, 2021 [log-scaled] \\
    $Unemployment\_rate$ & & Unemployment rate, 2021  \\
    $Pct\_Adults\_Less\_High$ & & Percent of adults with less than a high school diploma, 2017-21 \\
    $Pct\_Adults\_Bachelor$ & & Percent of adults with a bachelor's degree or higher, 2017-21 \\
    $Economic\_typology$ & & County economic types, 2015 edition. Non-overlapping economic-dependence county indicator \\
    $POP\_ESTIMATE$ & & Resident total population estimate, 7/1/2021 [log-scaled] \\
    $R\_NATURAL\_CHG$ & & Natural increase rate in period 7/1/2020 to 6/30/2021 \\
    $R\_NET\_MIG$ & & Net migration rate in period 7/1/2020 to 6/30/2021 \\ \bottomrule
    \end{tabular}%
    }
\end{table}%

\subsection{Implementation of IGWR with Global and Local Bandwidth} \label{section4.2}
\begin{figure}[h!]
\begin{center}
    \subfigure[Global Bandwidth Setting]{%
        \includegraphics[scale=0.2]{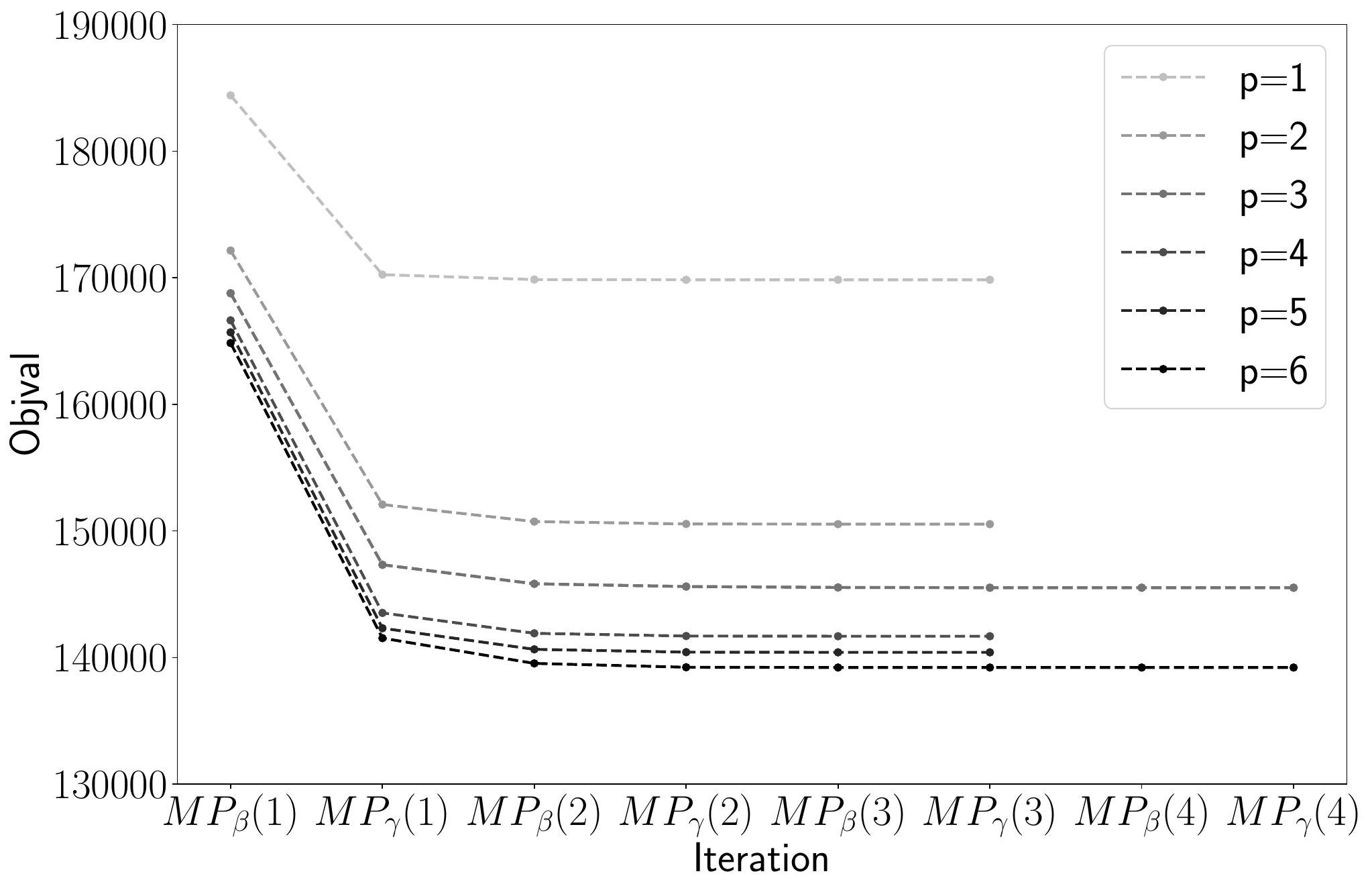} \label{example_bandwidth_global}
        }\qquad \quad
    \subfigure[Local Bandwidth Setting]{%
        \includegraphics[scale=0.2]{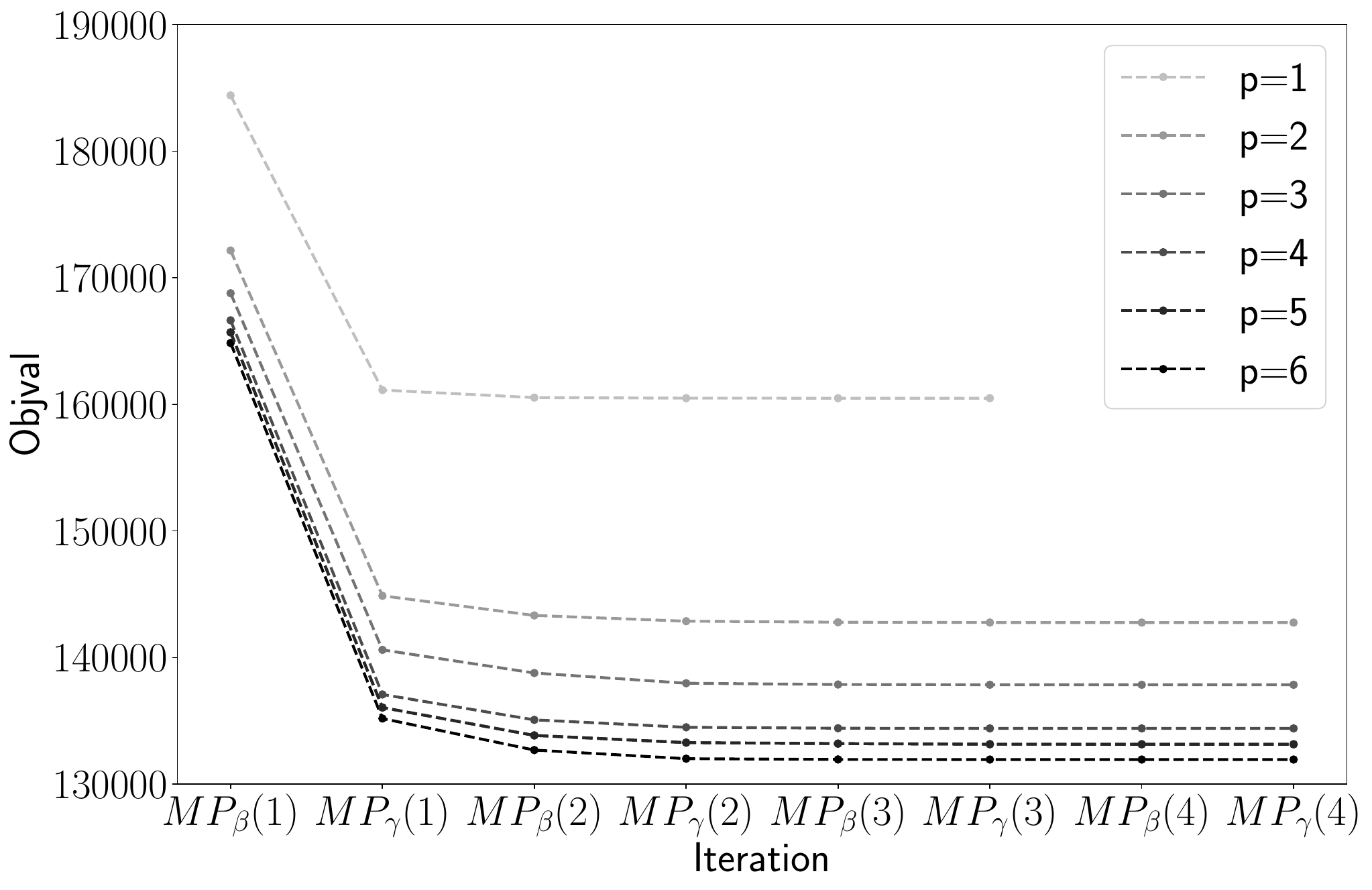} \label{example_bandwidth_local}
    }
\end{center}
\vspace{-0.5cm}
    \caption{IGWR Convergence Result for Georgia Dataset}
\label{ACS Convergence Result}
\end{figure}

This section details the implementation of the IGWR algorithm with both global and local bandwidth settings, denoted as IGWR-G and IGWR-L, respectively. Utilizing the Georgia dataset, we examine the performance for $p \in \{1,2,3,4,5,6\}$. Figure \ref{ACS Convergence Result} displays the objective function values across iterations, illustrating the algorithms' convergence. The horizontal axis denotes iterations, labeled with $MP_{\beta}(t)$ and $MP_{\gamma}(t)$, which represent the objective function values at iteration $t$ of Algorithm \ref{algo_myalgo}, as detailed in Lines \ref{algo_myalgo_3} and \ref{algo_myalgo_4}. As expected given the result in Section \ref{section3.2}, both global and local settings show non-increasing objective function values over iterations for all $p$ values. In general, the first few iterations show the greatest improvements, and we use the remaining iterations to fine-tune the model. 

\begin{figure}[h!]
\begin{center}
    \subfigure[Global Bandwidth Setting]{%
        \includegraphics[height=3.5cm]{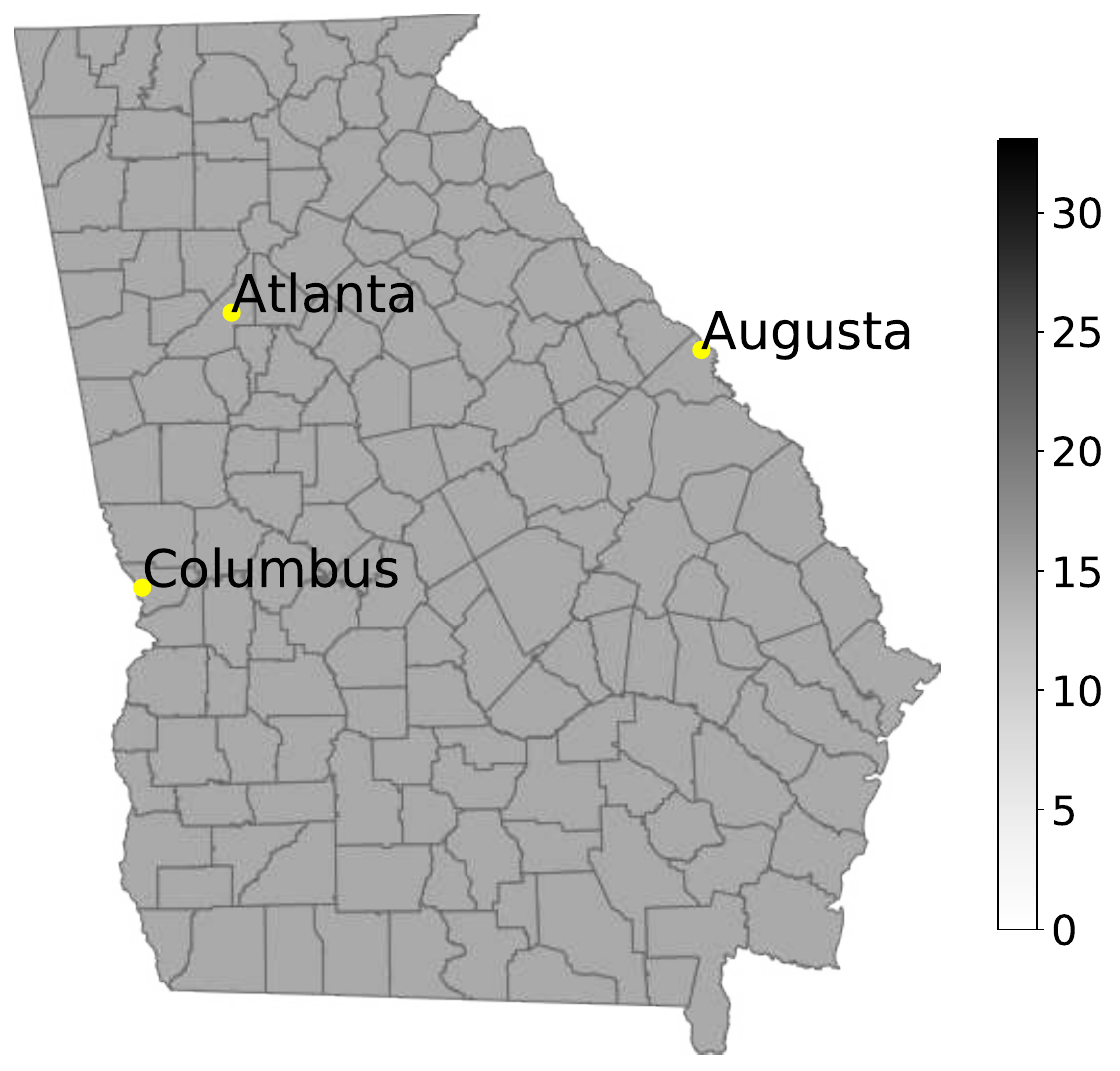} \label{georgia_bandwidth_global}
        }\qquad \quad
    \subfigure[Local Bandwidth Setting]{%
        \includegraphics[height=3.5cm]{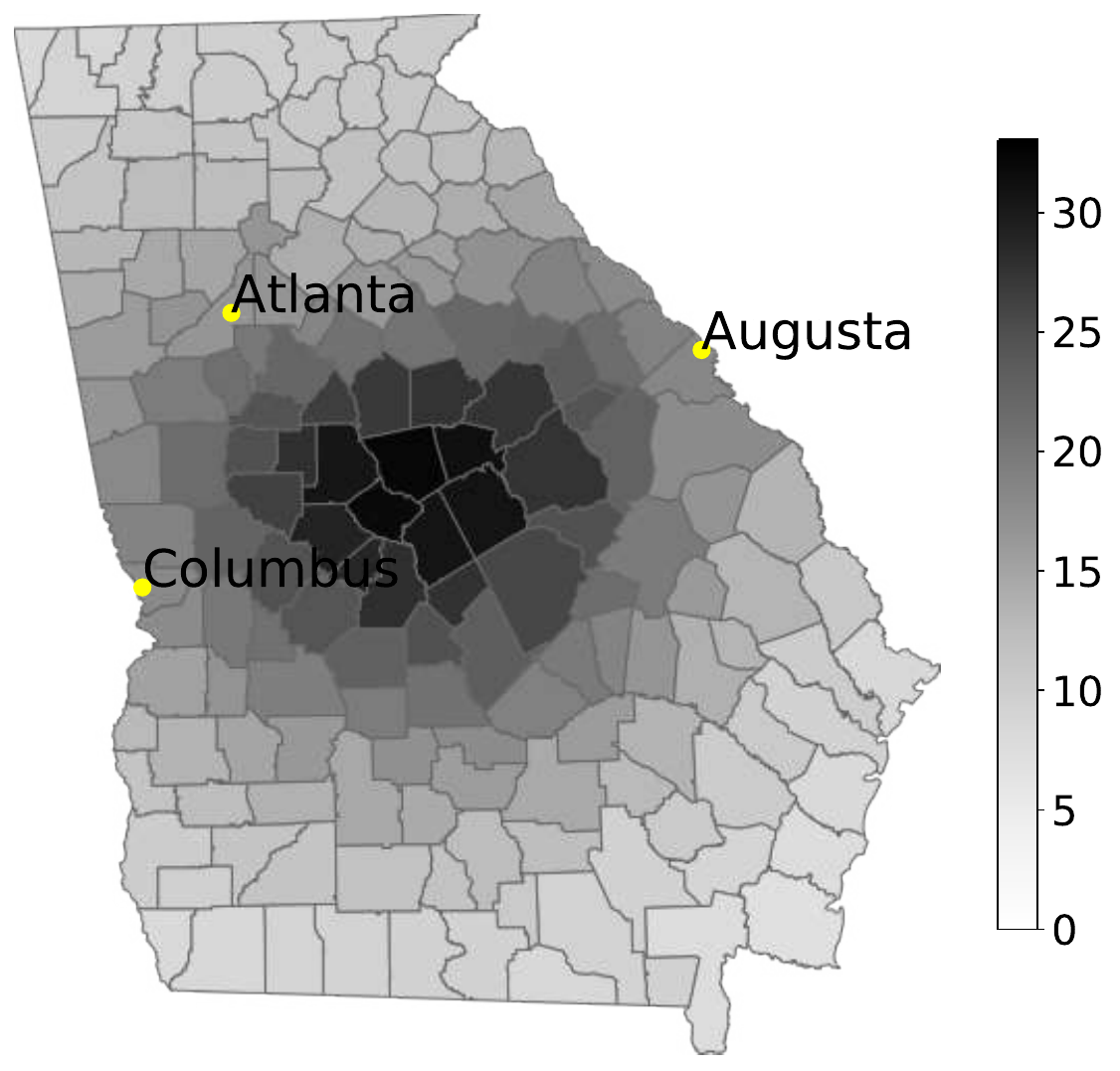} \label{georgia_bandwidth_local}
    }
\end{center}
\vspace{-0.5cm}
    \caption{IGWR-G and IGWR-L Bandwidths for Georgia Dataset}
\label{Bandwidth Comparison}
\end{figure}

In Table \ref{tab:table4}, we present the objective function values (Obj), residual sum of squares (RSS), R square ($R^2$), and $R^2_{\text{adj}}$ values for each bandwidth setting and $p \in \{1,2,3,4,5,6\}$. In the selected variables column, the ``G" and ``L" marks represent the selected IVs for IGWR-G and IGWR-L, respectively. The subset selection result indicates both algorithms select identical subsets for all $p$ values and the global and local settings do not make any difference in the subset selection decision. Overall, both IGWR-G and IGWR-L algorithms perform similarly with slight variations of objective values or RSS. In the Georgia dataset, the objective function values of IGWR-L are lower than those of IGWR-G, whereas IGWR-G has lower RSS values than IGWR-L. This is expected because the local model can obtain tailored optimal parameters for each focal point whereas the global model needs to consider all focal points simultaneously. The larger RSS values of IGWR-L may imply that the model is overfitted. However, the local model still provides valuable insight into spatially varying patterns. In Figure \ref{Bandwidth Comparison}, we present the estimated bandwidths of the two models. Recall that a higher $\gamma_o$ implies that the nearer locations have larger weights. As illustrated in Figure \ref{Bandwidth Comparison}, we can interpret this as the “closer to the center of Georgia, the more information from the nearby area is used.”

\begin{table}[h!]
  \centering
  \small
  \caption{IGWR-G and IGWR-L Results for Georgia Dataset}
  \scalebox{0.8}{
    \begin{tabular}{c|cccccc|cccc|cccc}
    \toprule
          & \multicolumn{6}{c}{Selected variables $^\dagger$}                & \multicolumn{4}{|c}{IGWR-G} & \multicolumn{4}{|c}{IGWR-L} \\ \midrule
    $p$     & FB    & Pop   & Rural & Eld   & Black & Pov   & Obj   & RSS   &  $R^2$ & $R^2_{\text{adj}}$  & Obj   & RSS   & $R^2$ & $R^2_{\text{adj}}$ \\ \cmidrule{1-1} \cmidrule(lr){2-7} \cmidrule(lr){8-11} \cmidrule(lr){12-15} 
1     & G L     &       &       &       &       &      & 169837 & 2020  & 0.606 & 0.603 & 160486 & 2054  & 0.599 & 0.597\\
    2     & G L     & G L    &       &       &       &       & 150535 & 1592  & 0.690 & 0.686 & 142759 & 1631  & 0.682 & 0.678 \\
    3     & G L   & G L    & G L    &       &       &       & 145514 & 1479  & 0.711 & 0.706 & 137846 & 1509  & 0.706 & 0.700 \\
    \textbf{4}     & \textbf{G L}     & \textbf{G L}    & \textbf{G L}     & \textbf{G L}    &       &       & \textbf{141682} & \textbf{1393}  & \textbf{0.728} & \textbf{0.721} &\textbf{134400} & \textbf{1419}  & \textbf{0.723} & \textbf{0.716}\\
    5     & G L    & G L    & G L    & G L     & G L     &       & 140409 & 1358  & 0.735 & 0.726 & 133148 & 1397  & 0.728 & 0.719\\
    6     & G L    & G L    & G L    & G L     & G L     & G L     & 139208 & 1325  & 0.742 & 0.731 & 131941 & 1340  & 0.739 & 0.728 \\ \bottomrule
    \end{tabular}%
    }
  \label{tab:table4}
  \scalebox{0.8}{
  \begin{tabular}{p{15.5cm}}
\begin{scriptsize}$\dagger$ We shortened the column names given space limitations. The full original column names are $PctFB$, $TotPop90$, $PctRural$, $PctEld$, $PctBlack$, and $PctPov$. The boldface indicates the selected subset cardinality.\end{scriptsize}
\end{tabular}
}
\end{table}%

\begin{figure}[h!]
\begin{center}
    \subfigure[IGWR-G]{%
        \includegraphics[height=4.0cm]{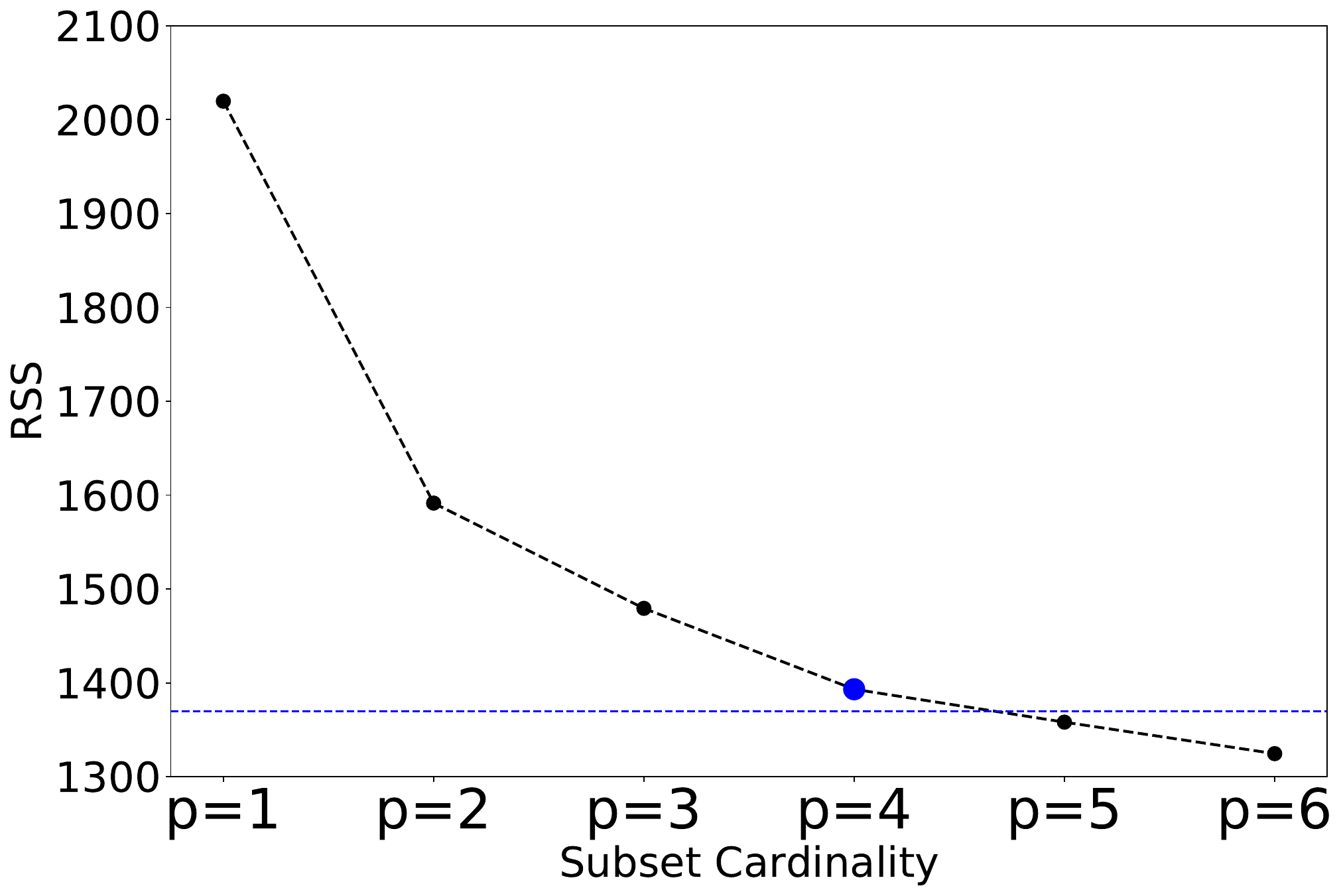} \label{scree_georgia_global}
        } \qquad
    \subfigure[IGWR-L]{%
        \includegraphics[height=4.0cm]{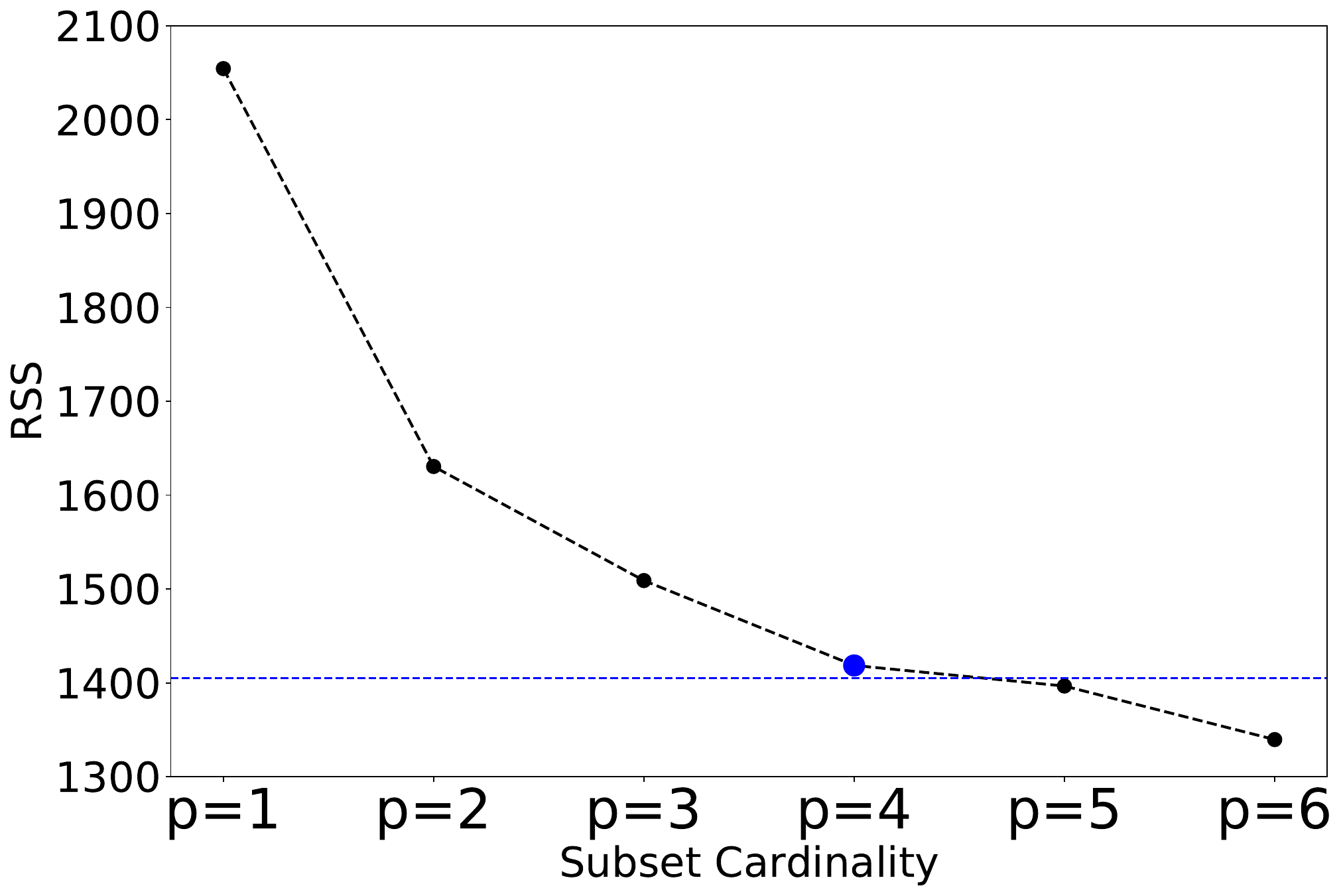} \label{scree_georgia_local}
    }
\end{center}
\vspace{-0.5cm}
    \caption{IGWR-G and IGWR-L Line Plots for Georgia Dataset}
\label{Scree plot georgia}
\end{figure}

We can check RSS to determine a reasonable subset cardinality. For both models, we obtain the best RSS values when $p = 6$. However, the improvements in RSS diminish as $p$ increases. Hence, we also check line plots of RSS over the number of selected IVs. Note that, as the number of selected variables increases, RSS may not decrease while the objective function value decreases monotonically. This is because the objective function minimizes the WSSE not the RSS. To determine the best subset using a line plot of RSS, we suggest two rules. First, if RSS monotonically decreases, find the elbow point. Second, if RSS does not decrease with an added variable, select the previous subset. In Figure \ref{Scree plot georgia}, the line plots present the RSS values over $p$, where the blue dots and lines indicate the selected $p$ value and cutting point for RSS. As the line plots show downward curves for both global and local settings, we check the elbow point and determine that $p=4$ is the best subset cardinality for IGWR-G and IGWR-L.

\vspace{-0.3cm}

\subsection{Georgia Dataset Case study: Comparing IGWR with Benchmarks Models} \label{section4.3}

In this section, we use the Georgia dataset to compare the performances of the proposed and benchmark models. Different models employ different subset selection procedures. For our models (IGWR-G and IGWR-L), we check the performances for $p = 4$ (selected as best in Section \ref{section4.2}). The FS approach, following \cite{fotheringham2013demographic} and \cite{gollini2015gwmodel}, stops adding variables when the AICc score begins to decrease. It selects subsets in the same order as shown in Table \ref{tab:table4}, with AICc beginning to decrease at $p=4$. In the GWL, the number of variables that turn to zero varies for each focal point, ranging from 0 to 5. The average of the number of variables is 3.75, which means GWL selects approximately $p \approx 4$. BGWR and MGWR are not capable of selecting a subset, and all IVs are used. Table \ref{tab:table5} presents the RSS, $R^2$, and $R^2_{\text{adj}}$ values for each model.

\begin{table}[htbp]
  \centering
  \small
  \setlength{\tabcolsep}{2pt}
	\caption{Goodness of Fit Comparisons for All Models for Georgia Dataset}
	\label{tab:table5}

    \scalebox{0.9}{
    \begin{tabular}{c | c | c | c | c | c | c}
    \toprule
          & IGWR-G & IGWR-L & FS & BGWR &     MGWR &       GWL \\ \hline
    Fit measures $\backslash \mbox{ } p$ &  $p=4$   &  $p=4$   & $p=4$  & All    & All  & $p \approx 4$ \\ \hline
    RSS   & 1393  &  1419  & 1351  & 1477     & 1031   & 1322 \\
    $R^2$ & 0.728 & 0.723 & 0.736 & 0.712  & 0.798 & 0.742 \\ 
    $R^2_{\text{adj}}$ & 0.721 & 0.716 & 0.730 & 0.700  & 0.815 & 0.735 \\ \bottomrule
    \end{tabular}%
    }
\end{table}%

The result in Table \ref{tab:table5} supports the competitive performance of three global subset selection methods (IGWR-G, IGWR-L, and the FS approach), compared to BGWR. MGWR exhibits the highest explanatory power, followed by GWL. However, high explanatory power alone does not fulfill the primary objective of GWR models, which is to accurately capture local variations without overfitting. Overfitting can result in non-interpretable patterns due to extreme variability in the estimated local coefficients. Consequently, this section focuses on visualizing and comparing the spatial patterns identified by different models to assess their ability to depict realistic geographical variations.

\begin{figure}[h!]
\begin{center}
    \subfigure[IGWR-G]{%
        \includegraphics[height=3.0cm]{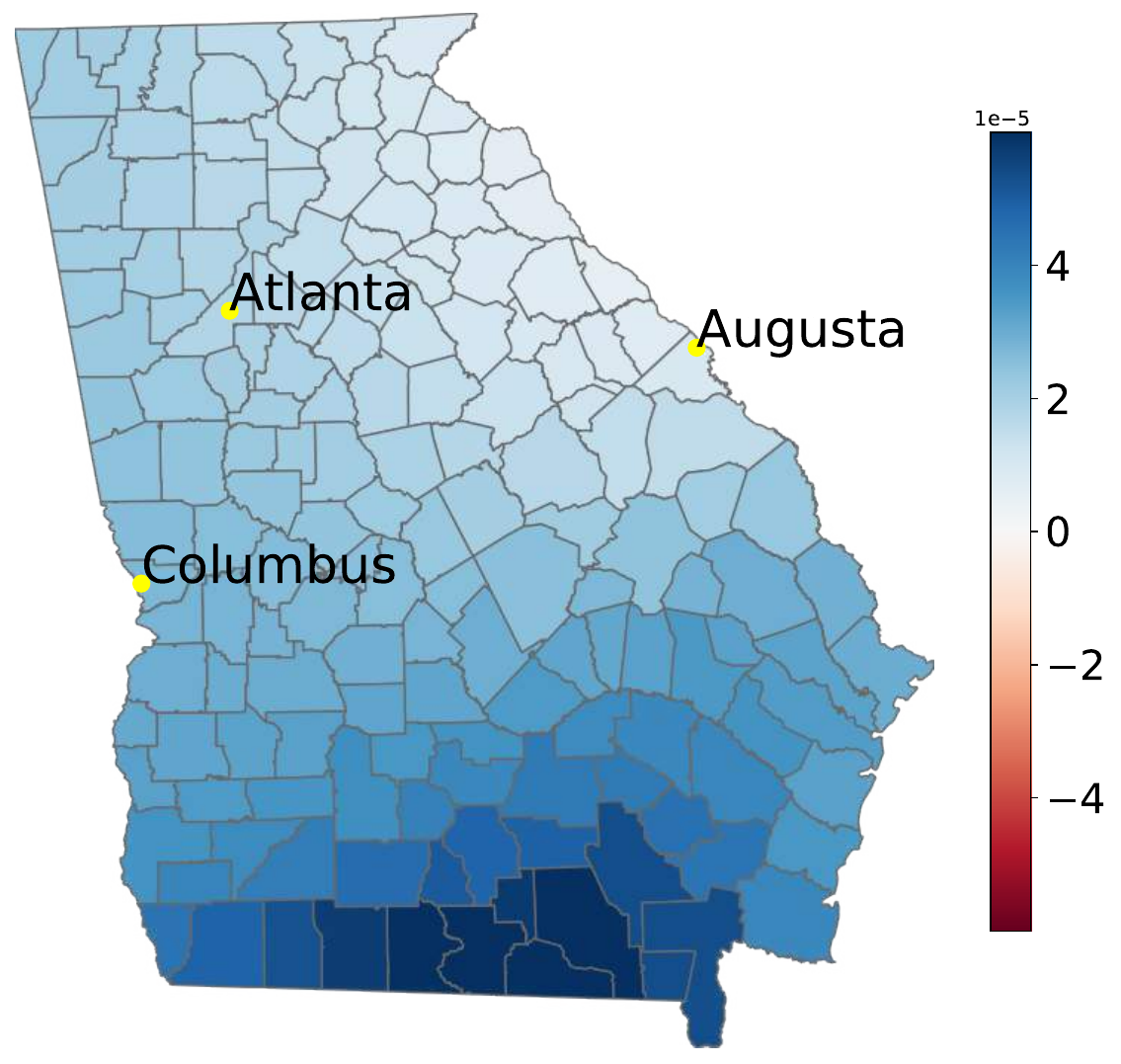} \label{coeff1_global}
        }
    \subfigure[IGWR-L]{%
        \includegraphics[height=3.0cm]{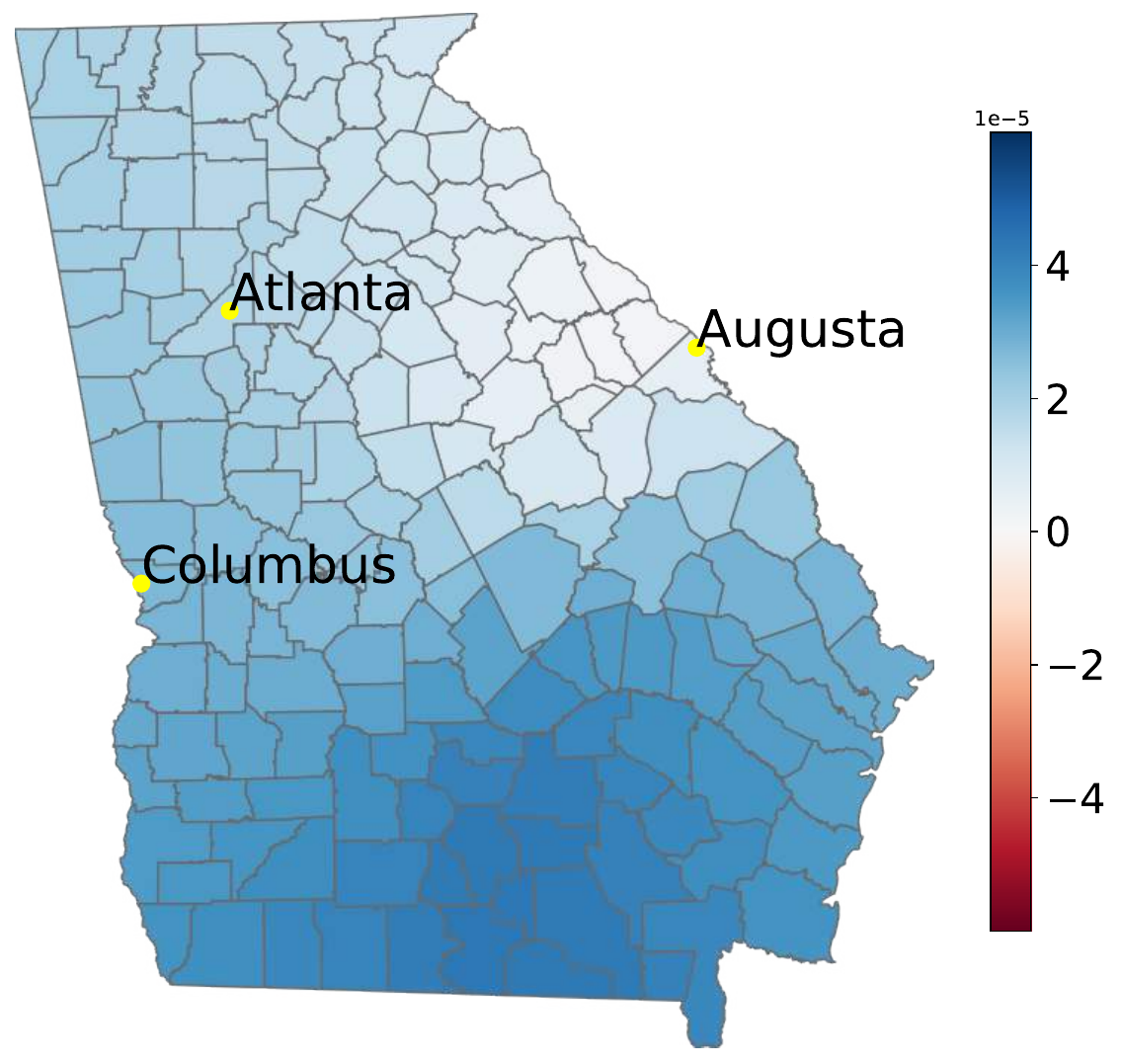} \label{coeff1_local}
    }
    \subfigure[FS]{%
        \includegraphics[height=3.0cm]{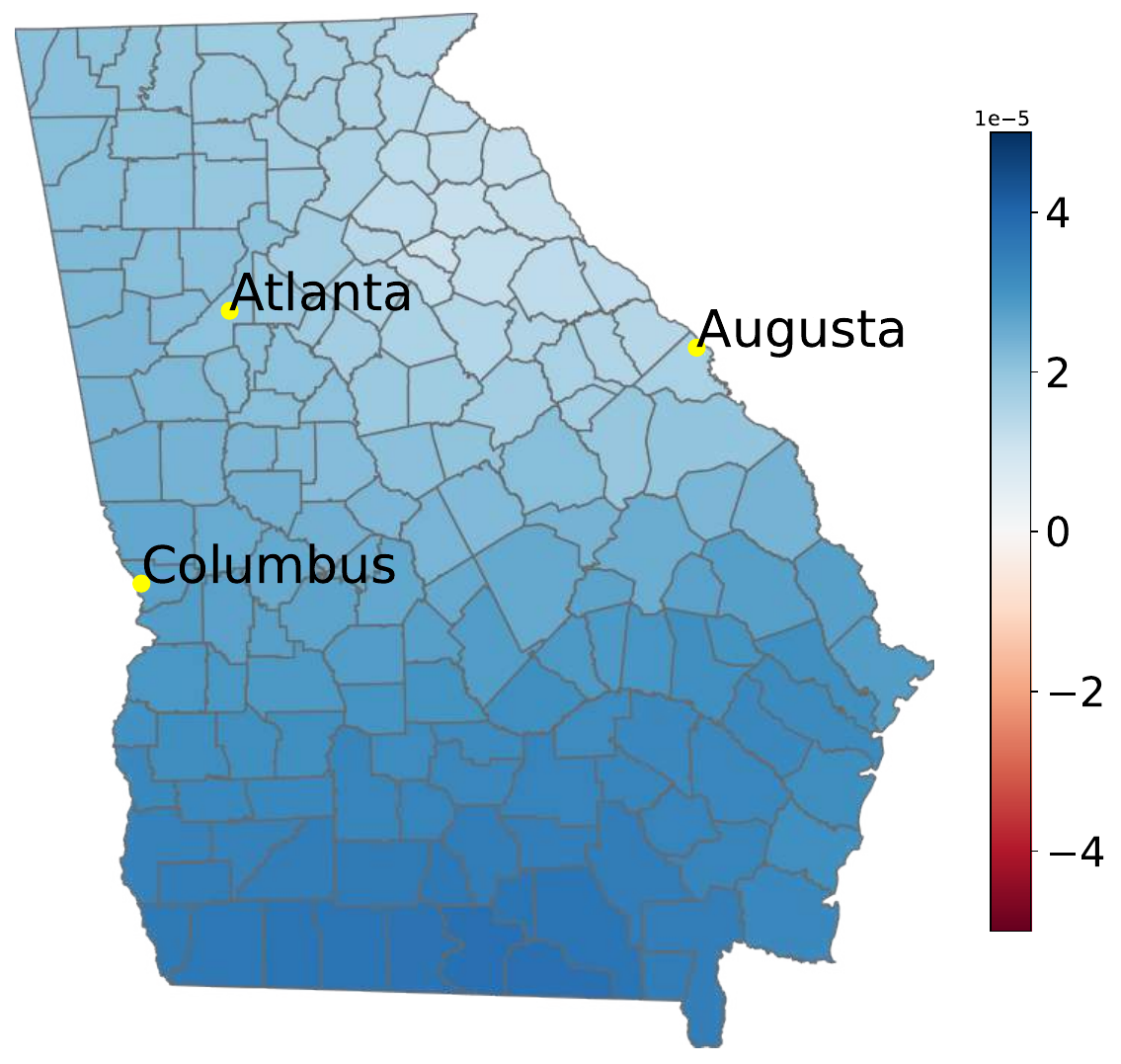} \label{coeff1_forward}
    }
\end{center}
\begin{center}
    \subfigure[BGWR]{%
        \includegraphics[height=3.0cm]{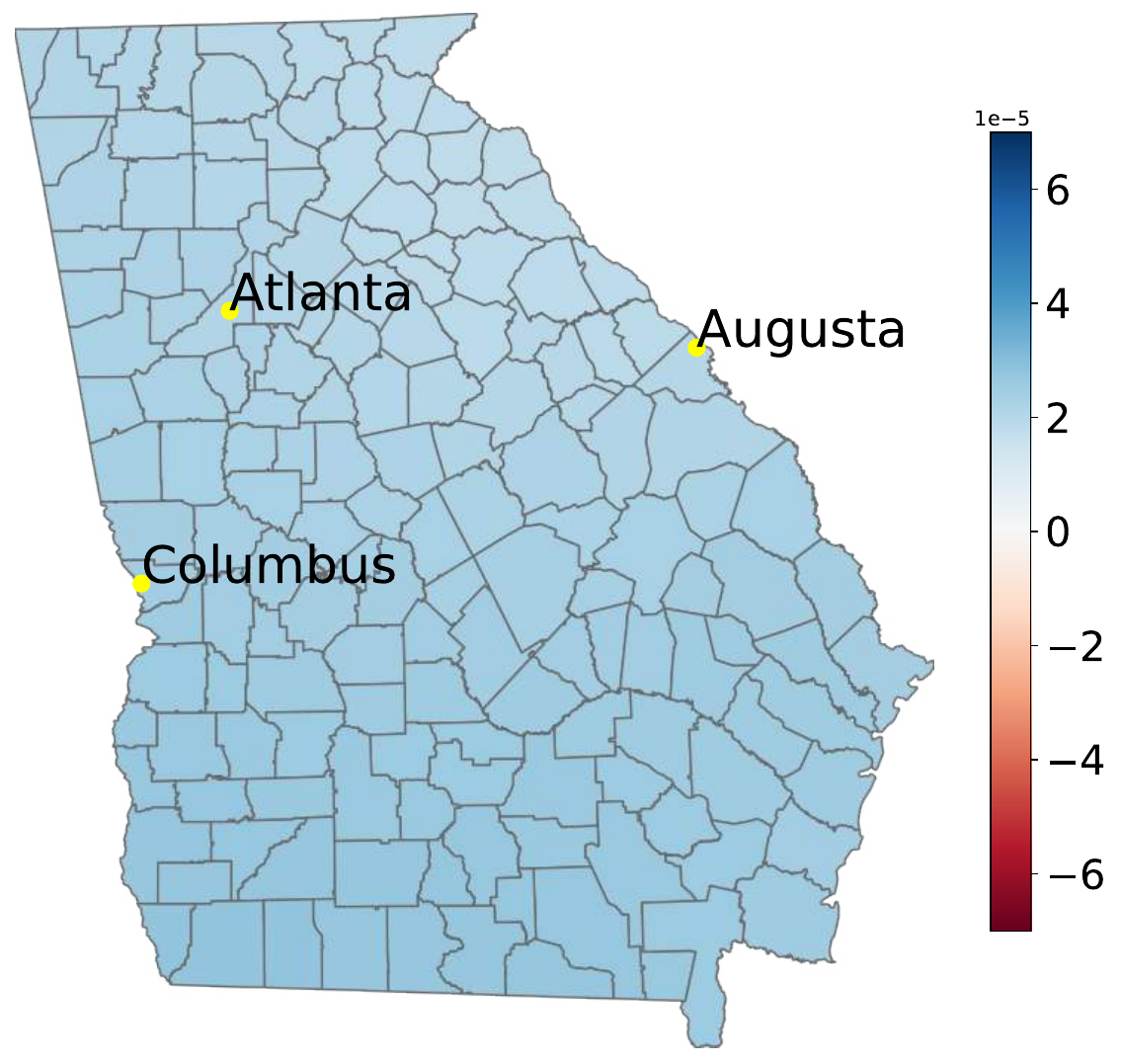} \label{coeff1_gwr}
        }
    \subfigure[MGWR]{%
        \includegraphics[height=3.0cm]{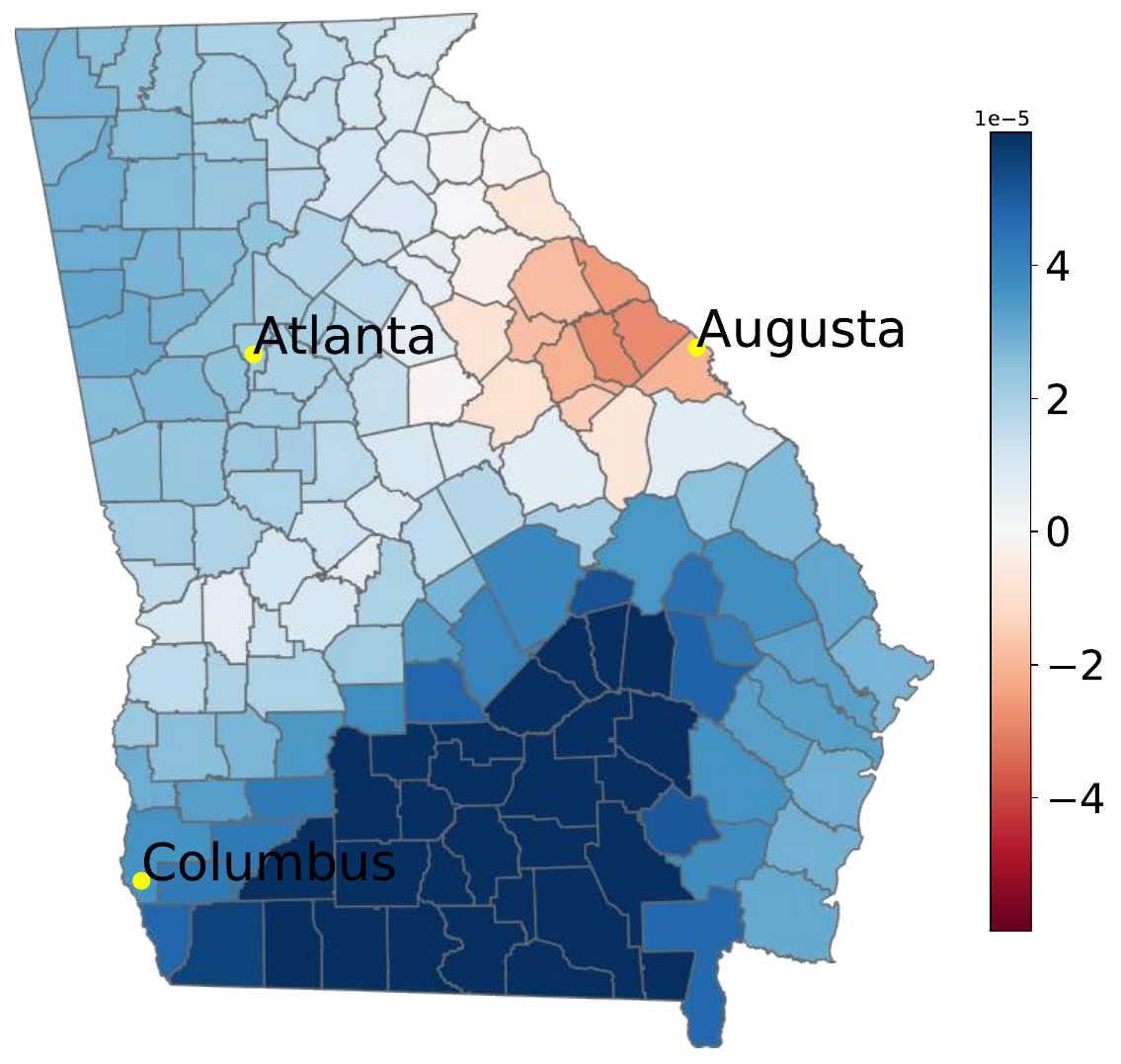} \label{coeff1_mgwr}
        }
    \subfigure[GWL (X: zero)]{%
        \includegraphics[height=3.0cm]{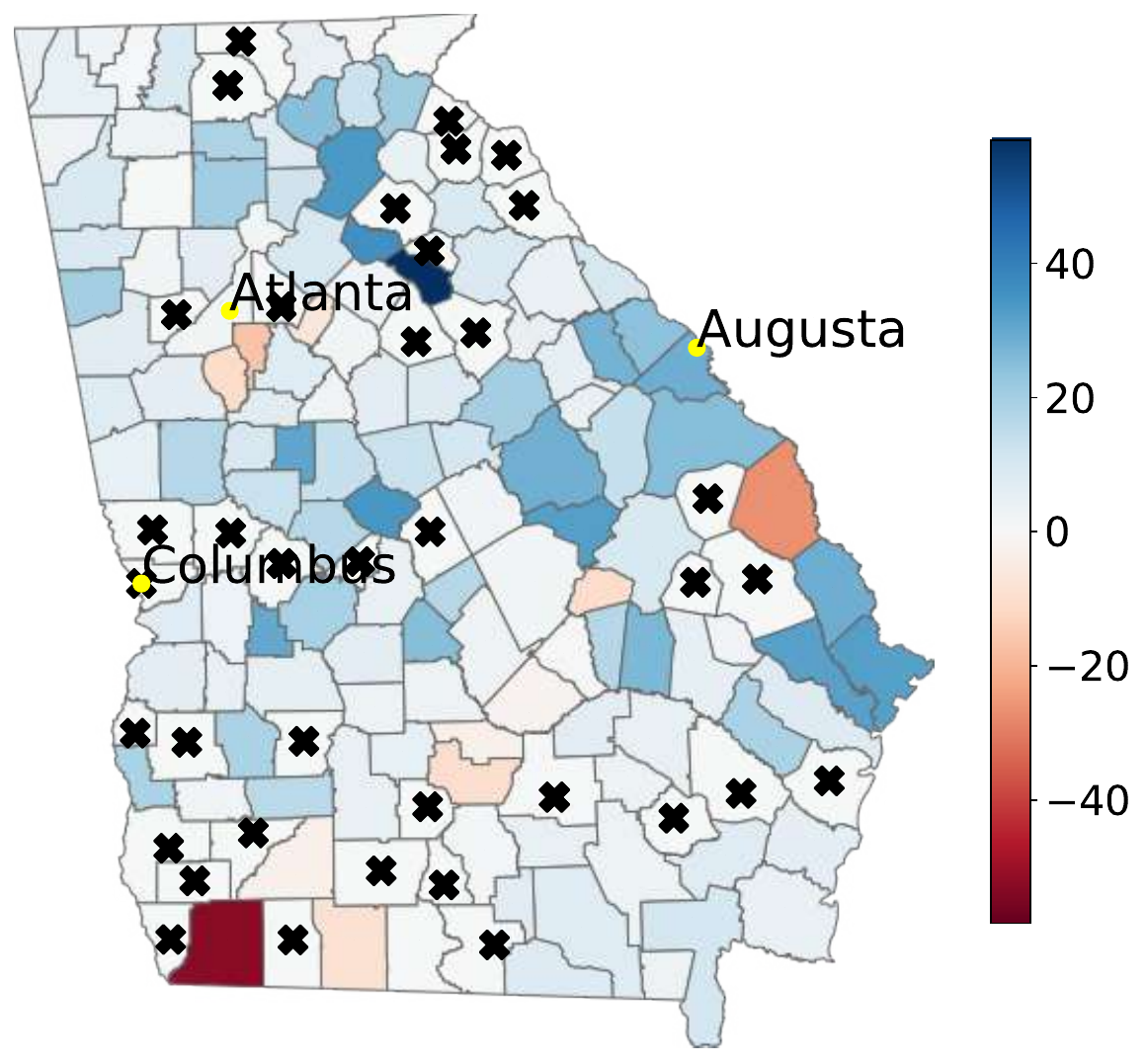} \label{coeff1_gwl}
        }
\end{center}
\vspace{-0.5cm}
    \caption{Coefficients of All Models for TotPop90 of Georgia Dataset}
\label{Comparison of TotPop90}
\end{figure}

\begin{figure}[h!]
\begin{center}
    \subfigure[IGWR-G]{%
        \includegraphics[height=3.0cm]{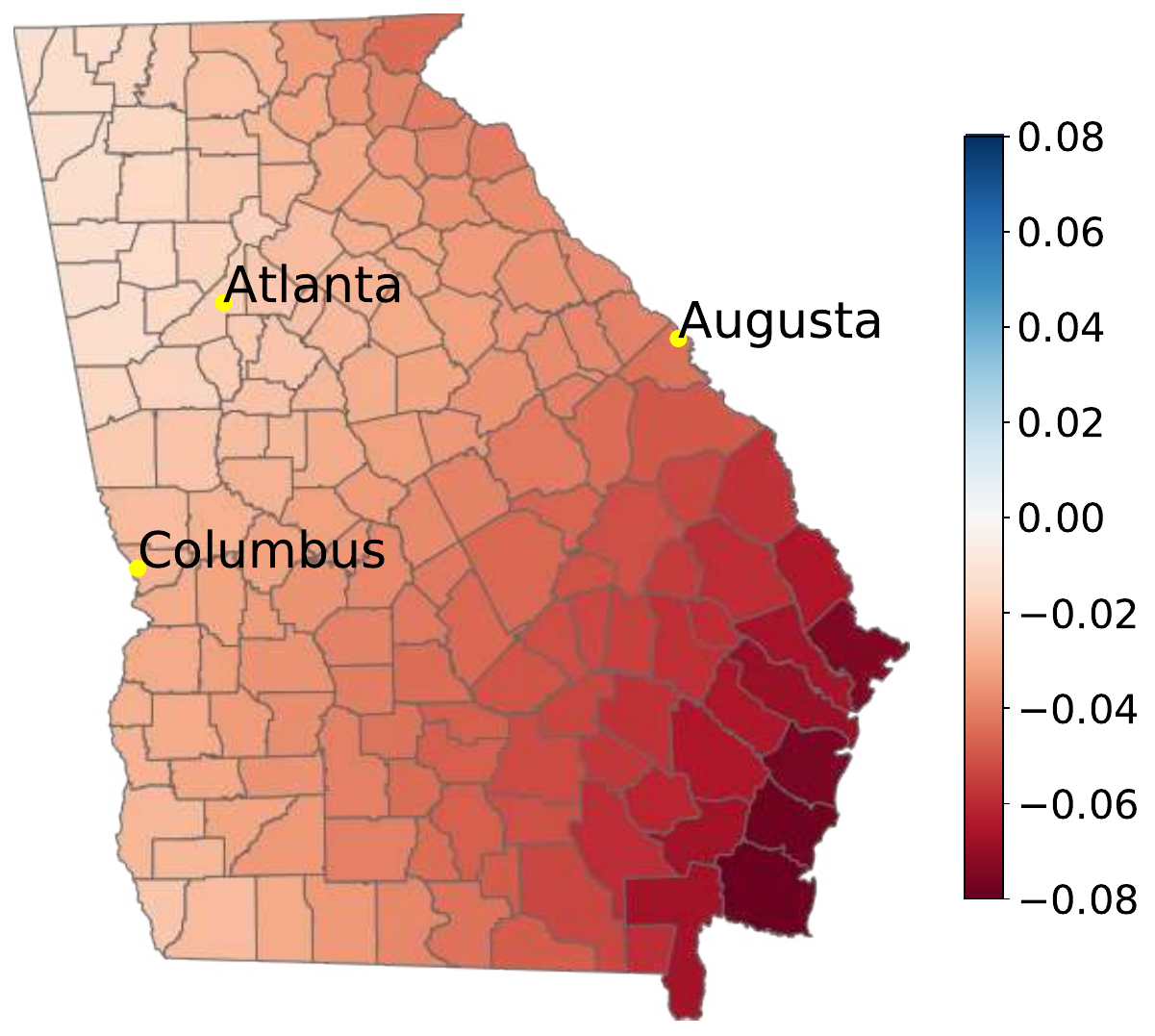} \label{coeff2_global}
        }
    \subfigure[IGWR-L]{%
        \includegraphics[height=3.0cm]{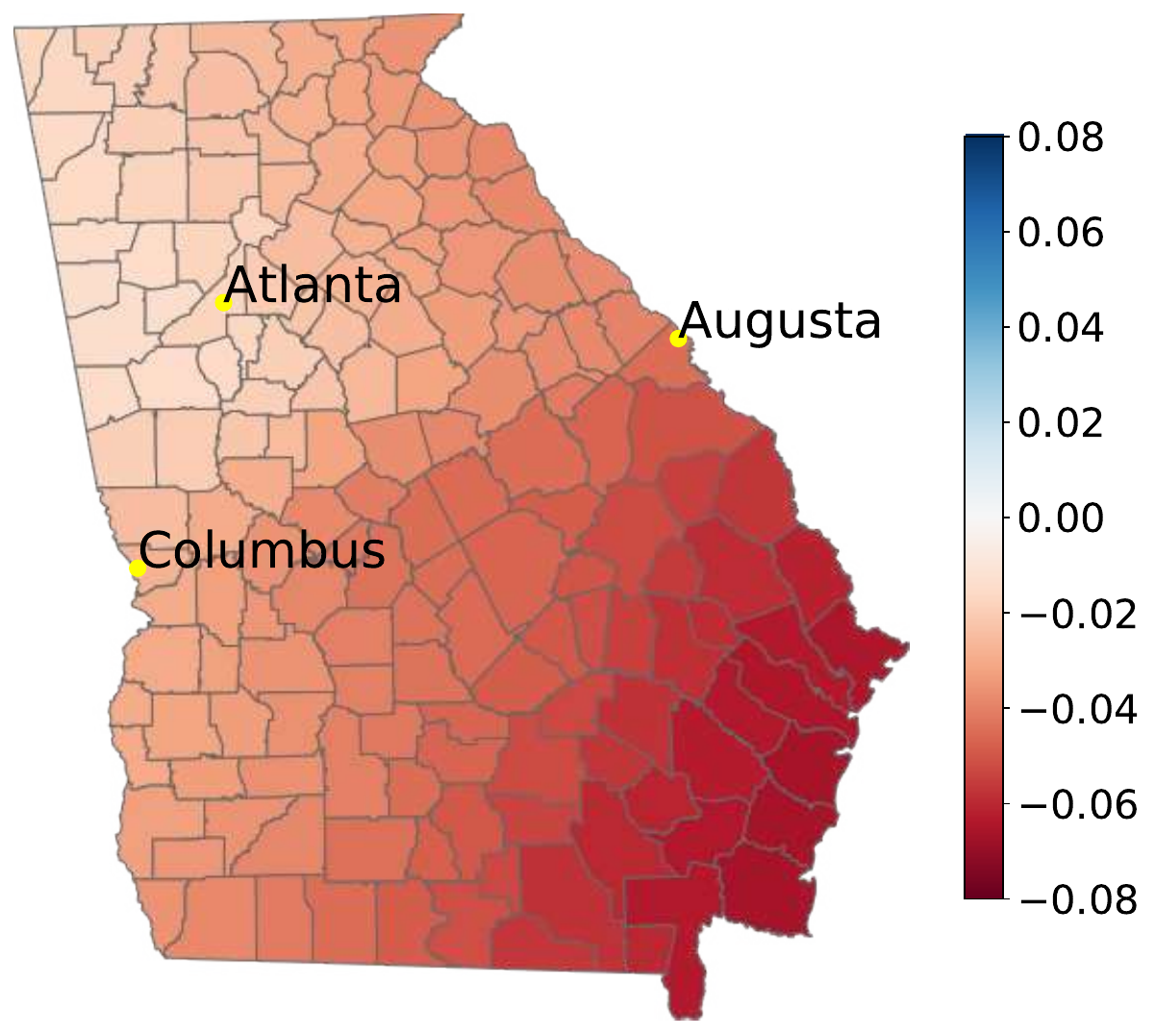} \label{coeff2_local}
    }
    \subfigure[FS]{%
        \includegraphics[height=3.0cm]{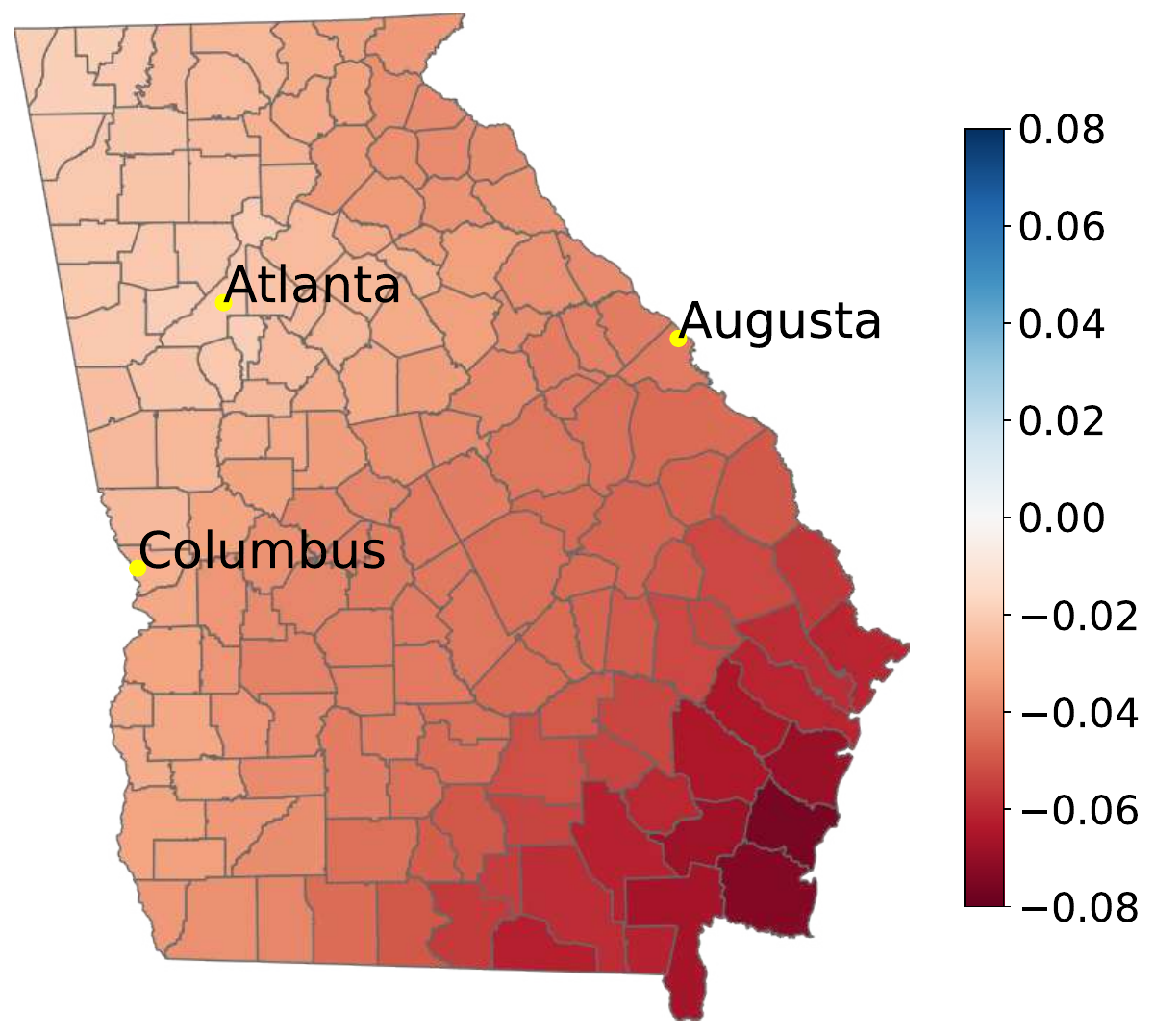} \label{coeff2_forward}
    }
\end{center}
\begin{center}
    \subfigure[BGWR]{%
        \includegraphics[height=3.0cm]{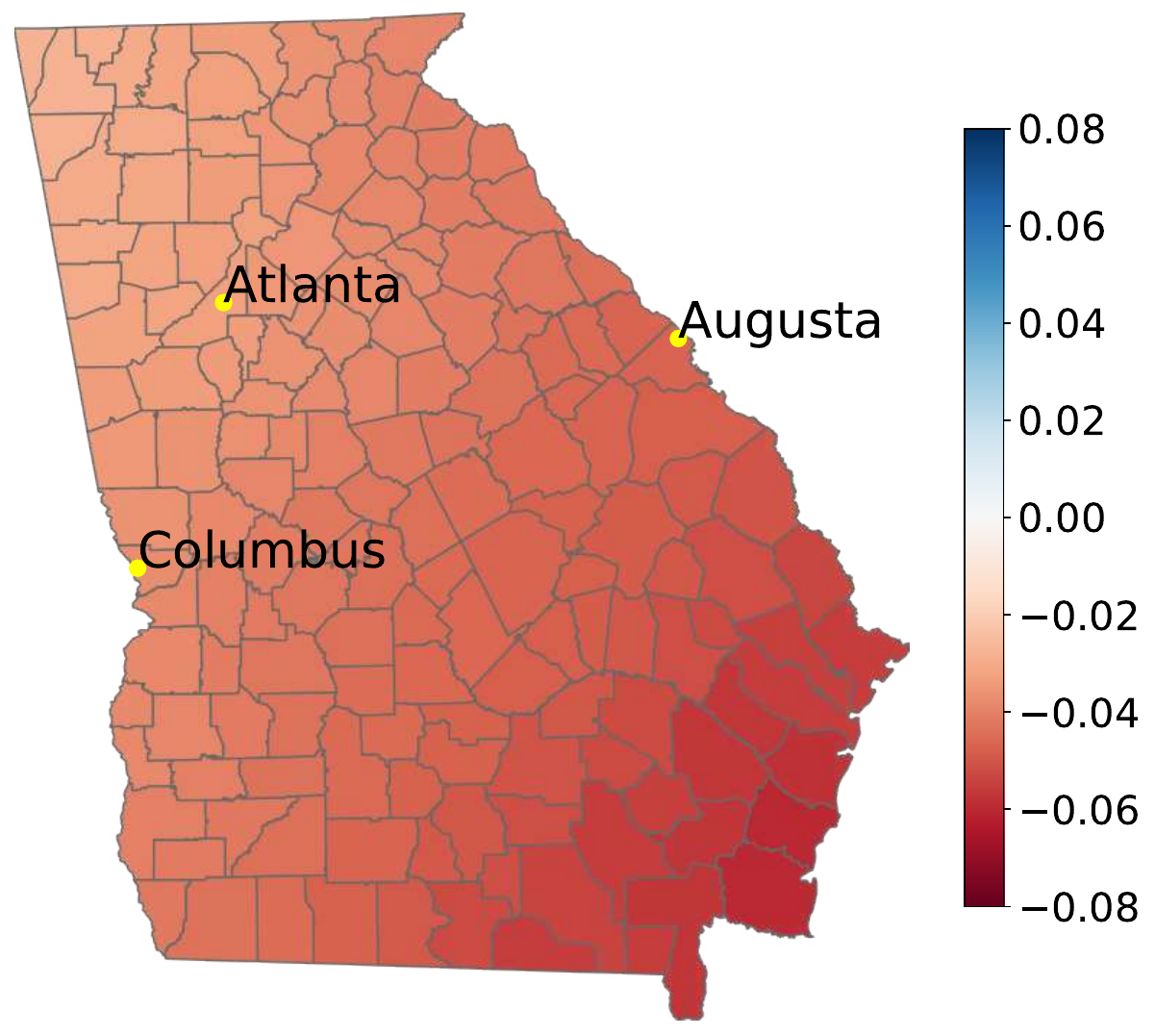} \label{coeff2_gwr}
        }
    \subfigure[MGWR]{%
        \includegraphics[height=3.0cm]{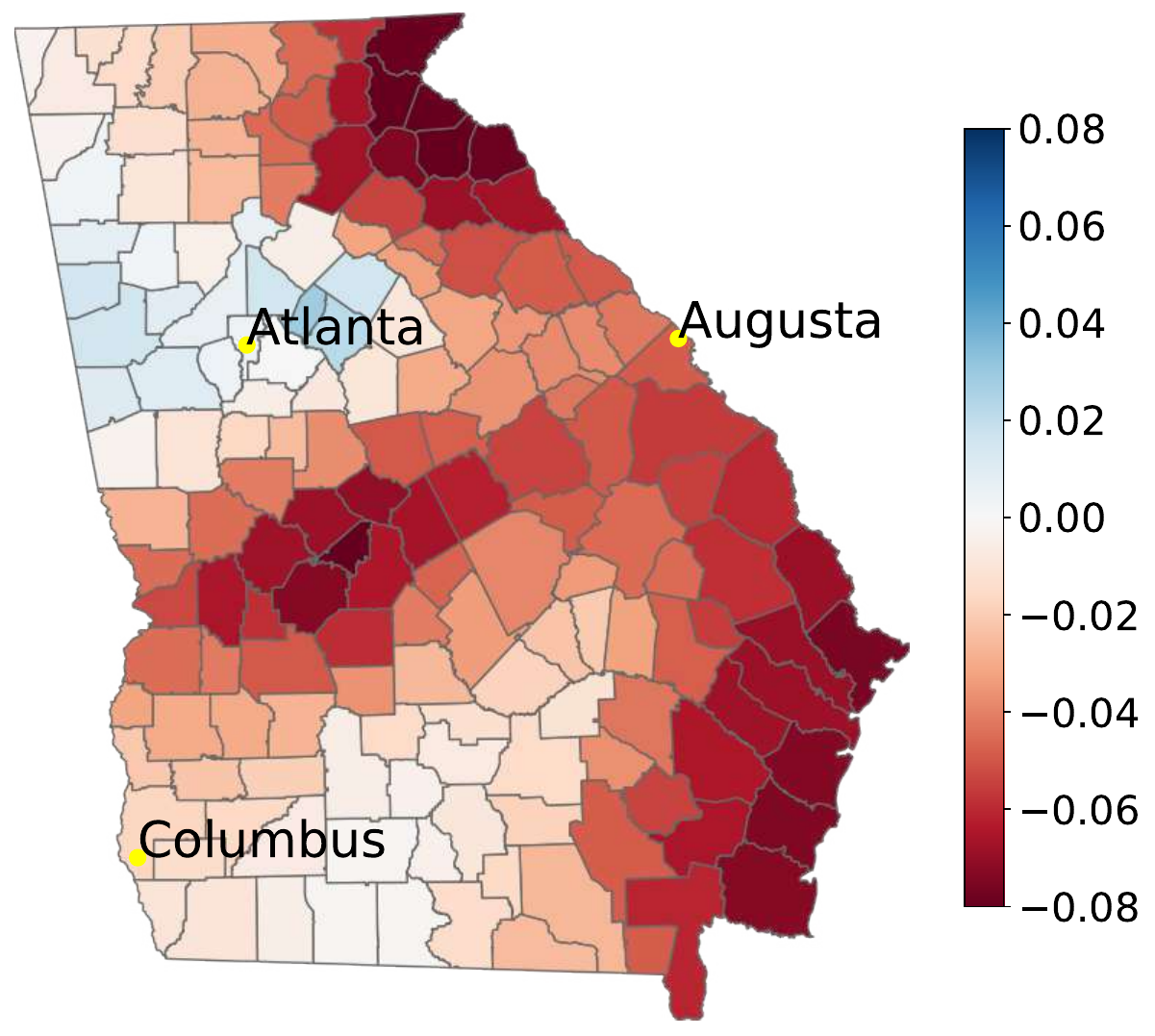} \label{coeff2_mgwr}
        }
    \subfigure[GWL (X: zero)]{%
        \includegraphics[height=3.0cm]{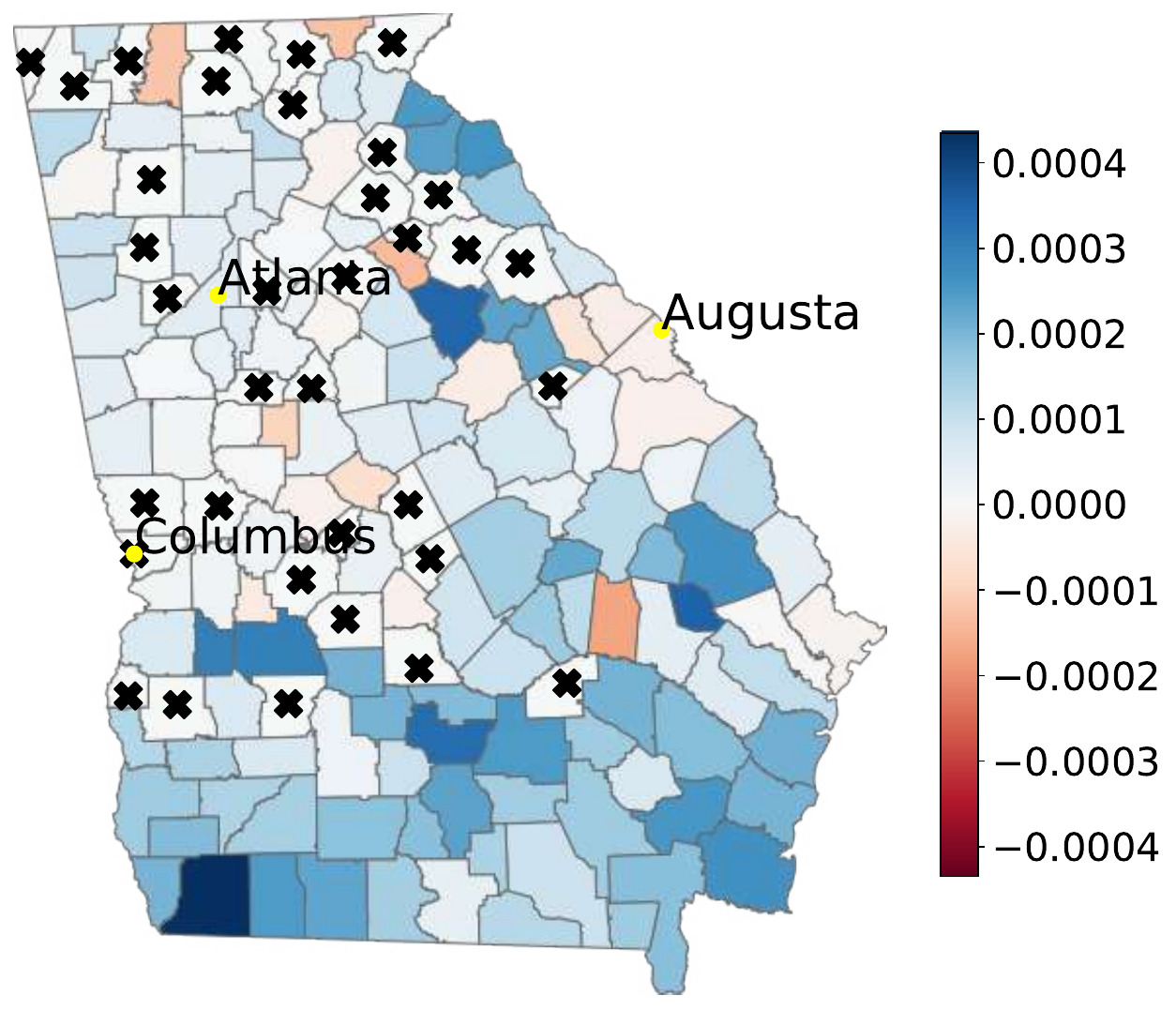} \label{coeff2_gwl}
        }
\end{center}
\vspace{-0.5cm}
    \caption{Coefficients of All Models for PctRural of Georgia Dataset}
\label{Comparison of PctRural}
\end{figure}

\begin{figure}[h!]
\begin{center}
    \subfigure[IGWR-G]{%
        \includegraphics[height=3.0cm]{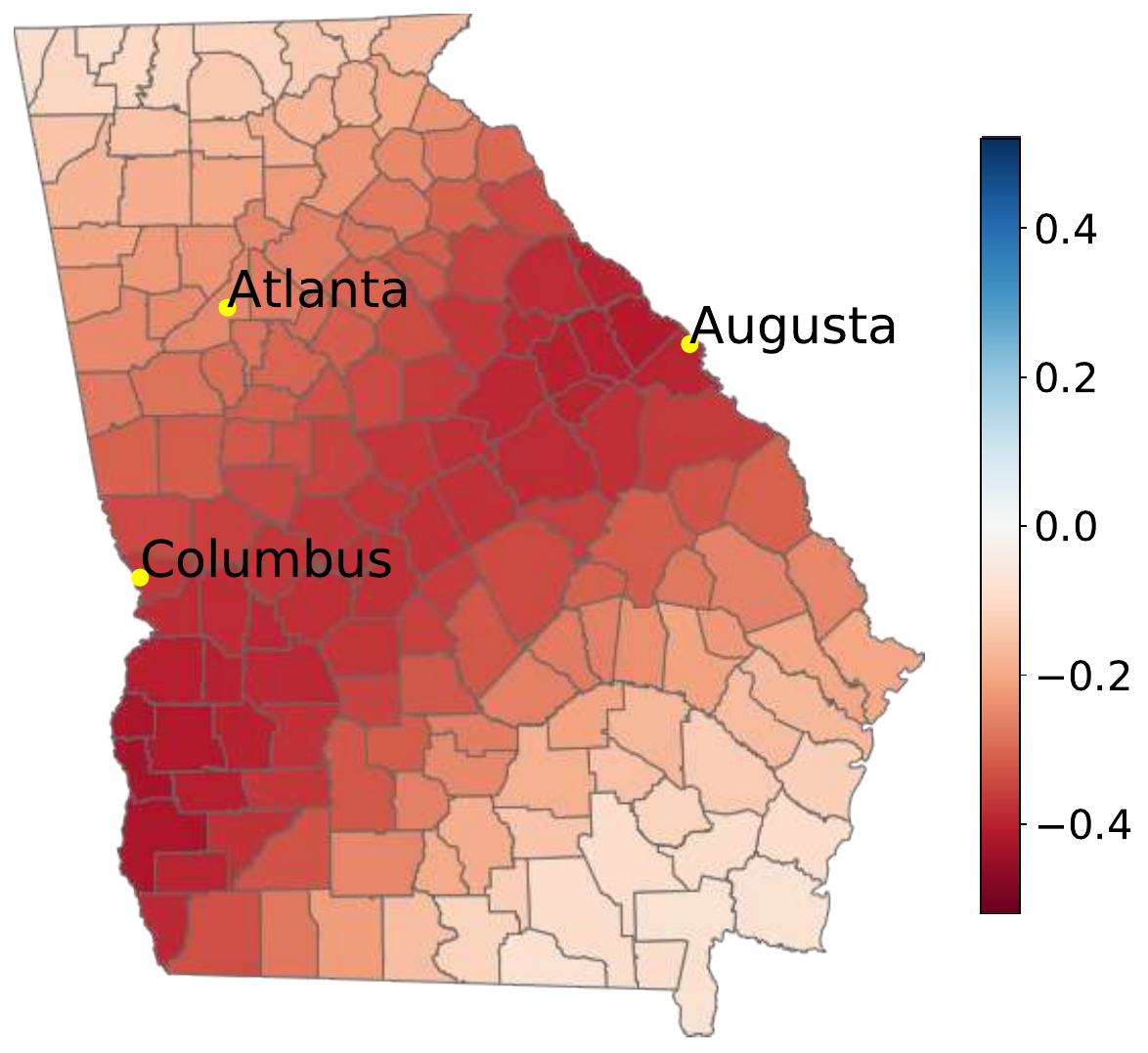} \label{coeff3_global}
        }
    \subfigure[IGWR-L]{%
        \includegraphics[height=3.0cm]{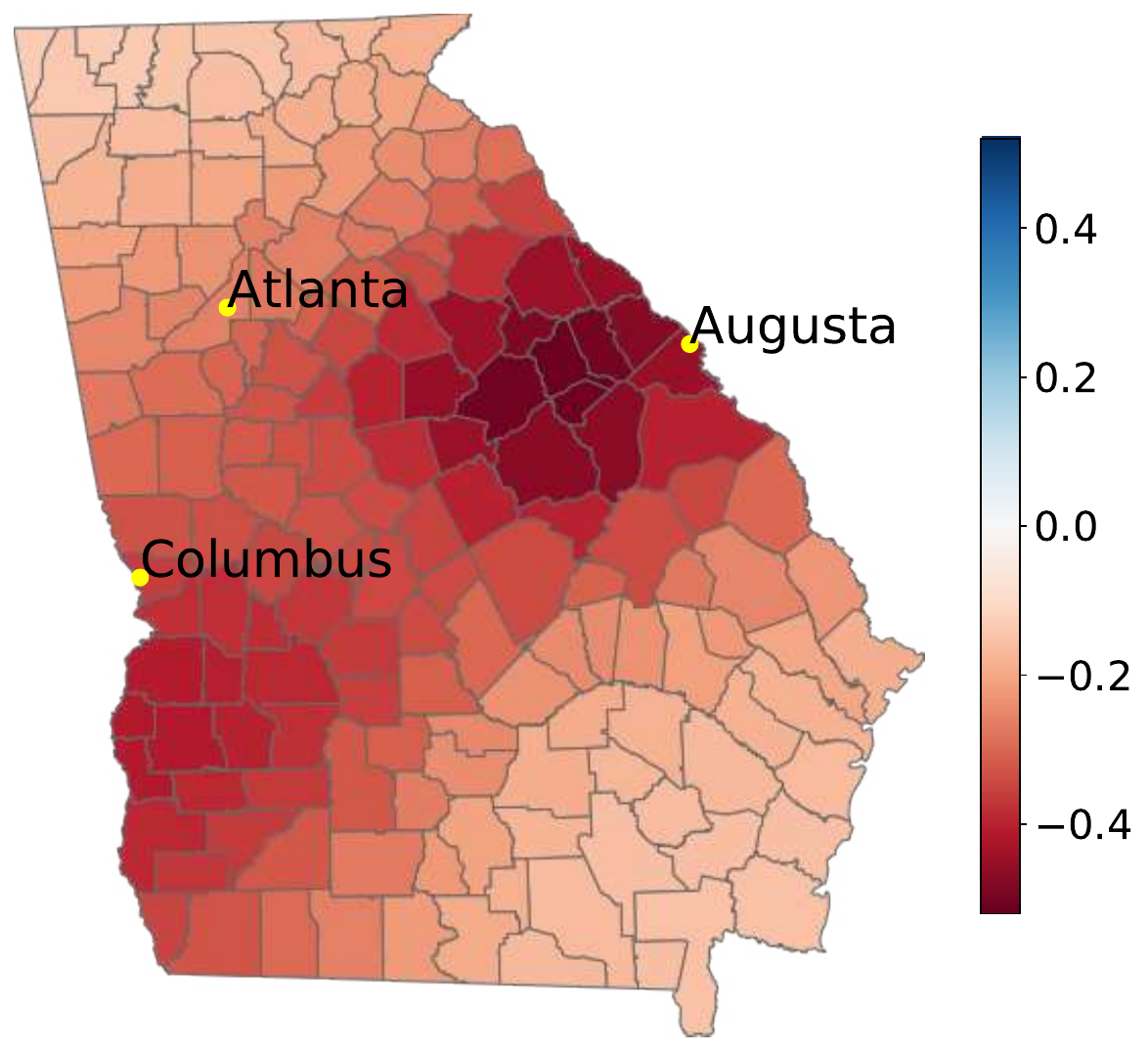} \label{coeff3_local}
    }
    \subfigure[FS]{%
        \includegraphics[height=3.0cm]{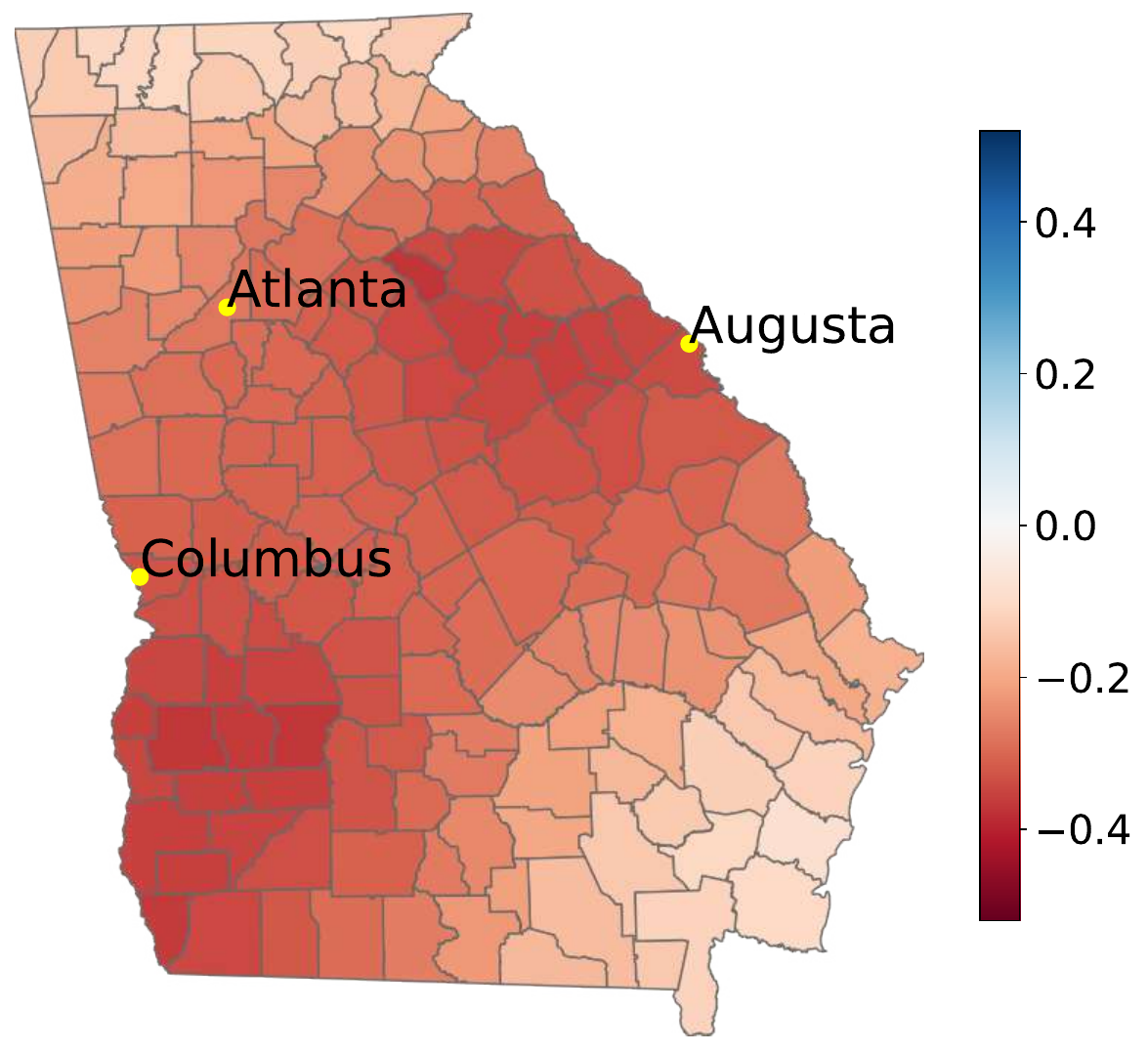} \label{coeff3_forward}
    }
\end{center}
\begin{center}
    \subfigure[BGWR]{%
        \includegraphics[height=3.0cm]{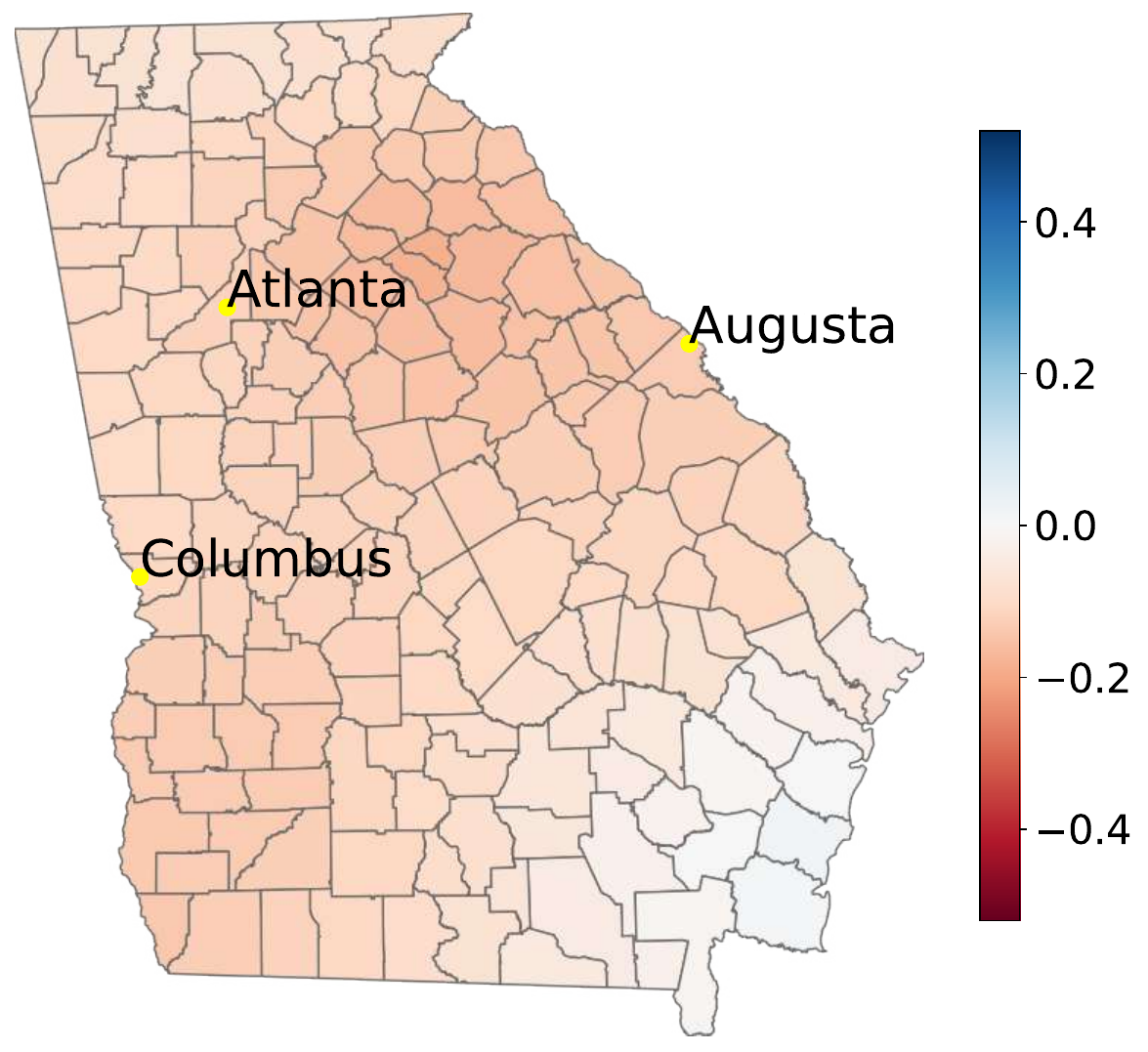} \label{coeff3_gwr}
        }
    \subfigure[MGWR]{%
        \includegraphics[height=3.0cm]{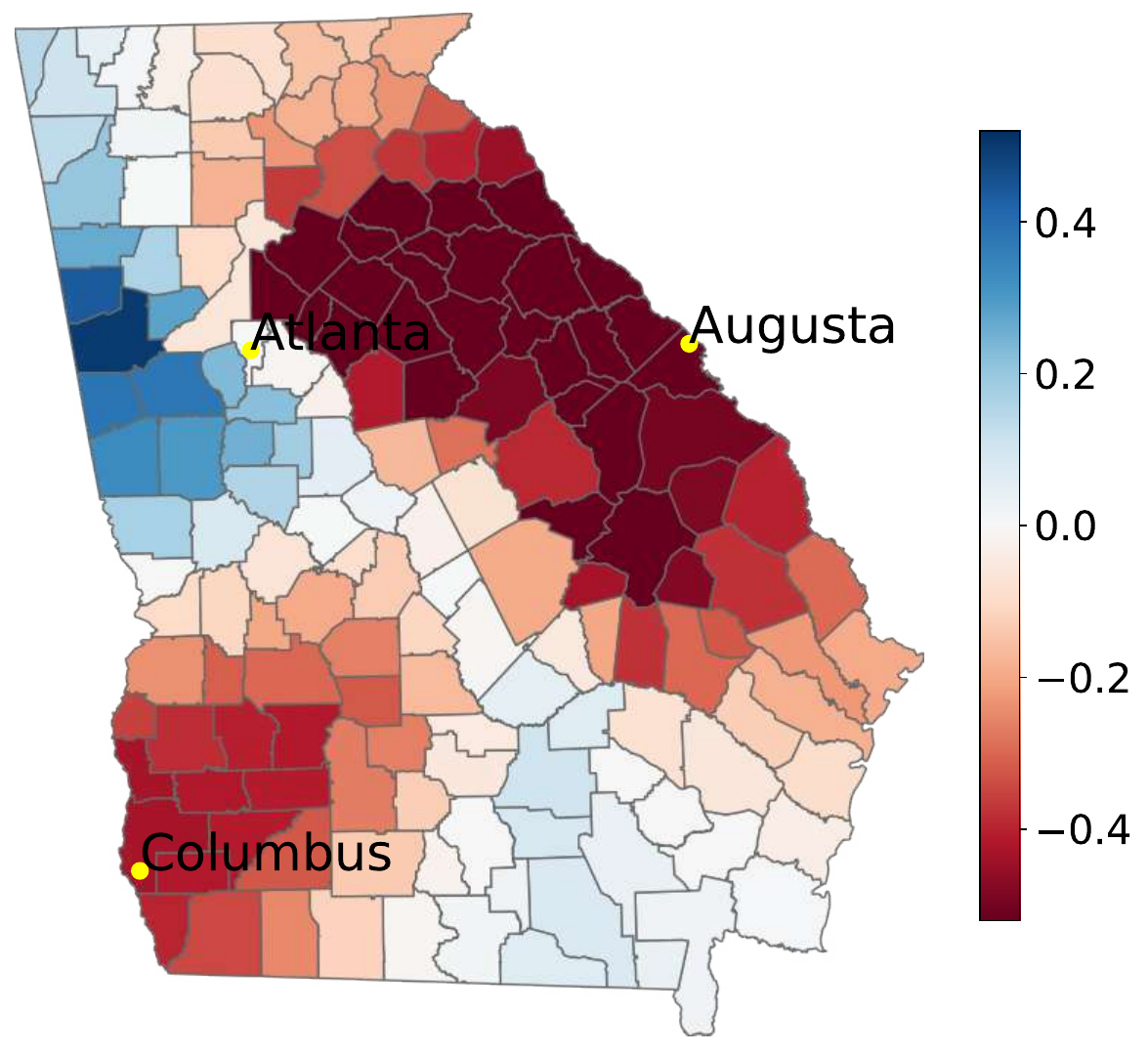} \label{coeff3_mgwr}
        }
    \subfigure[GWL (X: zero)]{%
        \includegraphics[height=3.0cm]{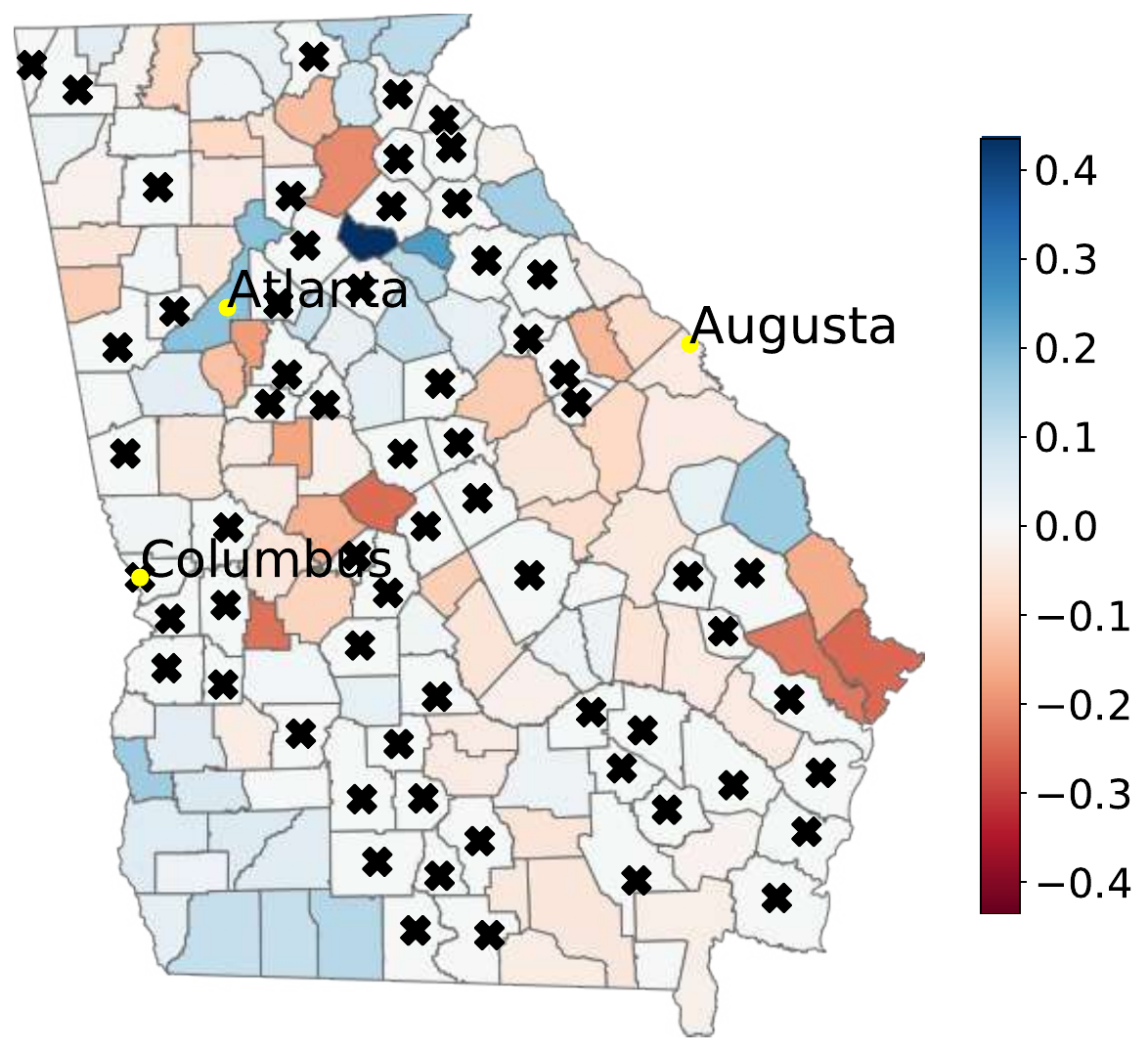} \label{coeff3_gwl}
        }
\end{center}
\vspace{-0.5cm}
    \caption{Coefficients of All Models for PctEld of Georgia Dataset}
\label{Comparison of PctEld}
\end{figure}

\begin{figure}[h!]
\begin{center}
    \subfigure[IGWR-G]{%
        \includegraphics[height=3.0cm]{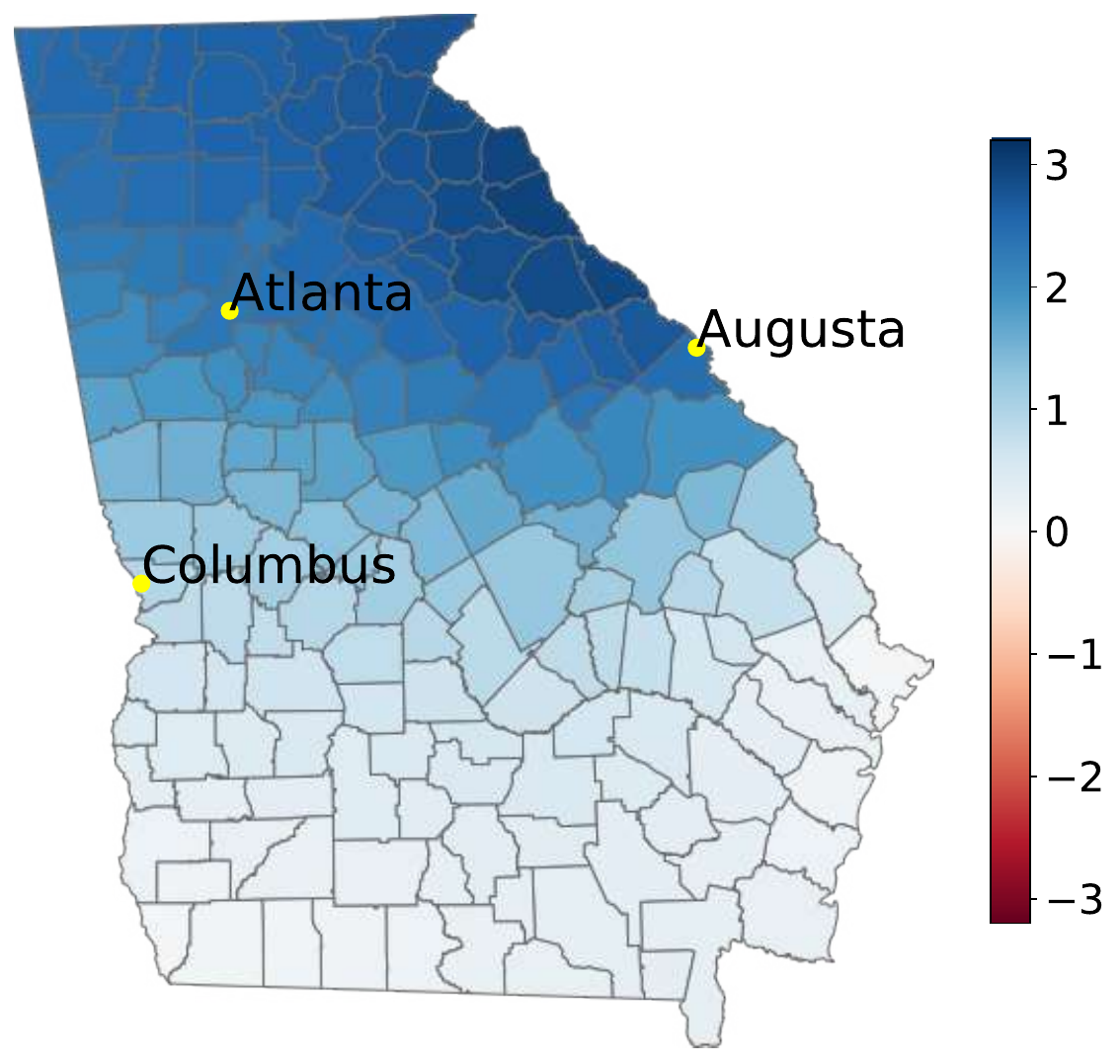} \label{coeff4_global}
        }
    \subfigure[IGWR-L]{%
        \includegraphics[height=3.0cm]{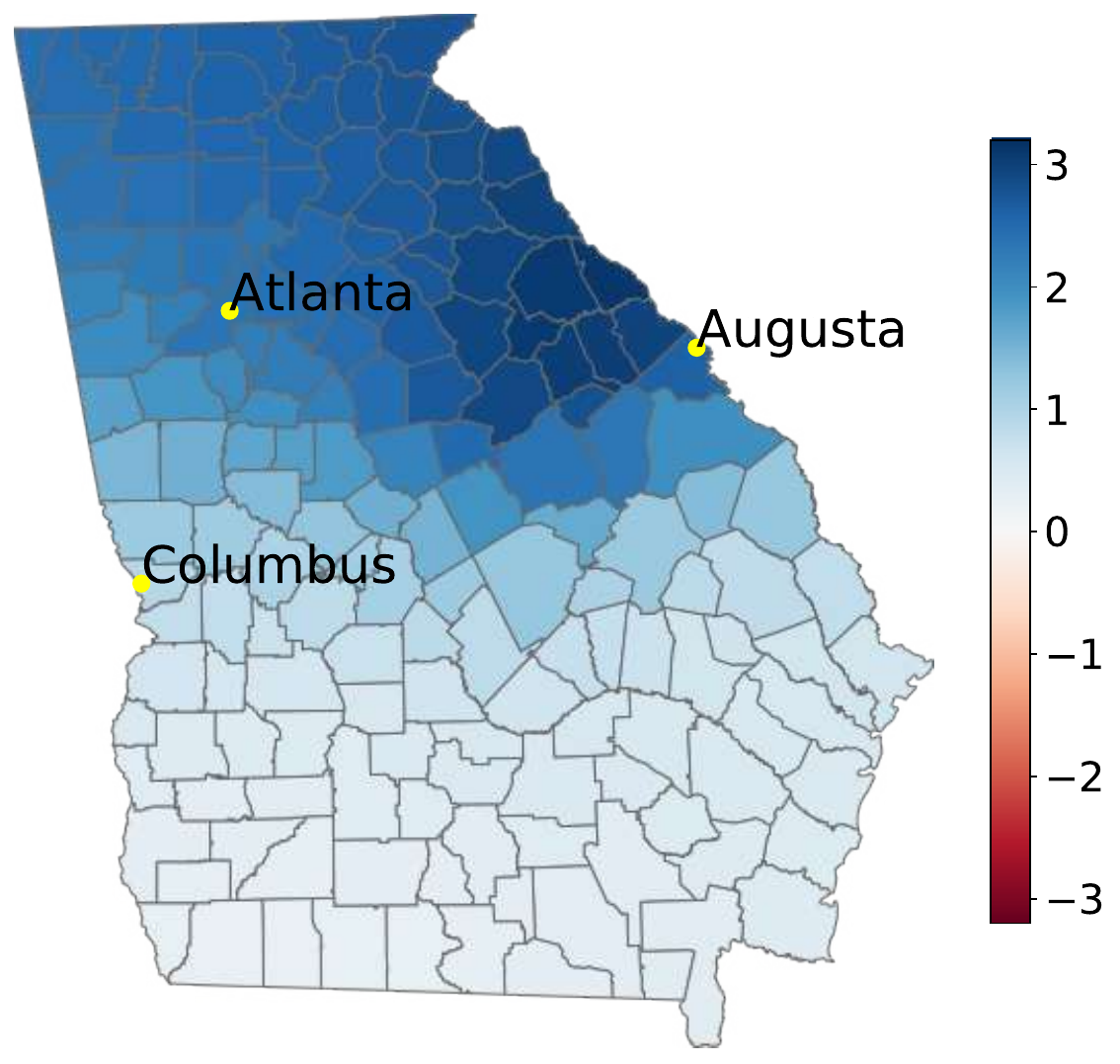} \label{coeff4_local}
    }
    \subfigure[FS]{%
        \includegraphics[height=3.0cm]{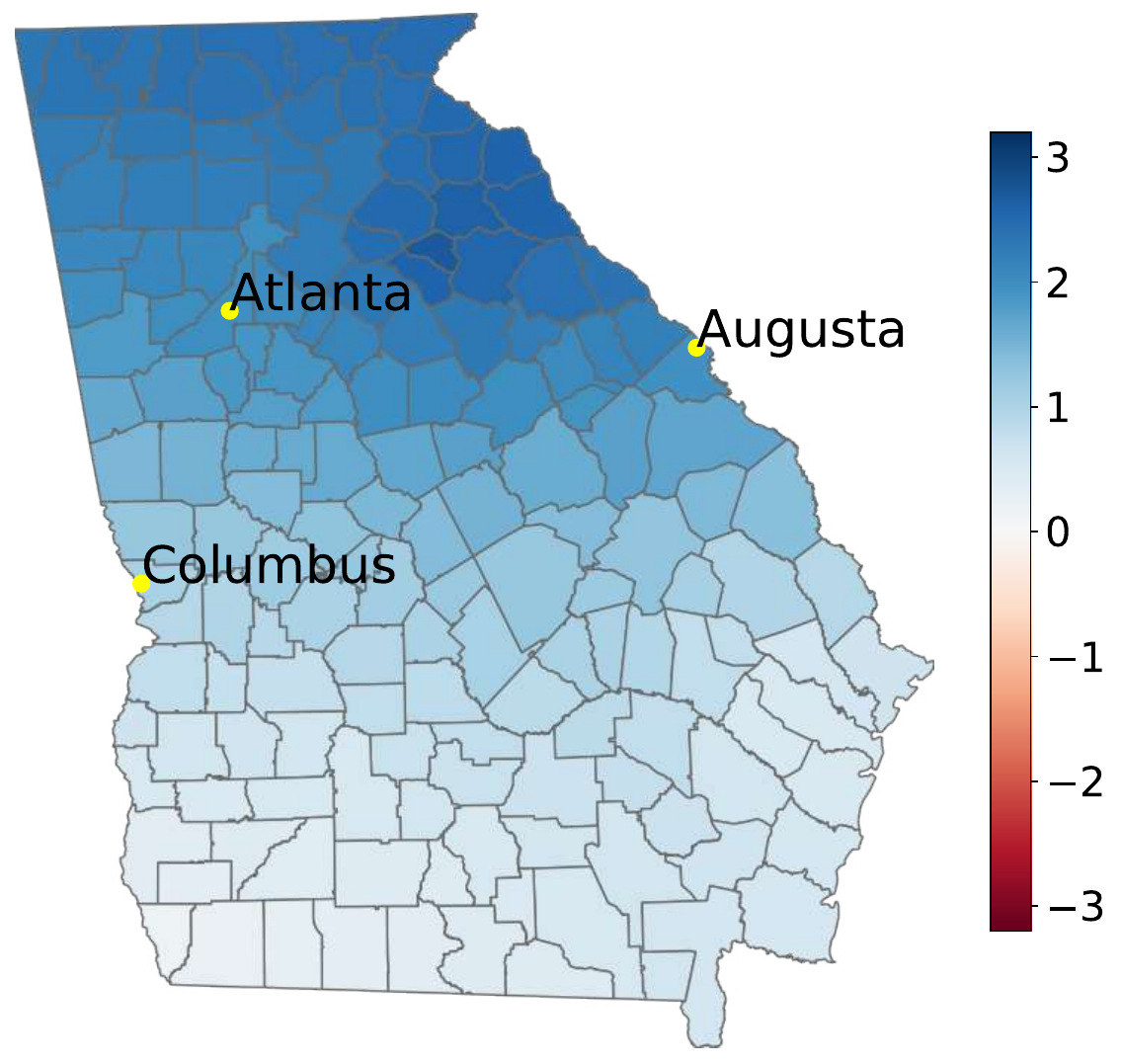} \label{coeff4_forward}
    }
\end{center}
\begin{center}
    \subfigure[BGWR]{%
        \includegraphics[height=3.0cm]{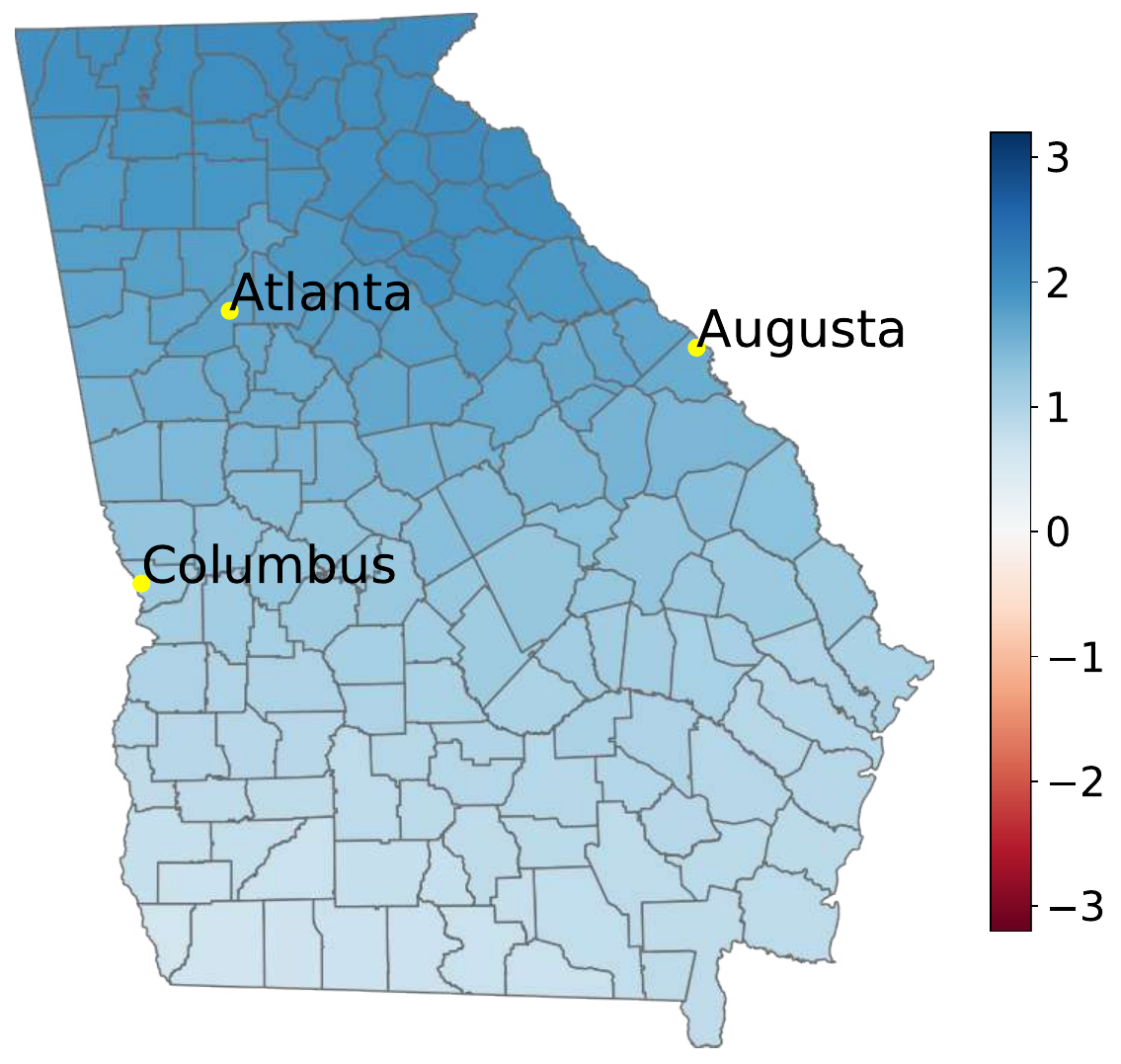} \label{coeff4_gwr}
        }
    \subfigure[MGWR]{%
        \includegraphics[height=3.0cm]{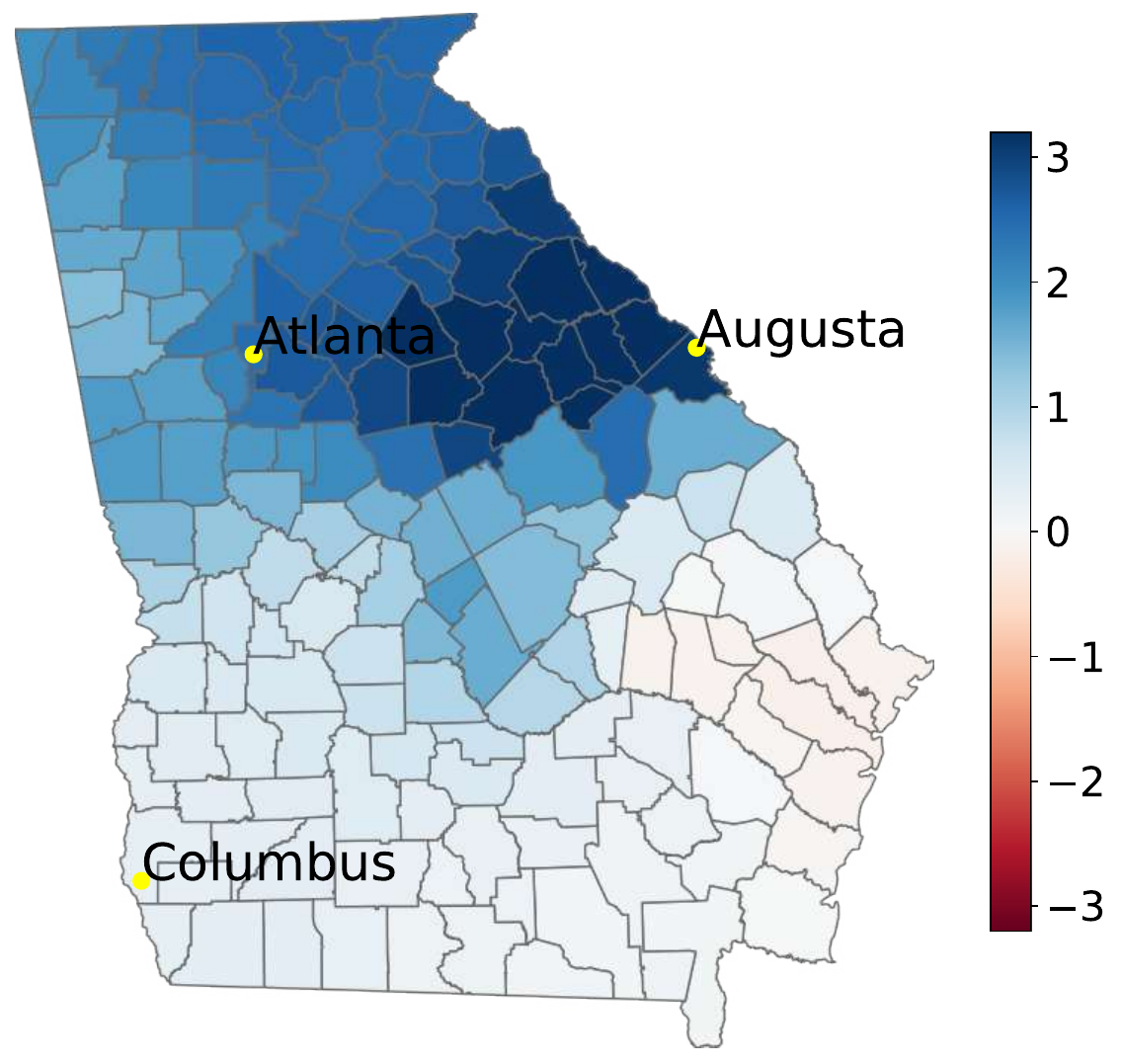} \label{coeff4_mgwr}
        }
    \subfigure[GWL (X: zero)]{%
        \includegraphics[height=3.0cm]{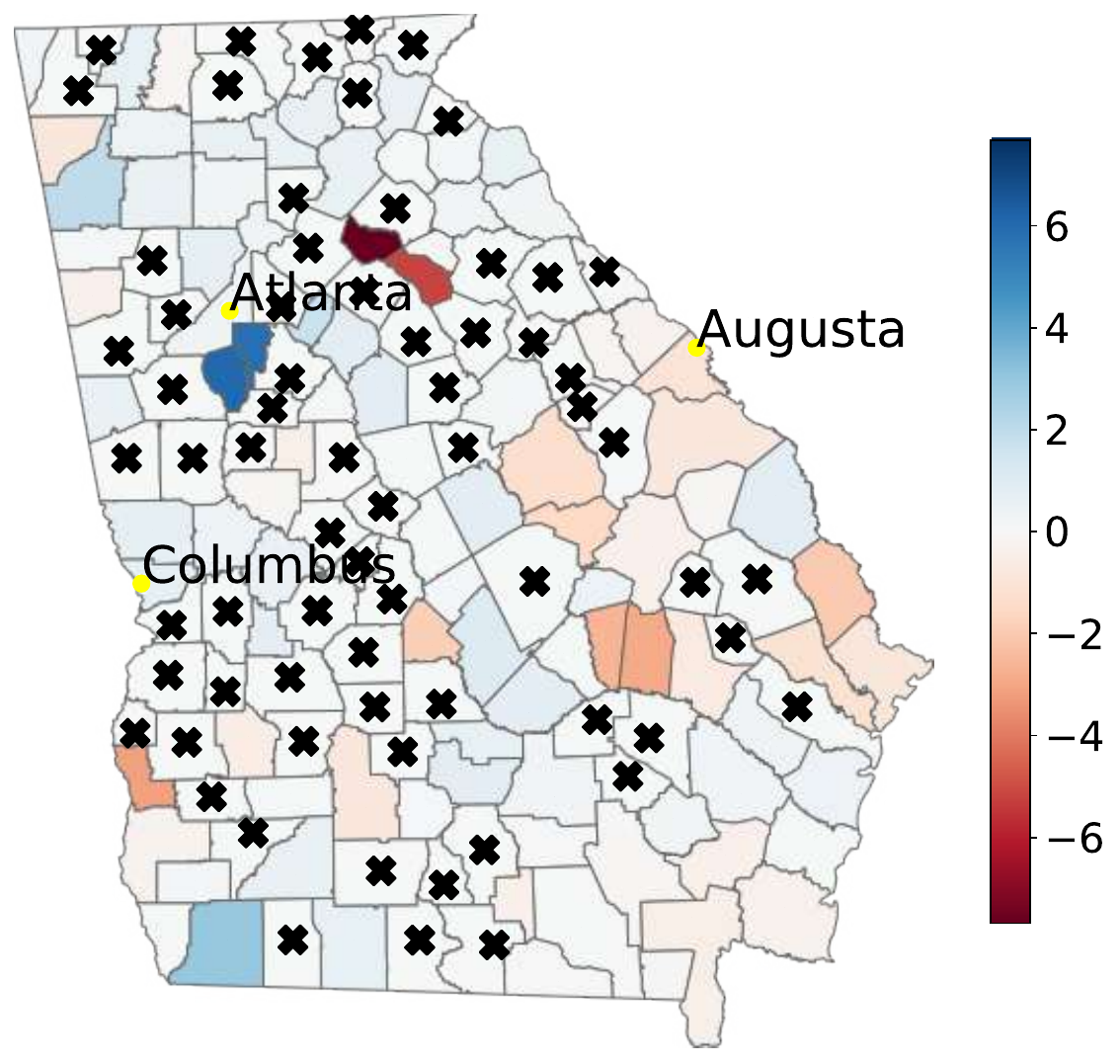} \label{coeff4_gwl}
        }
\end{center}
\vspace{-0.5cm}
    \caption{Coefficients of All Models for PctFB of Georgia Dataset}
\label{Comparison of PctFB}
\end{figure}

We present the spatial patterns for four key variables: \(TotPop90\), \(PctRural\), \(PctFB\), and \(PctEld\), analyzed using six different models: IGWR-G, IGWR-L, FS with \(p = 4\), GWR, MGWR with all IVs, and GWL with \(p \approx 4\). Figures \ref{Comparison of TotPop90} through \ref{Comparison of PctFB} compare these patterns. The IGWR-G, IGWR-L, and FS models display similar spatial distributions. BGWR, while also showing a comparable pattern, tends to be less distinct. In contrast, MGWR presents markedly different and more pronounced patterns, highlighting either an exaggeration or a divergence from consensus observed in other models. GWL, noted for its variability, exhibits highly fluctuating spatial patterns, further underscoring the need for careful model selection and validation to ensure the reliability of local regression analyses.

For example, in Figure \ref{Comparison of TotPop90}, the coefficients in \ref{coeff1_global} - \ref{coeff1_forward} increase toward the southern region, whereas the coefficients in \ref{coeff1_gwr} show little spatial variation. MGWR also displays increasing coefficients toward the southern region but with noticeable exaggeration. In Figure \ref{Comparison of PctRural}, the coefficients in \ref{coeff2_global} - \ref{coeff2_gwr} decrease toward the southeastern region, while the coefficients in \ref{coeff2_mgwr} reveal an extreme pattern near southeastern region and different trends near north and center regions. This exaggeration is more evident in Figure \ref{Comparison of PctEld}. Overall, when MGWR produces a pattern similar to those of IGWR, FS, and BGWR, it tends to show either an extremely exaggerated pattern or a pattern that differs significantly from the others. GWL returns heterogeneous subsets and shows volatile spatial patterns through Figures \ref{Comparison of TotPop90} - \ref{Comparison of PctFB}, where the coefficients shrunken to zero appear as black crosses. This is because GWL locally executes variable selection, which makes capturing spatial patterns difficult. Although we cannot definitively determine which patterns represent the true spatial variation, IGWR and FS are recommended for capturing smoothly varying patterns in the IVs.

In summary, our IGWR algorithm and the FS approach generally select variable subsets that not only demonstrate robust explanatory power but also maintain a stable geographically varying pattern. The BGWR, though similar in explanatory power, tends to produce less distinct patterns compared to our model and FS. MGWR, while superior in fitness measures, often exhibits exaggerated or divergent trends relative to other models. Conversely, GWL struggles both in explanatory power and in capturing reasonable spatial patterns, highlighting the limitations of local subset selection. It is essential to recognize that the interpretability of spatial patterns can vary across datasets, requiring visual inspections alongside model selection to validate the model's accuracy.

\subsection{Ohio Dataset Case Study: Exploring Differences in Global Subset Selection} \label{section4.4}

In this section, we compare the global subset selection approaches (IGWR and FS) using the Ohio art and culture dataset to show that these two methods can lead to different subsets. In Section \ref{section4.3}, we demonstrated the effectiveness of the global subset selection methods, where the subset selection results were identical for both IGWR and FS. However, this is not always the case, as the FS approach is a heuristic and uses a stepwise procedure, while IGWR aims to identify the optimal subset. The Ohio dataset contains 15 IVs, with some of the IVs being highly correlated or nearly duplicates. For example, the correlation matrix in Figure \ref{heatmap_corr_ohio} shows that many IVs are highly correlated including \textit{totpop} and \textit{tothse}, and \textit{medhinc} and \textit{meanhinc}. Thus, to improve interpretability and reduce collinearity, subset selection becomes a necessary step for building a regression model.

The subset selection results for the Ohio dataset with $p=1,2,\cdots,10$ are shown in Table \ref{tab:table7}. In the selected variables columns, the ``G", ``L", and ``F" marks reflect selected IVs for IGWR-G, IGWR-L, and FS, respectively. We also report RSS, $R^2$, and $R^2_{\text{adj}}$ for IGWR and FS. The results show that the selected subsets for the Ohio dataset differ between IGWR and FS, contrasting with the results for the Georgia dataset where both approaches produce the same subset selection results. Additionally, in the FS approach, once an IV is selected, it remains included as $p$ increases, whereas this is not always the case with IGWR. IGWR also demonstrates better RSS results than the FS approach in the Ohio dataset. The findings suggest that complementary facilities (\textit{comp}), the percentage of the population aged 65 and over (\textit{age65}), total population (\textit{pop}), and the percentage of the white population (\textit{race\_white}) are the most frequently included IVs.

\begin{table}[htbp]
  \centering
  \small
  \setlength{\tabcolsep}{1.5pt}
  \caption{IGWR-G and IGWR-L Results for Ohio Dataset} \label{tab:table7}
  \scalebox{0.8}{
    \begin{tabular}{c|ccccccccccccc|ccc|ccc|ccc}
    \toprule
          & \multicolumn{13}{c}{Selected variables$^\dagger$}                                & \multicolumn{3}{|c}{IGWR-G} & \multicolumn{3}{|c}{IGWR-L} & \multicolumn{3}{|c}{FS} \\ \midrule
    $p$     & comp  & age65 & pov   & pop   & white  & minc & medc & grad & bach & unemp & age18 & inc200k & hse & RSS & $R^2$ & $R^2_{\text{adj}}$ & RSS & $R^2$ & $R^2_{\text{adj}}$ & RSS & $R^2$ & $R^2_{\text{adj}}$ \\ \cmidrule{1-1} \cmidrule{2-14} \cmidrule{15-17} \cmidrule{18-20} \cmidrule{21-23}
    1     & GLF   &       &       &       &       &   &    &       &       &       &       &      & & 1275 & 0.417 &  0.410 & 1272  & 0.419 & 0.412 & 1382 &  0.369 & 0.361 \\
    2     & GLF   & GL\phantom{F} &       &  \phantom{GL}F     &       &   &    &       &       &       &       &       & & 1168 & 0.466 & 0.454 & 1149 & 0.475 & 0.462  & 1348 & 0.385 & 0.370 \\
    3     & GLF    & GL\phantom{F}     & GL\phantom{F} &   \phantom{GL}F   &   \phantom{GL}F    &   &    &      &    &       &       &       & & 1080 & 0.506 & 0.489 & 1040 & 0.525 & 0.508 & 1299 & 0.407 & 0.386 \\
    4     & GLF    & GL\phantom{F} &      & GLF & GLF &    &    &       &       &   \phantom{GL}F    &       &       & & 1028 & 0.530 & 0.508 &  996 & 0.545 & 0.523   &  1284 & 0.413 & 0.385 \\
    5     &  \phantom{GL}F     & GL\phantom{F} & GL\phantom{F}    & GLF &  GLF    & GLF &  &  GL\phantom{F}   &   &   \phantom{GL}F   &       &       & & 921 & 0.579 & 0.553 & 883 & 0.596 & 0.571   & 1247 & 0.430 & 0.396 \\
    6     & G\phantom{L}F    & GL\phantom{F}    & GL\phantom{F} & GLF & GLF & GLF   &  \phantom{GL}F & \phantom{G}L\phantom{F}    &        & \phantom{GL}F    &       &       & & 914 & 0.582 & 0.551   &  831 & 0.620 & 0.592    &  1200 &  0.452 & 0.411 \\
    7     & G\phantom{L}F    & GLF     & GL\phantom{F}     & GLF     & GLF     &    GLF &  \phantom{GL}F  &     & GL\phantom{F}     & \phantom{G}LF     &       &       & & 839 & 0.616 & 0.583    & 732   & 0.665 & 0.636 & 1171 & 0.465 & 0.418 \\
    8     & \textbf{GLF}     & \textbf{GLF}     & \textbf{GLF}     & \textbf{GLF}     & \textbf{GLF}     &  \textbf{GLF} & \phantom{GL}F   &     & \textbf{GL}\phantom{F}    & \textbf{GLF}     &       &       & & \textbf{757} & \textbf{0.654} & \textbf{0.619}    & \textbf{711} & \textbf{0.675} & \textbf{0.642}  & \textbf{1149} & \textbf{0.489}  & \textbf{0.430} \\
    9     & GLF    & GLF & GLF & GLF & GLF &    GLF & \phantom{GL}F &  & GLF    & GLF   &    GL\phantom{F}   &     & & 736 & 0.664 & 0.625 & 690 & 0.684 & 0.648   &  1119 & 0.489 & 0.430 \\
    10     & GLF    & GLF & GLF & GLF & GLF &   GLF  & \phantom{GL}F &     & GL\phantom{F}    & GLF   &    GLF   & GL     & F & 722 & 0.670 & 0.627 & 676 & 0.691 & 0.651  &  1109 & 0.493  & 0.427 \\    \bottomrule
    \end{tabular}%
  }
\scalebox{0.8}{
\begin{tabular}{p{16cm}}
\begin{scriptsize}$\dagger$ We shortened the column names given space limitations. The full original column names are: $fct\_complement$, $age\_GE65$, $econ\_poverty$, $totpop$, $race\_white$, $educ\_grad$, $meanhinc$, $medhinc$, $econ\_unemp$, $age\_GE18$, $income\_GE200k$, $tothse$. The omitted original features were not selected by the algorithm.\end{scriptsize}
\end{tabular}}
\end{table}%
\vspace{-0.5cm}

\begin{figure}[H]
\begin{center}
    \subfigure[IGWR-G]{%
        \includegraphics[height=4.0cm]{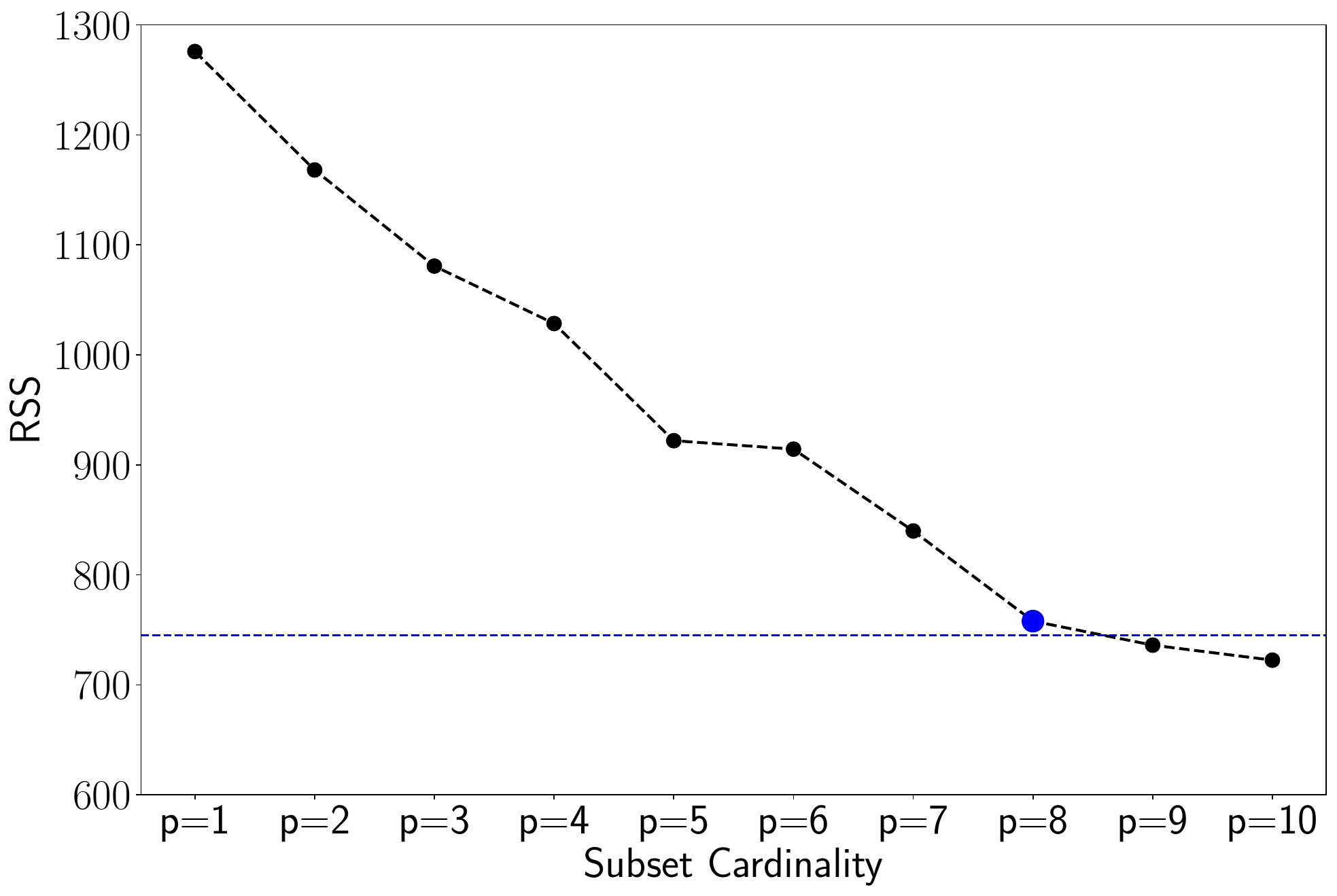} \label{scree_ohio_global}
        }\qquad
    \subfigure[IGWR-L]{%
        \includegraphics[height=4.0cm]{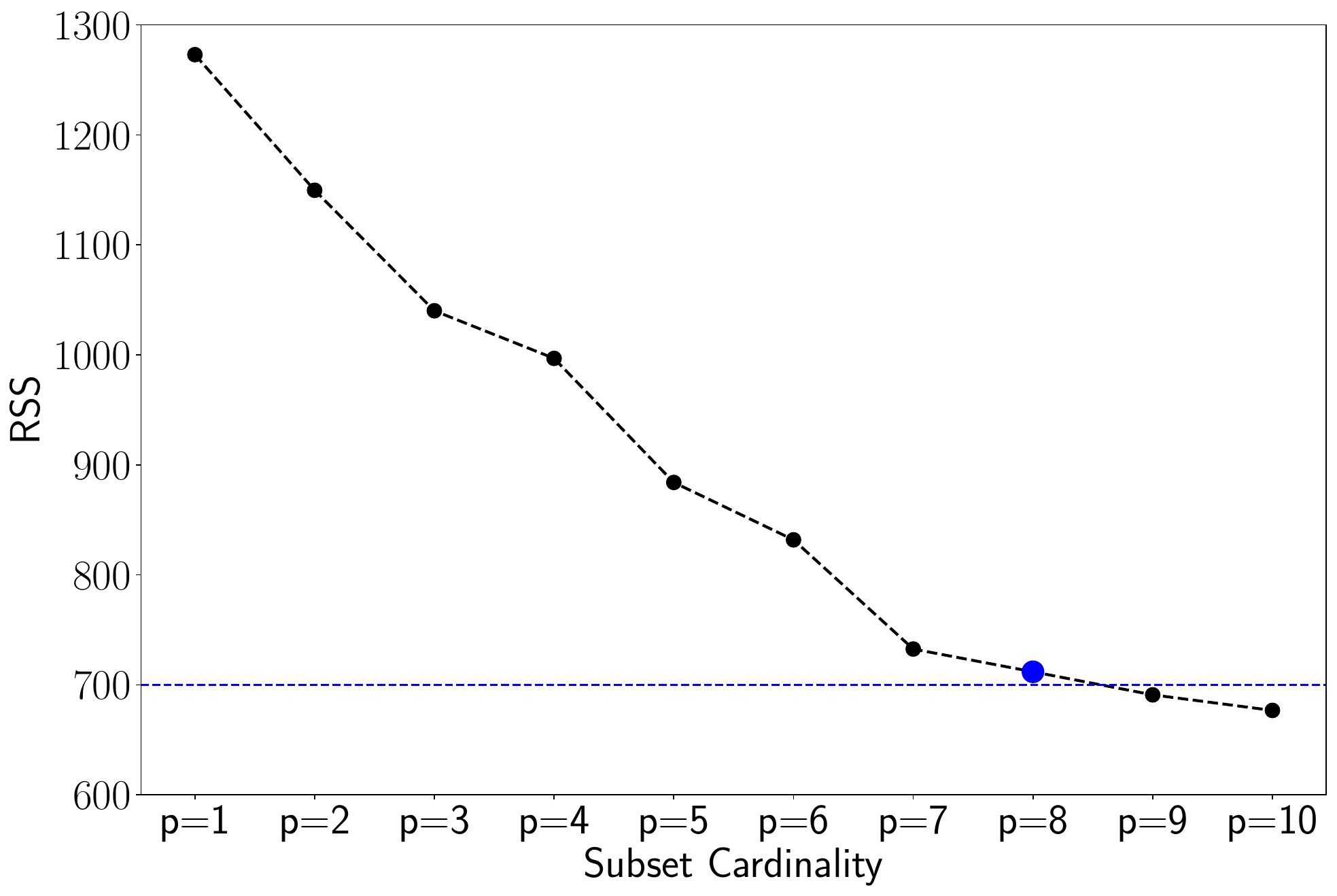} \label{scree_ohio_local}
    }
\end{center}
\vspace{-0.5cm}
    \caption{IGWR-G and IGWR-L Line Plots for Ohio Dataset}
\label{Scree Plot Ohio}
\end{figure}
\vspace{-0.3cm}
Figure \ref{Scree Plot Ohio} presents the line plots of RSS over the number of IVs for the two algorithms, IGWR-G and IGWR-L. Note that the maximum number of IVs selected by IGWR is 10 due to Constraint \eqref{mae_derive1_e}, which prevents the inclusion of highly correlated IVs. To determine the best subset cardinality, we applied the same rule as in Section \ref{section4.2}, and $p=8$ is the best cardinality for both global and local bandwidth settings in Figure \ref{Scree Plot Ohio}. The FS approach selected only one IV when using AIC \citep{fotheringham2013demographic} and three IVs when using AICc \citep{gollini2015gwmodel}. In terms of the RSS elbow point, the best cardinality for the FS approach is also $p=8$. The detailed results of FS including these plots are available in the online supplement.

Among the global subset selection approaches—IGWR-G, IGWR-L, and FS—we recommend using IGWR-G for most practical applications. The global setting of IGWR-G provides consistent estimated effects across multiple regions, whereas the local setting of IGWR-L may result in overfitting or less consistent results over regions. The FS approach is heuristic and may not always find the best subset for each cardinality $p$.

\begin{figure}[h!]
\begin{center}
    \subfigure[Totpop]{%
        \includegraphics[height=2.8cm]{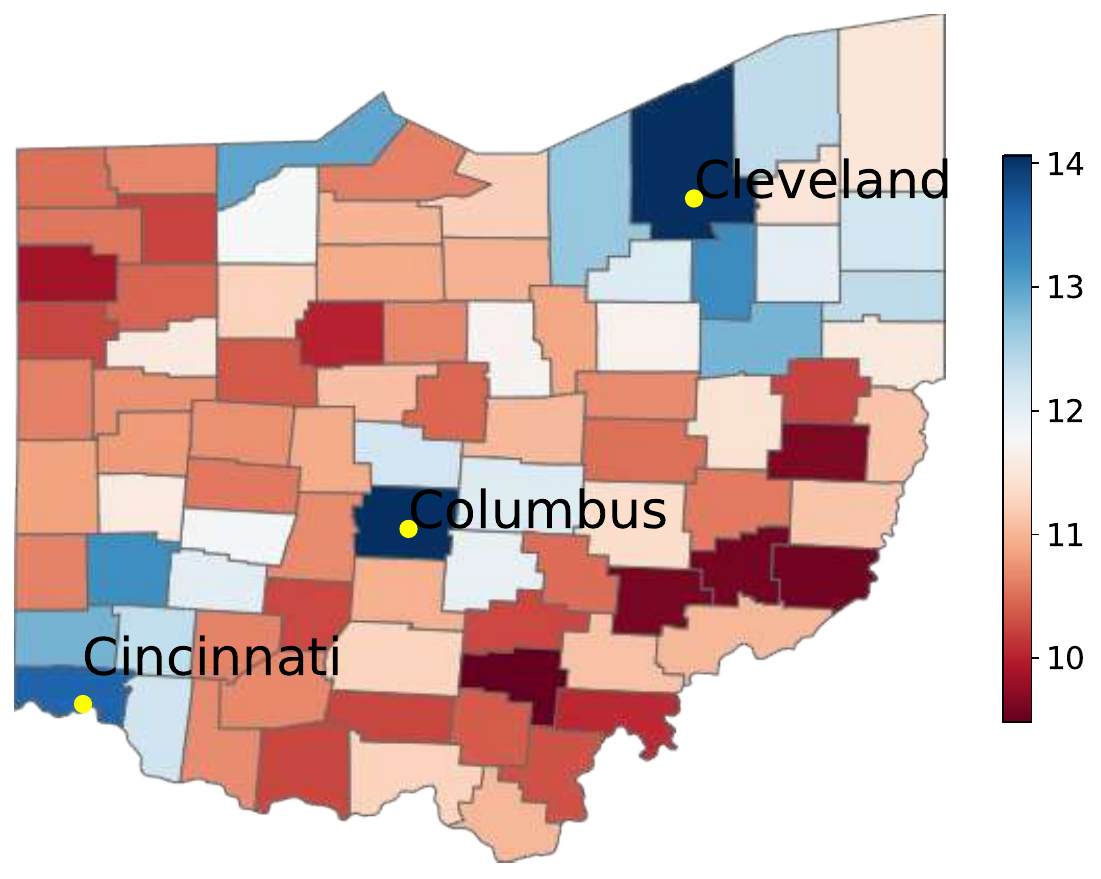} \label{hmap_totpop}
        }
    \subfigure[Fct\_complement]{%
        \includegraphics[height=2.8cm]{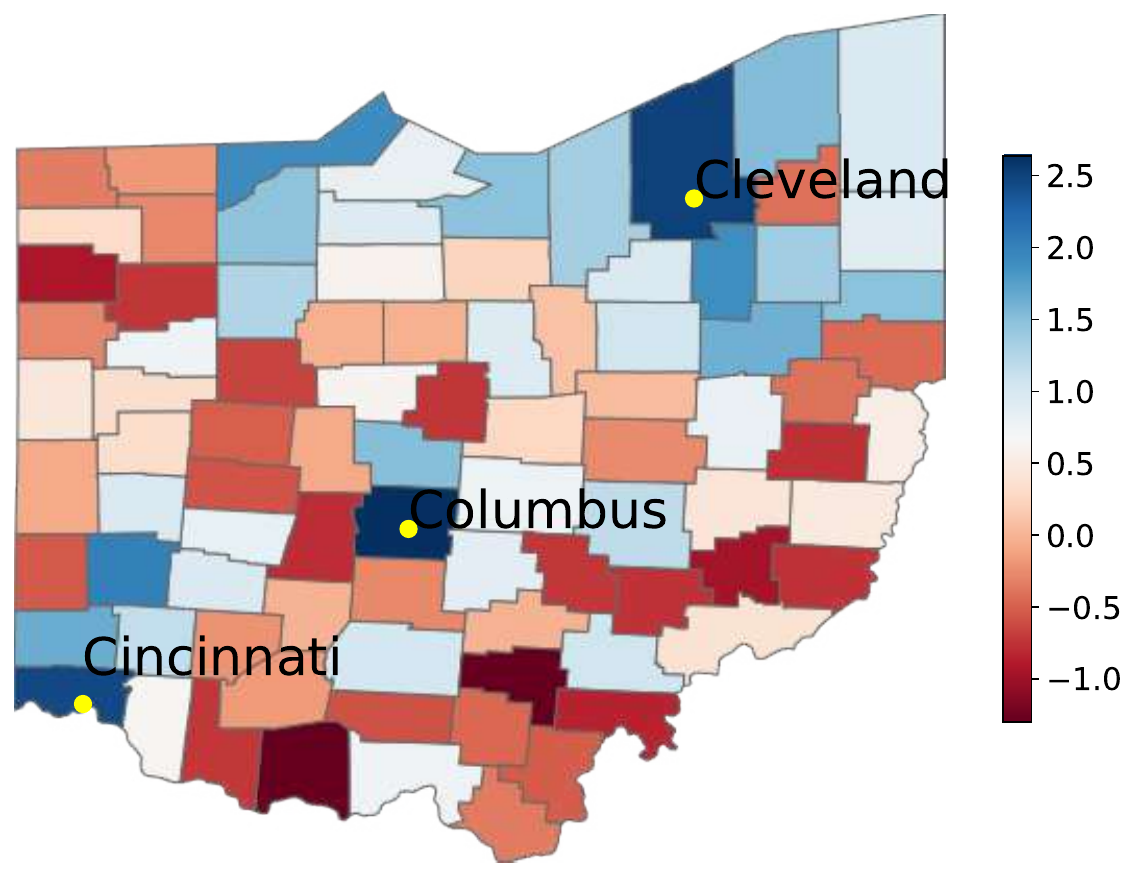} \label{hmap_fct_complement}
        }
    \subfigure[Meanhinc]{%
        \includegraphics[height=2.8cm]{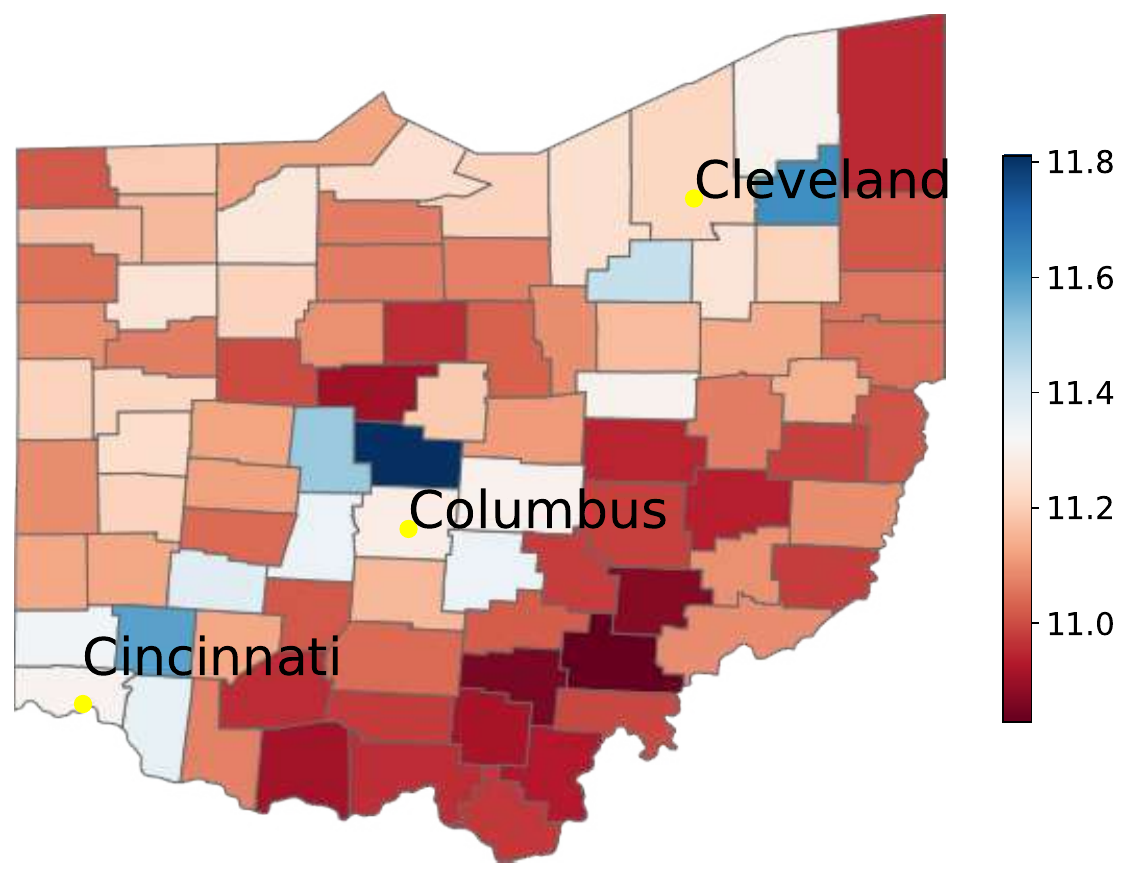} \label{hmap_MEANHINC}
        } 
    \subfigure[Educ\_bach]{%
        \includegraphics[height=2.8cm]{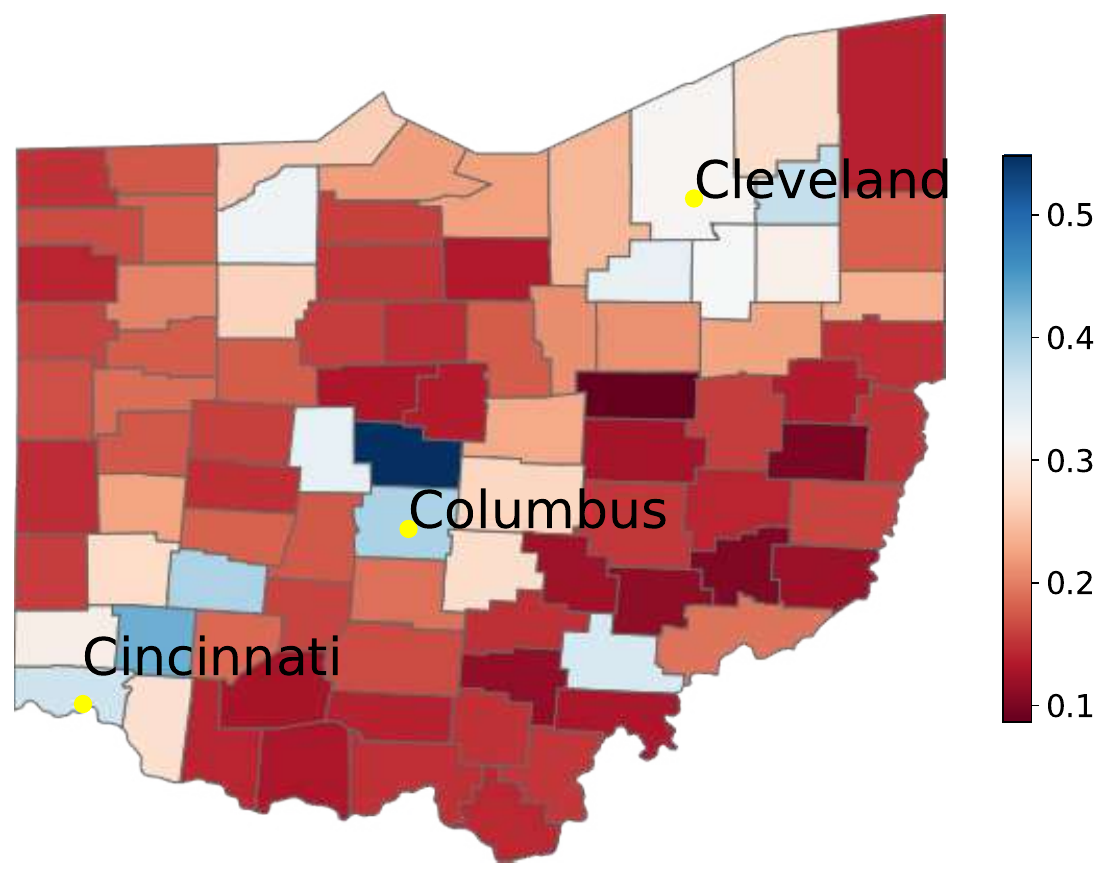} \label{hmap_educ_bach}
        }
\end{center}
\vspace{-1.0cm}
\begin{center}
    \subfigure[Econ\_unemp]{%
        \includegraphics[height=2.8cm]{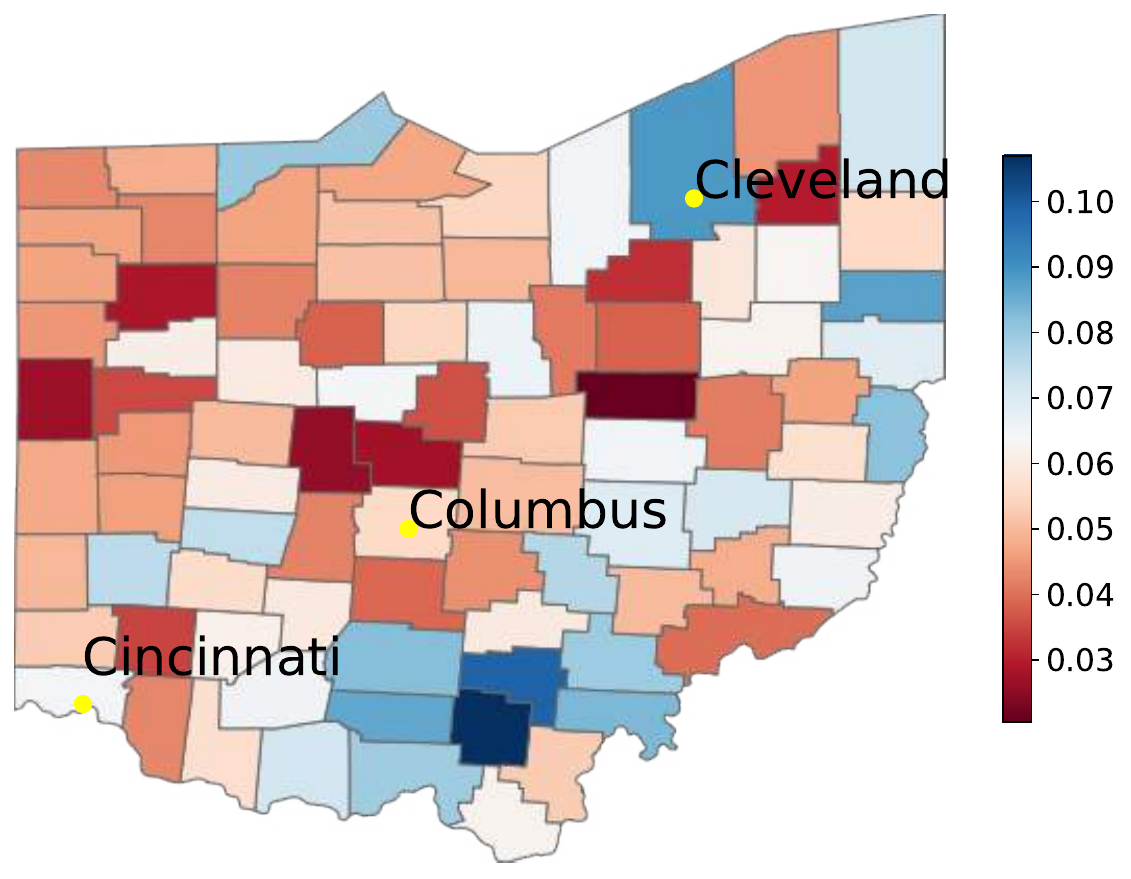} \label{hmap_econ_unemp}
        }
    \subfigure[Econ\_poverty]{%
        \includegraphics[height=2.8cm]{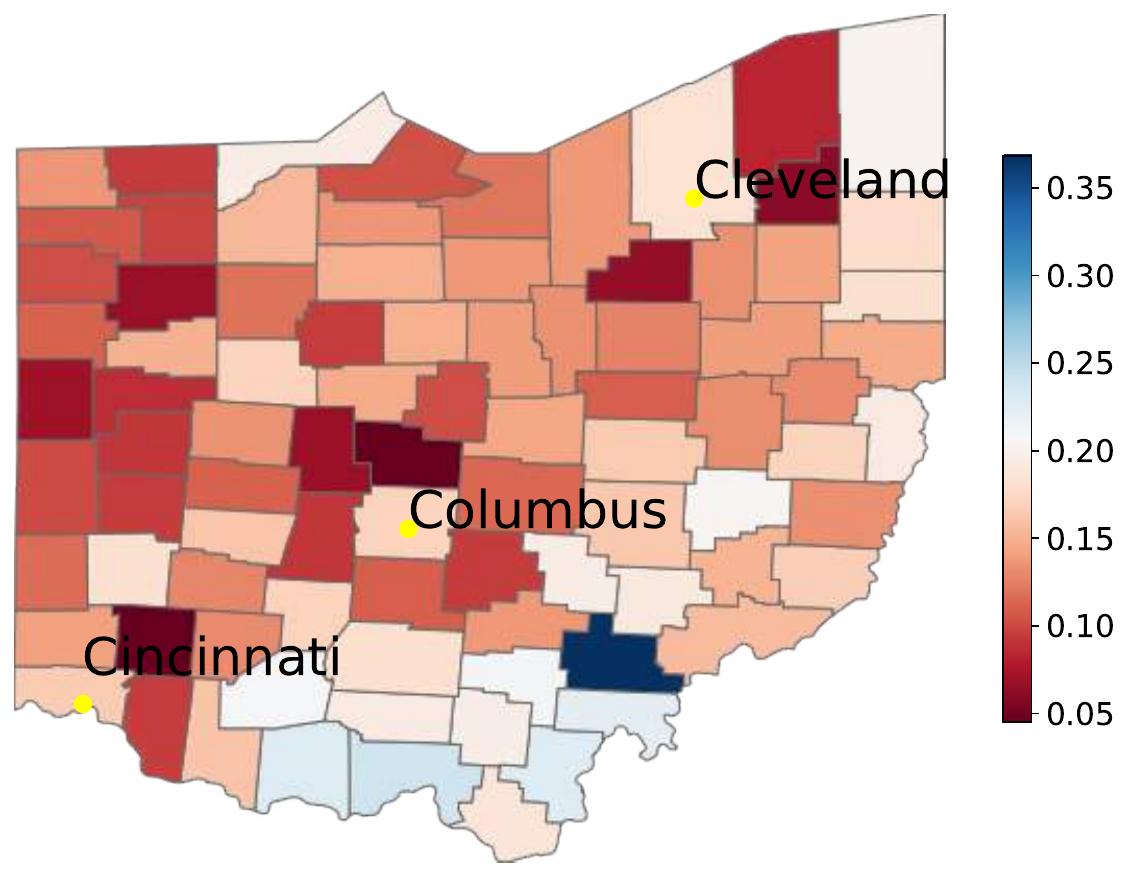} \label{hmap_econ_poverty}
        }
    \subfigure[Race\_white]{%
        \includegraphics[height=2.6cm]{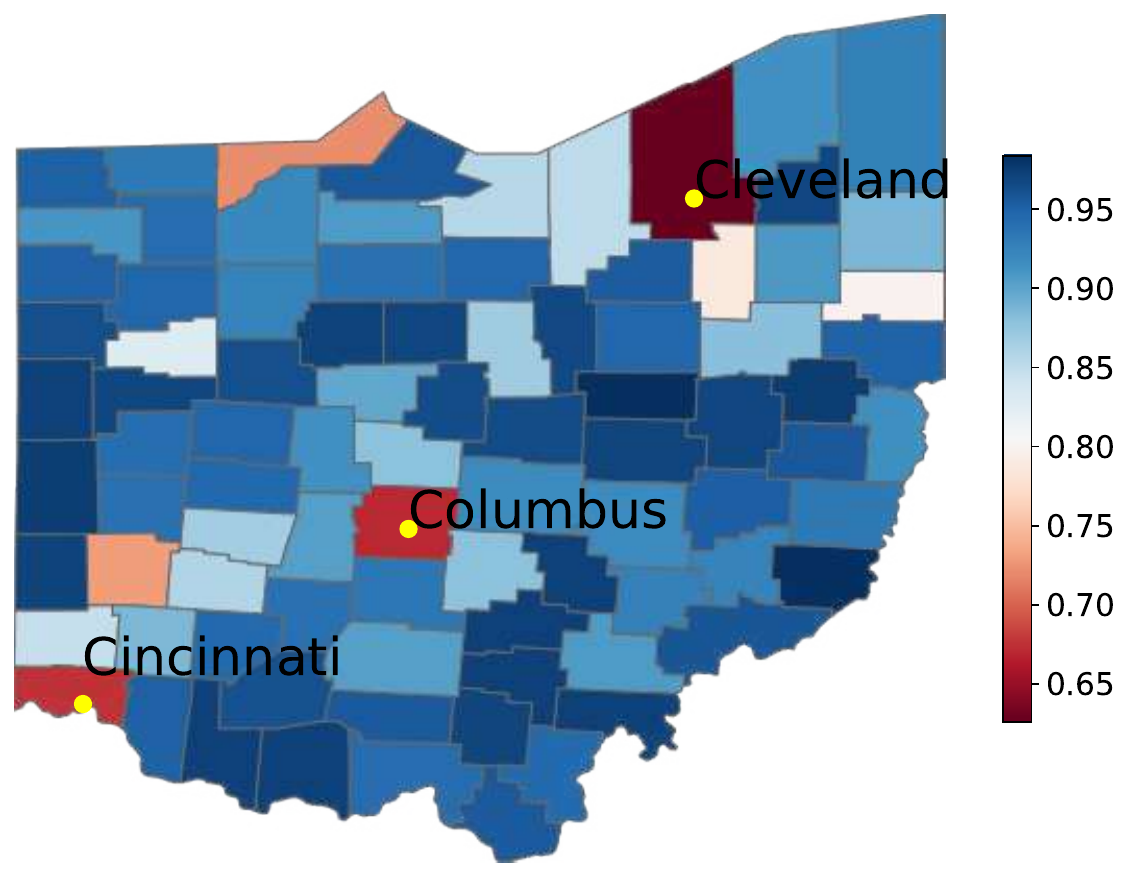} \label{hmap_race_white}
        }
    \subfigure[Age\_GE65]{%
        \includegraphics[height=2.8cm]{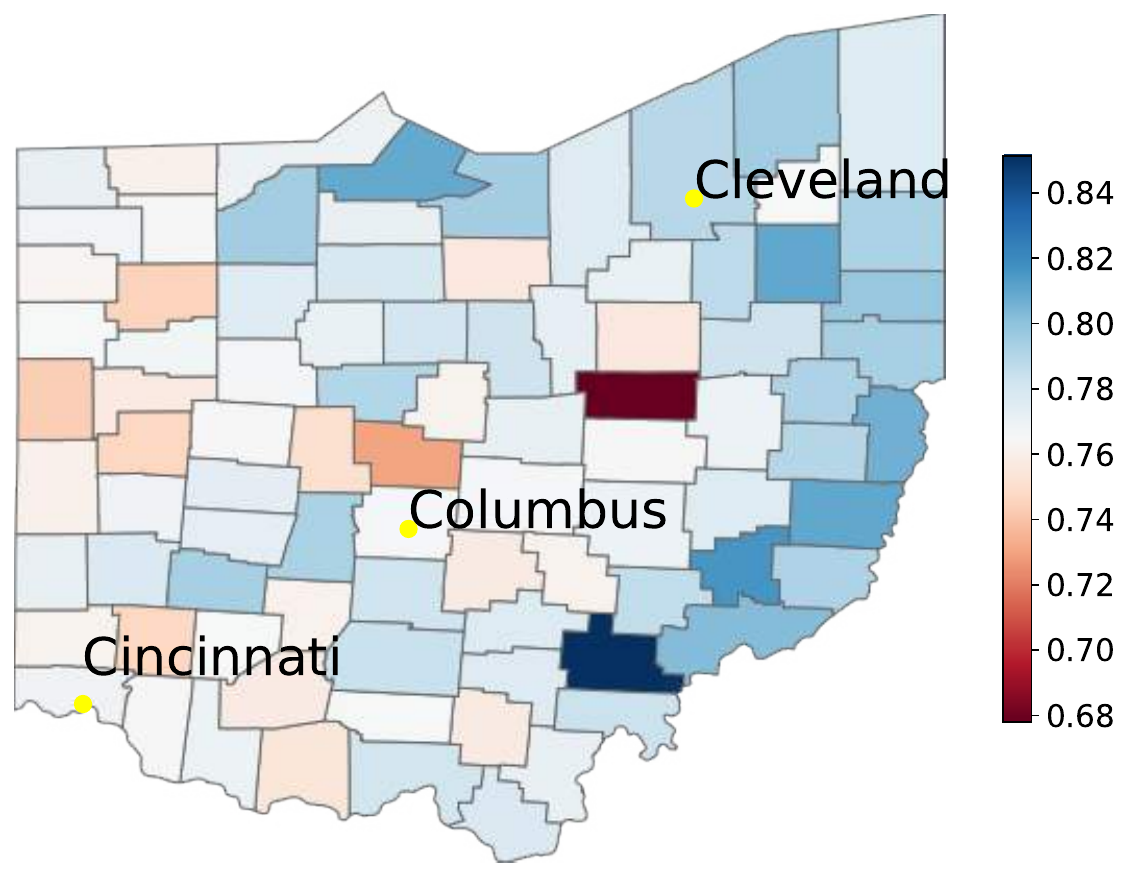} \label{hmap_age_GE65}
        }
\end{center}
\vspace{-0.5cm}
    \caption{Heatmaps of Selected IVs for Ohio Dataset }
\label{hmaps of IVs}
\end{figure}

\begin{figure}[h!]
\begin{center}
    \subfigure[Totpop]{%
        \includegraphics[height=2.8cm]{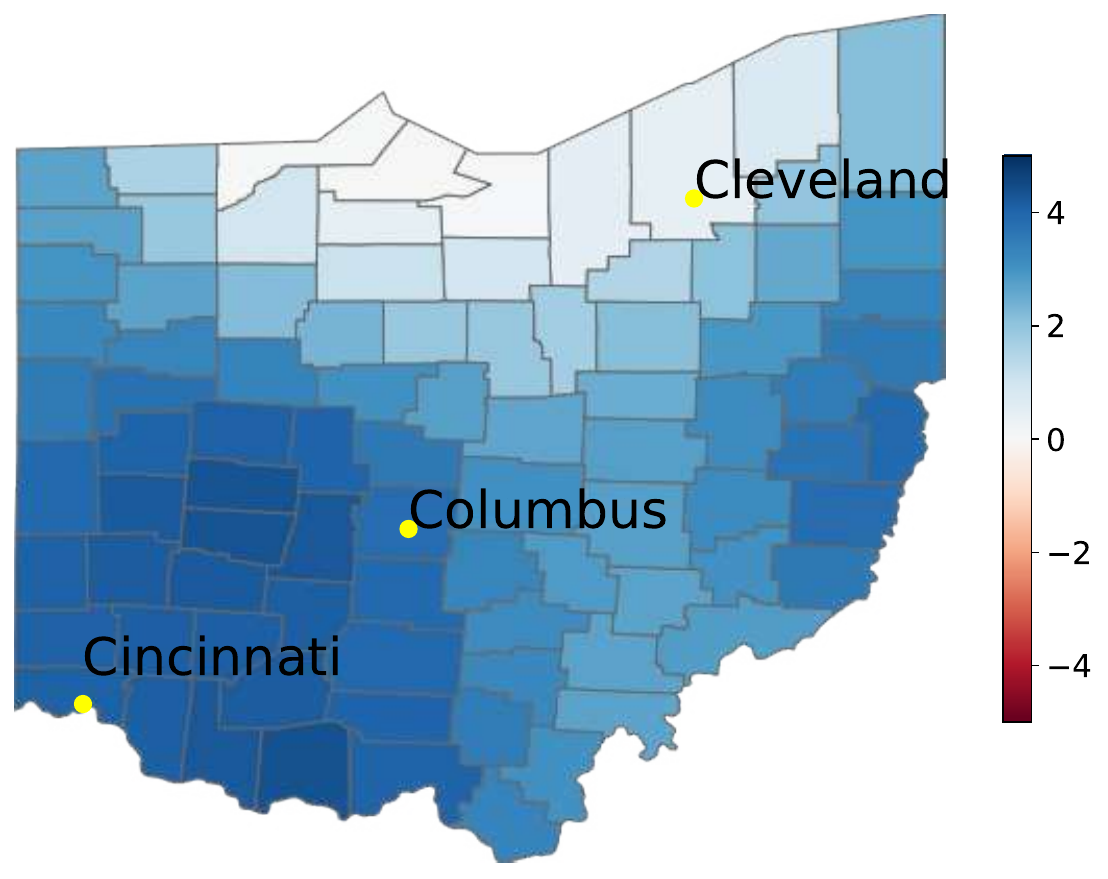} \label{ohio_coeff1_global}
        }
    \subfigure[Fct\_complement]{%
        \includegraphics[height=2.8cm]{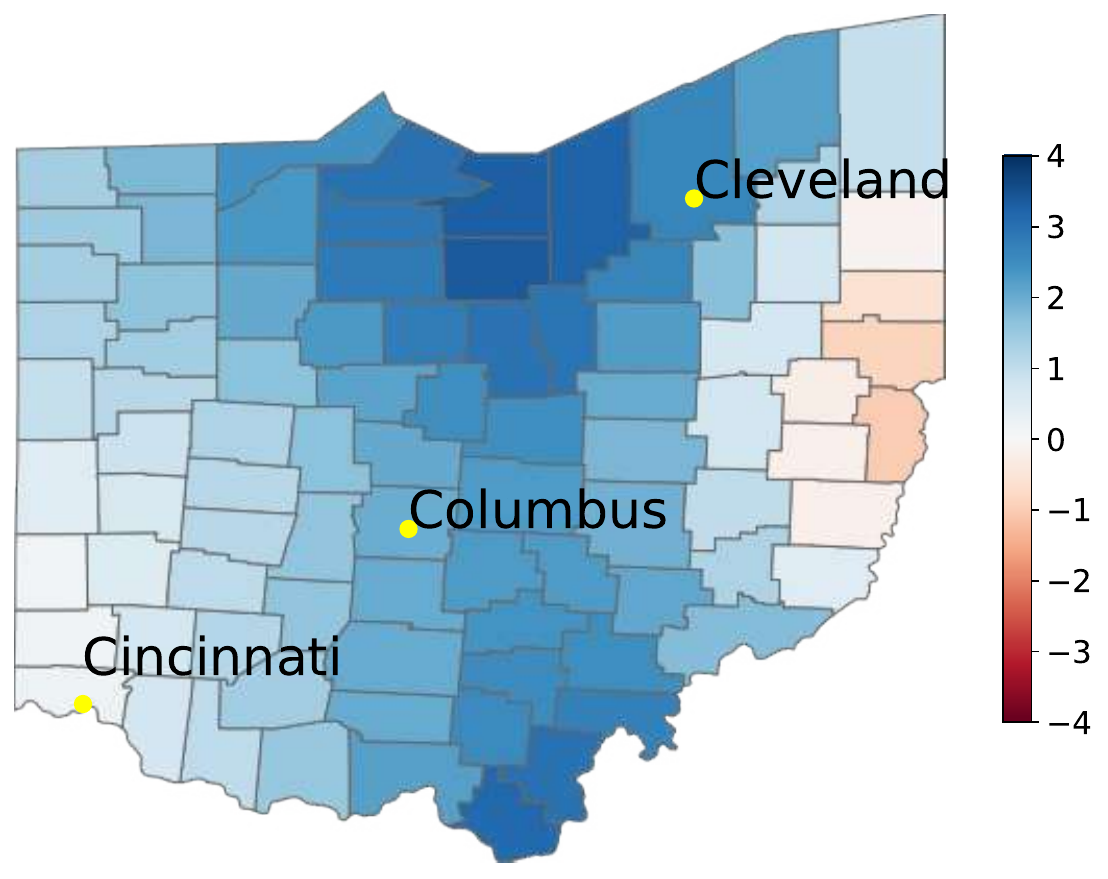} \label{ohio_coeff2_global}
        }
    \subfigure[Meanhinc]{%
        \includegraphics[height=2.8cm]{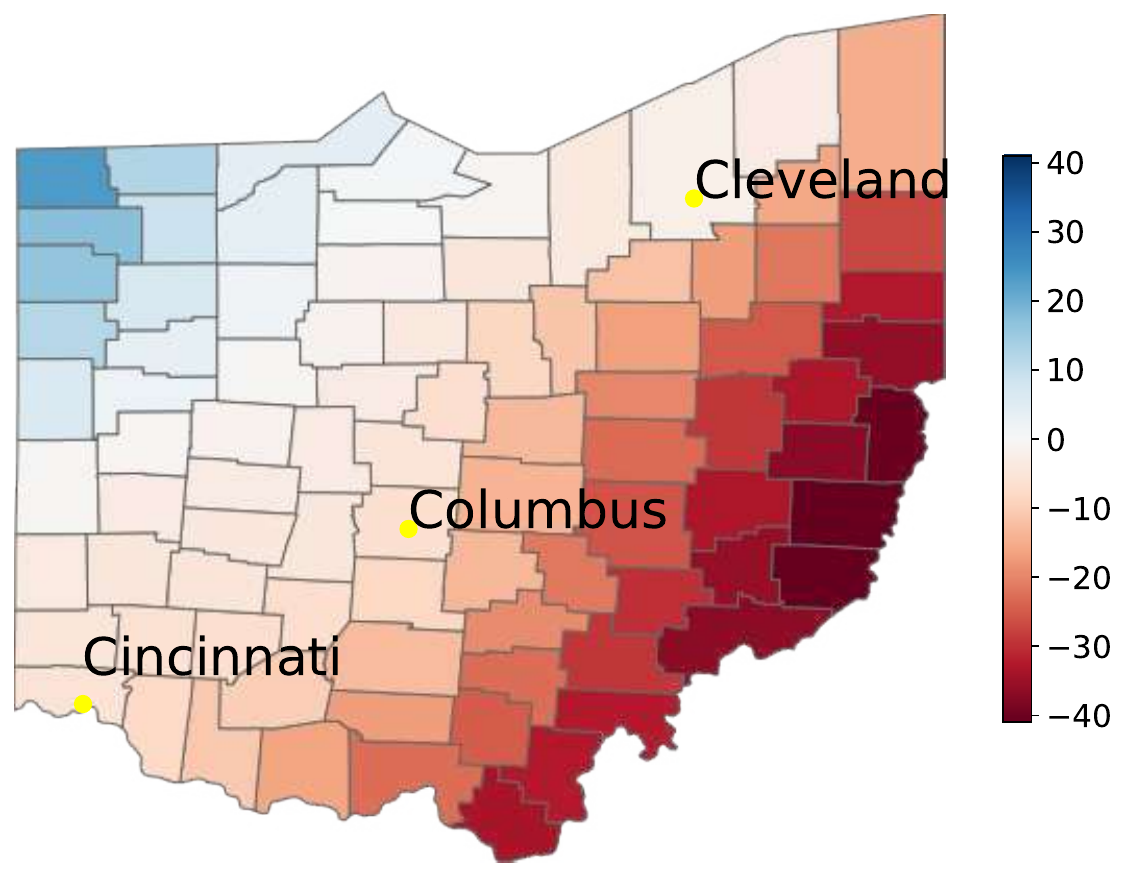} \label{ohio_coeff3_gwr}
        } 
    \subfigure[Educ\_bach]{%
        \includegraphics[height=2.8cm]{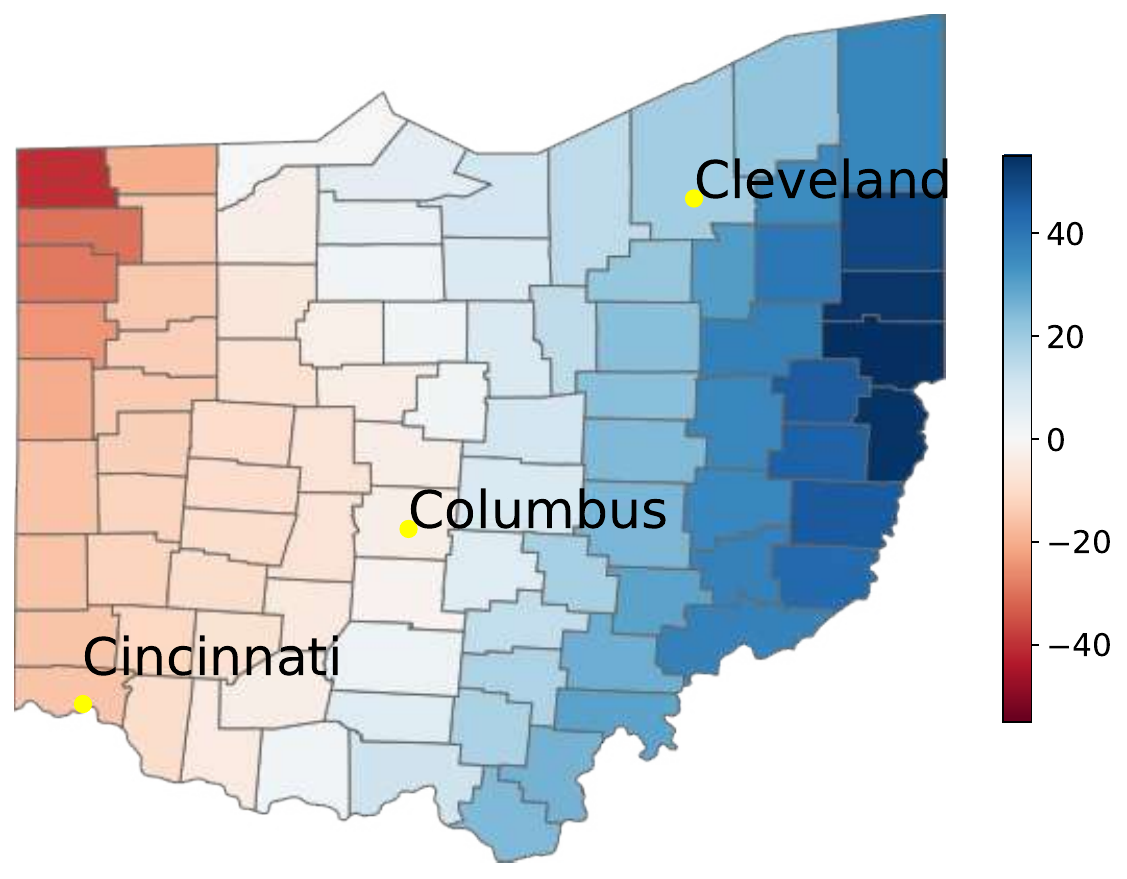} \label{ohio_coeff4_gwr}
        }
\end{center}
\vspace{-1.0cm}
\begin{center}
    \subfigure[Econ\_unemp]{%
        \includegraphics[height=2.8cm]{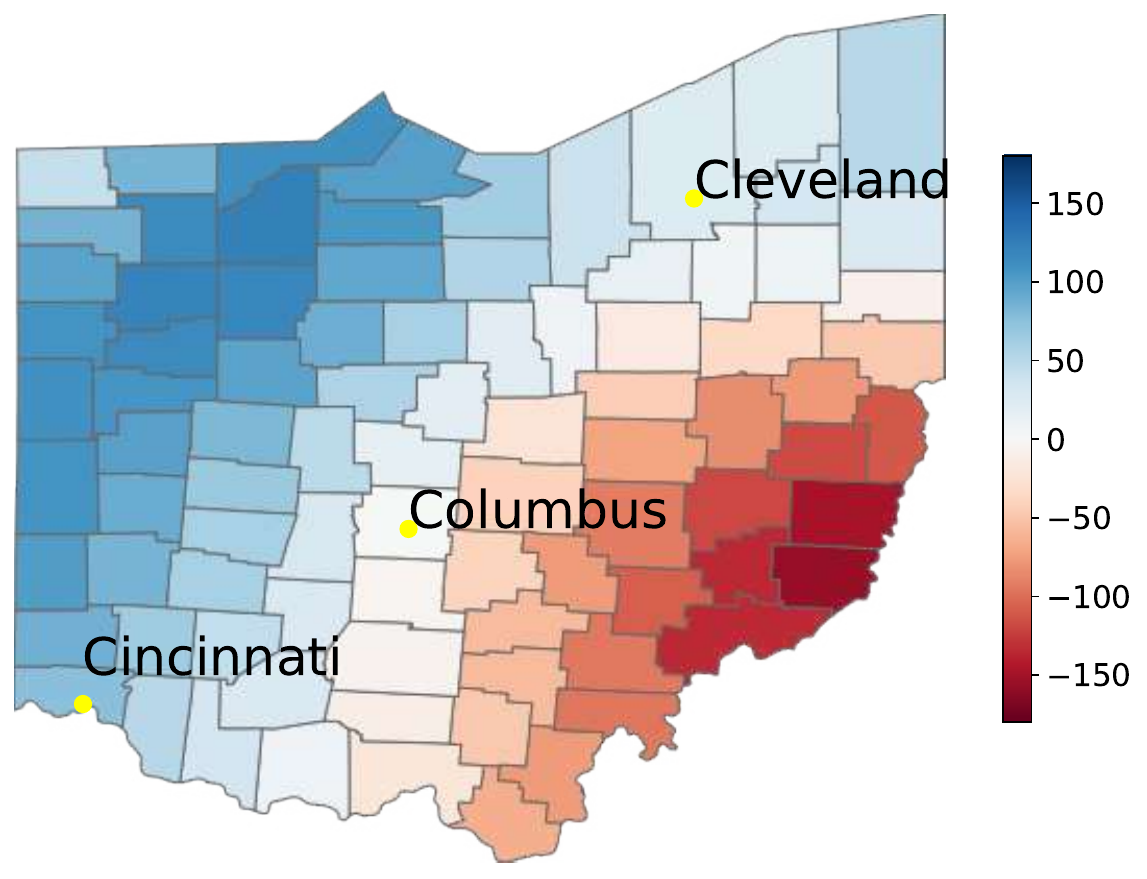} \label{ohio_coeff5_global}
        }
    \subfigure[Econ\_poverty]{%
        \includegraphics[height=2.8cm]{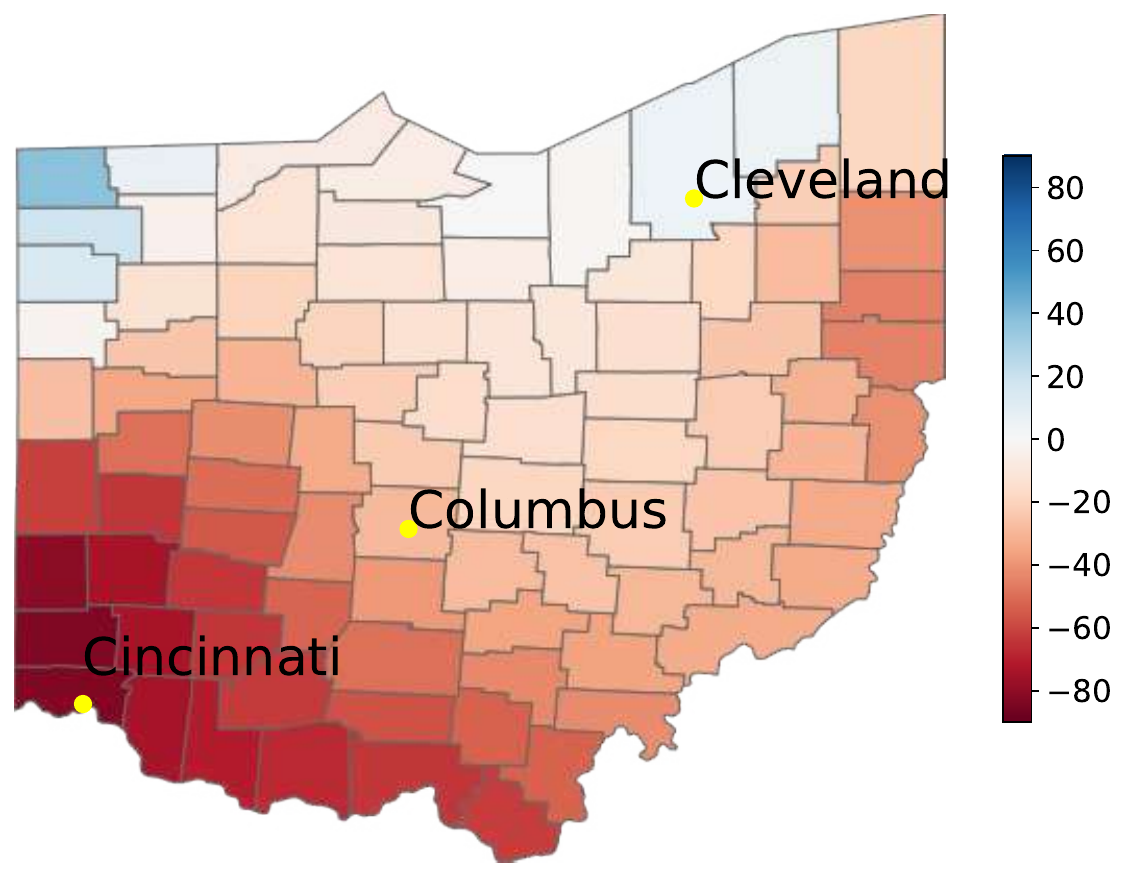} \label{ohio_coeff6_global}
        }
    \subfigure[Race\_white]{%
        \includegraphics[height=2.8cm]{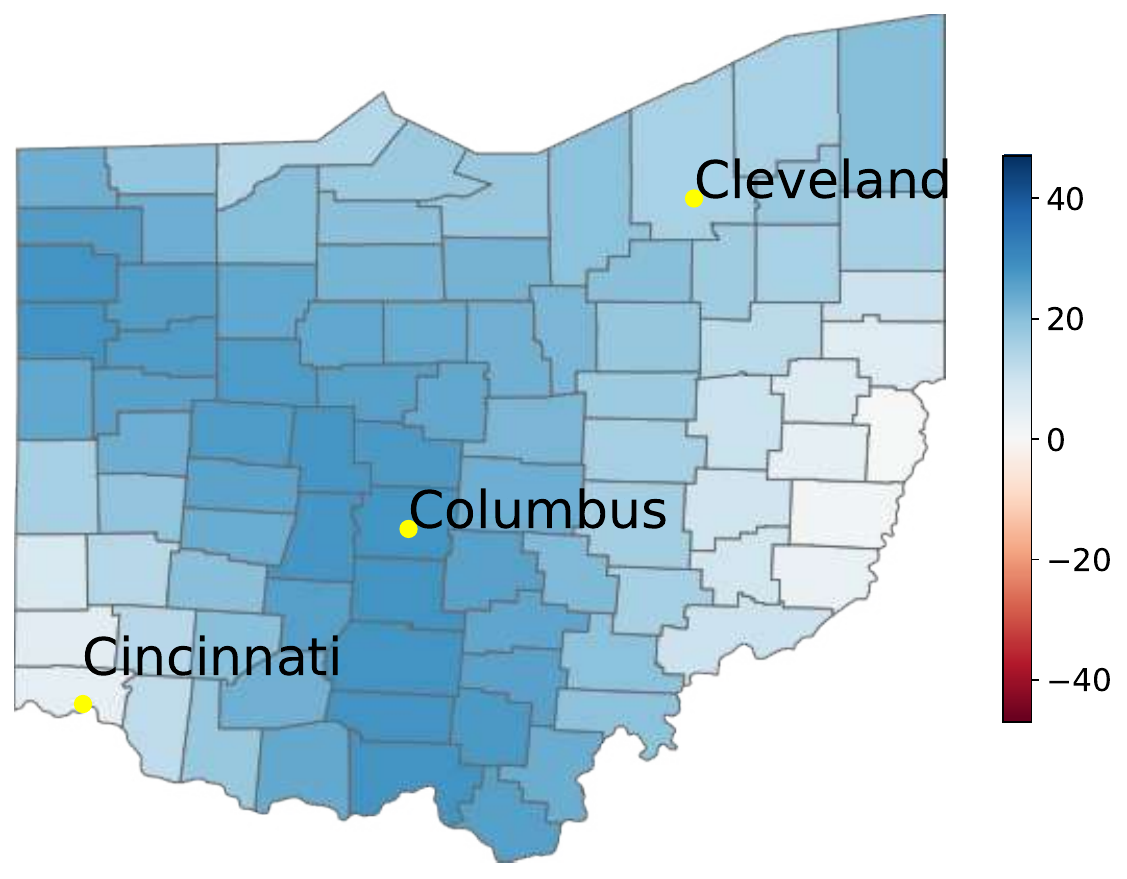} \label{ohio_coeff7_global}
        }
    \subfigure[Age\_GE65]{%
        \includegraphics[height=2.8cm]{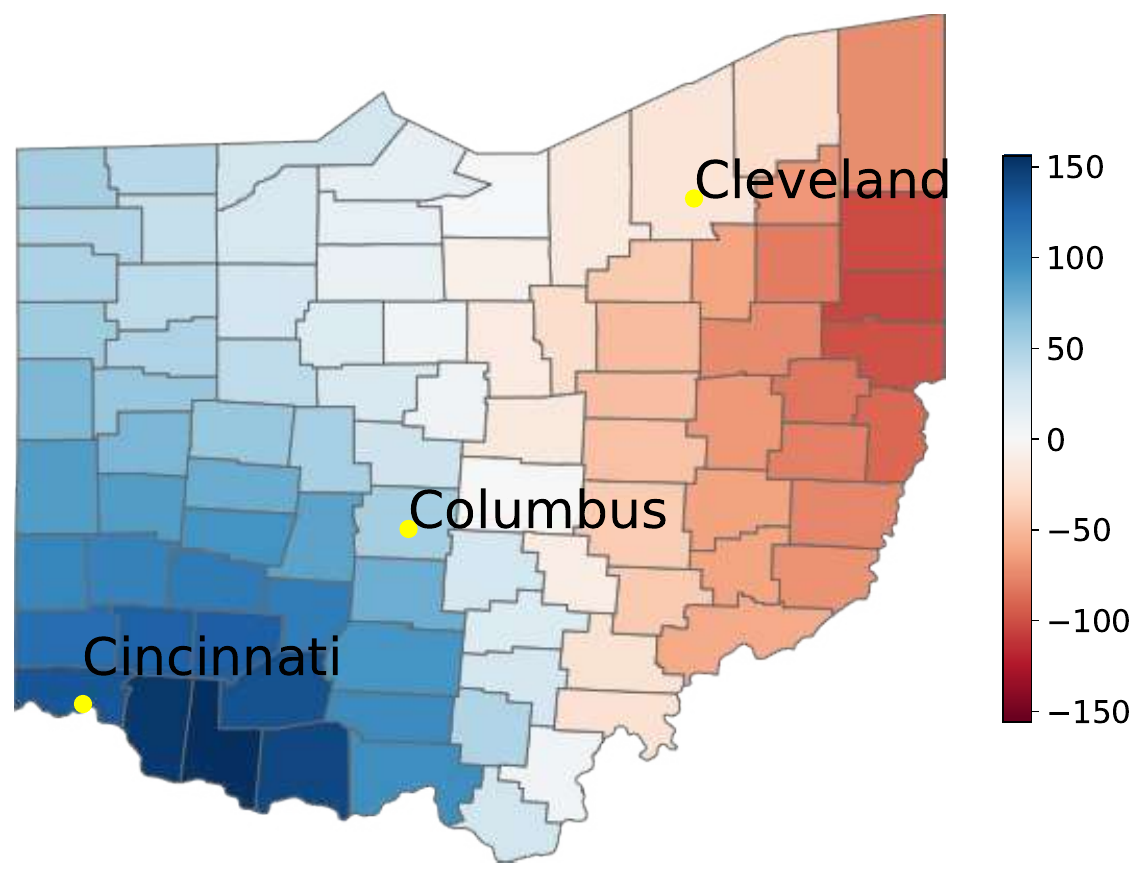} \label{ohio_coeff8_global}
        }
\end{center}
\vspace{-0.5cm}
    \caption{Spatially Varying Coefficient for Ohio Dataset (IGWR-G, $p=8$)}
\label{Comparison of coeffcients}
\end{figure}

In Figures \ref{hmaps of IVs} and \ref{Comparison of coeffcients}, we present the heatmaps of the variables and spatially varying coefficients, respectively, for the global bandwidth setting (IGWR-G) with $p=8$. In Figure \ref{ohio_coeff1_global}, the coefficient of the total population has a positive effect and tends to be high in the southwestern region, where Columbus and Cincinnati are located. This implies that the total population strongly and positively affects the average expenditure per organization in the southwestern region. The other coefficients have their own spatially varying pattern. For example, in Figure \ref{ohio_coeff7_global}, the effect of the white population is positive and tends to increase when it is closer to the center region, near Columbus. The effect of poverty rates is mostly negative and tends to be high in the southwestern region in Figure \ref{ohio_coeff6_global}. The observed positive effects of the total population, white race, and complement facilities and the negative effects of the poverty rate align with the existing literature \citep{bone2021engages}. However, the negative effects of mean household income and the mixed-sign effects of education level, unemployment rate, and senior rate require further explanation. Typically, household income, education level, and senior rate are likely to have positive effects, while the unemployment rate is likely to be negative.

Before discussing these effects, we emphasize the importance of performing the comparison locally, considering neighboring counties, rather than at a global level. This is because the nearby neighboring counties have larger weights and a greater influence on the focal county. The pronounced negative effects of mean household income on the southeastern border, as depicted in Figure \ref{ohio_coeff3_gwr}, indicate that an increase in income reduces the DV. This implies that low-income counties have higher per-organization expenditures than the neighboring counties in that region. This observation is supported by the presence of multiple counties with zero expenditure but higher mean household income than their neighboring counties in the southeastern border, as illustrated in Figures \ref{heatmap_ohio} and \ref{hmap_MEANHINC}. Similarly, the negative effects of education level in the western region shown in Figure \ref{ohio_coeff4_gwr} align with the trend that counties with lower education levels have higher per-organization expenditures than their neighboring counties in that region, as Figures \ref{heatmap_ohio} and \ref{hmap_educ_bach} show. Conversely, the positive effects of the unemployment rate in the northwestern region, presented in Figure \ref{ohio_coeff5_global}, suggest that counties with higher employment rates in that region have higher per-organization expenditures than their neighboring counties. Furthermore, the negative effects of the senior rate in the eastern region, as depicted in Figure \ref{ohio_coeff8_global}, indicate that counties with higher senior rates have lower per-organization expenditures than their neighboring counties.

In summary, we observe that the average expenditure per organization is not distributed according to the socio-economic and demographic characteristics in some regions of the state. However, this finding does not necessarily imply unfair distribution of expenditures, as organizations may serve visitors from other counties, states, or countries.

\subsection{Performance Comparison of Models Across Datasets} \label{section4.5}
To support the findings from Sections \ref{section4.2}-\ref{section4.4} and to generalize the characteristics of each model, we present comprehensive results across three datasets: Georgia, Ohio, and the US Census. The US Census dataset includes data from 20 selected states, bringing the total number of test cases to 22. Table \ref{tab:adj_R_comparison} presents the $R^2_{\text{adj}}$ values, subset cardinality, and average symmetric differences for these 22 test cases across all six models. The $R^2_{\text{adj}}$ metric reflects the explanatory power while penalizing the number of independent variables to prevent overfitting. The subset cardinality represents the number of selected variables for each dataset. The average symmetric difference measures the average symmetric difference between pairs of subsets across all local models using the following formula.
\begin{equation}
\text{avg\_symm\_diff} = \frac{\sum_{o, o' \in O} 
 \text{symm\_diff}(s_o,s_{o'})}{ {|O| \choose 2}},
\end{equation}
where $s_o$ is the selected subset for local model $o$ and \text{symm\_diff}($s_o$,$s_{o'}$) = $(s_o \setminus s_{o'}) \cup (s_{o'} \setminus s_o).$ For example, if there are three focal points with subsets $s_1$, $s_2$, and $s_3$, the symmetric difference is calculated as $\text{avg\_symm\_diff} = \frac{1}{3} \cdot (\text{symm\_diff}(s_1, s_2) + \text{symm\_diff}(s_1, s_3) + \text{symm\_diff}(s_2, s_3)).
$ In the average symmetric difference column, all methods except for GWL report symmetries in subset selection across all pairs of focal points. Conversely, GWL exhibits notable asymmetries in subset selection, with an average symmetric difference of 4.34 out of an average subset cardinality of 6.91. This substantial variation indicates that comparing spatial patterns for each IV is not feasible with GWL.

Subset cardinality is detailed for the global subset selection methods (IGWR-G, IGWR-L, and FS) and the local subset selection method, GWL. Note that BGWR and MGWR, which do not perform subset selection, are excluded from the subset cardinality column. For GWL, we follow the default settings of the GWL function in the \texttt{gwrr} package and calculate the average subset cardinality across all local models. For the global subset selection methods, subset cardinality for the Georgia and Ohio datasets is determined by identifying the elbow point based on RSS, as introduced in Section \ref{section4.3}, ensuring uniformity between IGWR and FS. For the US\_Census dataset, the subset cardinality is set to the closest integer to the average subset cardinality of GWL applied to this dataset. Notably, the subset cardinalities for both global and local subset selection methods are very similar, ensuring fairness in the comparisons.

\begin{figure}[h!]
\begin{center}
    \includegraphics[scale=0.5]{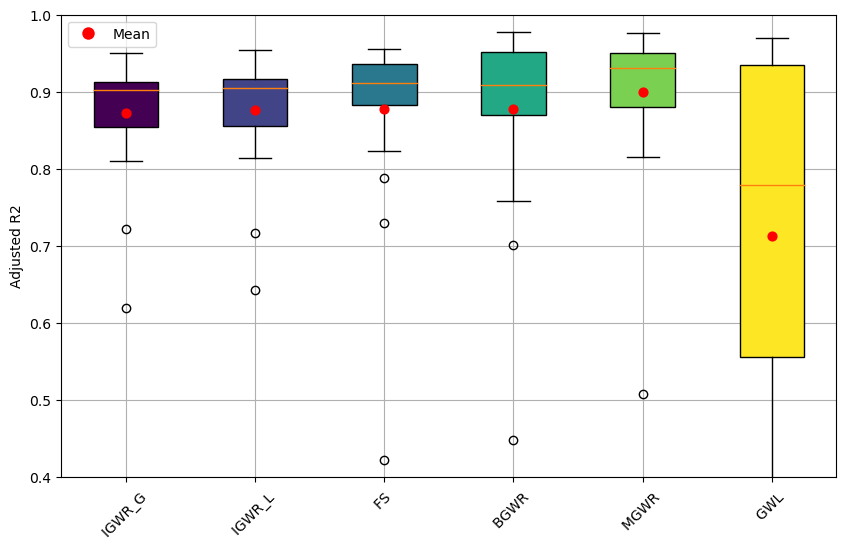} 
\end{center}
\vspace{-0.5cm}
    \caption{ Box Plots Comparing Adjusted $R^2$ Across Models}
\label{boxplots_adj_R2}
\end{figure}

Figure \ref{boxplots_adj_R2} illustrates the box plot of $R^2_{\text{adj}}$ for the different models. Based on Table \ref{tab:adj_R_comparison} and Figure \ref{boxplots_adj_R2}, the models are divided into three groups based on the performances. The first group, MGWR, achieves the best $R^2_{\text{adj}}$ values among all methods. The second group includes IGWR-G, IGWR-L, FS, and BGWR, all of which exhibit similar $R^2_{\text{adj}}$ values. Finally, the third group, GWL, demonstrates poor average performance compared to other methods, although it performs well on certain datasets. Although the MGWR model achieves the highest $R^2_{\text{adj}}$, this is primarily due to its use of different bandwidths for each IV. While this approach typically enhances explanatory power, it also leads to spatial patterns that differ from those of other models (IGWR, FS, and BGWR), as shown in Section \ref{section4.3}. These patterns often exaggerate or diverge from the consensus observed in the other models. Therefore, considering the drawbacks of MGWR and GWL, our focus shifts to comparing the second groups.

Within the second group, the global subset selection methods (IGWR-G, IGWR-L, and FS) produce $R^2_{\text{adj}}$ values comparable to BGWR in terms of both median (yellow line in Figure \ref{boxplots_adj_R2}) and mean values (red dot in Figure \ref{boxplots_adj_R2}). Furthermore, the first and third quartiles for the global subset selection methods are higher and lower, respectively, than those of BGWR, indicating that global subset selection provides more robust results. This also implies that using only a subset of the variables can perform better than using all variables.

The three global subset selection methods—IGWR-G, IGWR-L, and FS—exhibit similar performance. While the IGWR algorithm is based on a mathematical programming model, its performance is similar to the heuristic FS approach. Two key reasons explain this: (1) the IGWR algorithm does not directly minimize the WSSE but instead maximizes the log-likelihood function, and (2) minimizing the WSSE does not always result in the lowest RSS value. The second reason highlights a key distinction between best subset selection in MLR and global subset selection in GWR. For example, consider a dataset with 100 rows. In MLR, minimizing SSE during estimation is equivalent to minimizing RSS during evaluation, as both are calculated over the same 100 terms. In contrast, in GWR, the WSSE during estimation involves $100 \times 100$ terms, as it aggregates all local models using the weight matrix. Meanwhile, the RSS during evaluation is computed over only 100 terms, without accounting for the weight matrix. When the estimation and evaluation metrics align, as in MLR, a mathematical programming-based algorithm outperforms heuristic methods. However, this is not the case in GWR, which explains why the IGWR algorithm does not outperform the FS approach, despite their performances being closely matched.


\begin{table}[h!]
  \centering
  \small
    \caption{ Comparison of Explanatory Power, Subset Cardinality, and Symmetric Difference Across Models}
    \label{tab:adj_R_comparison}
    \scalebox{0.8}{
    \begin{tabular}{ccccccccccc}
    \toprule
    \multicolumn{1}{c}{\multirow{2}{*}{Dataset\_State}} & \multicolumn{6}{c}{Adjusted R squared} & \multicolumn{2}{c}{Subset Cardinality} & \multicolumn{2}{c}{Average Symmetric   Difference} \\ \cmidrule{2-11}
    \multicolumn{1}{c}{} & \multicolumn{1}{c}{IG} & \multicolumn{1}{c}{IL} & \multicolumn{1}{c}{FS} & \multicolumn{1}{c}{BG} & \multicolumn{1}{c}{MG} & \multicolumn{1}{c}{GW} & \multicolumn{1}{c}{IG, IL, FS} & \multicolumn{1}{c}{GW} & \multicolumn{1}{c}{IG, IL, FS, BG, MG} & \multicolumn{1}{c}{GW} \\ \midrule
    Georgia\_GA & 0.721 & 0.716 & 0.730 & 0.701 & 0.815 & 0.736 & 4 & 3.75 & 0.00 & 2.70 \\
    Ohio\_OH & 0.619 & 0.642 & 0.422 & 0.448 & 0.507 & 0.968 & 8 & 6.36 & 0.00 & 6.09 \\
    US\_census\_TX & 0.847 & 0.847 & 0.882 & 0.868 & 0.921 & 0.941 & 7 & 6.73 & 0.00 & 4.26 \\
    US\_census\_GA & 0.903 & 0.906 & 0.937 & 0.929 & 0.946 & 0.949 & 7 & 7.86 & 0.00 & 3.03 \\
    US\_census\_VA & 0.923 & 0.932 & 0.948 & 0.977 & 0.961 & 0.970 & 7 & 6.96 & 0.00 & 4.20 \\
    US\_census\_KY & 0.911 & 0.919 & 0.952 & 0.955 & 0.960 & 0.873 & 7 & 6.44 & 0.00 & 4.44 \\
    US\_census\_MO & 0.906 & 0.908 & 0.898 & 0.888 & 0.919 & 0.889 & 7 & 6.10 & 0.00 & 4.03 \\
    US\_census\_KS & 0.830 & 0.828 & 0.865 & 0.861 & 0.830 & 0.218 & 7 & 6.73 & 0.00 & 5.00 \\
    US\_census\_IL & 0.902 & 0.901 & 0.940 & 0.943 & 0.942 & 0.687 & 7 & 6.54 & 0.00 & 4.17 \\
    US\_census\_NC & 0.917 & 0.916 & 0.934 & 0.972 & 0.929 & 0.429 & 7 & 6.78 & 0.00 & 4.84 \\
    US\_census\_IA & 0.810 & 0.813 & 0.788 & 0.792 & 0.855 & 0.915 & 7 & 6.69 & 0.00 & 3.98 \\
    US\_census\_TN & 0.913 & 0.921 & 0.926 & 0.875 & 0.933 & 0.961 & 7 & 6.46 & 0.00 & 3.81 \\
    US\_census\_NE & 0.825 & 0.824 & 0.823 & 0.757 & 0.845 & 0.554 & 7 & 6.92 & 0.00 & 4.55 \\
    US\_census\_IN & 0.885 & 0.889 & 0.899 & 0.916 & 0.889 & 0.654 & 7 & 6.51 & 0.00 & 4.87 \\
    US\_census\_OH & 0.931 & 0.933 & 0.923 & 0.925 & 0.936 & 0.782 & 7 & 7.00 & 0.00 & 4.02 \\
    US\_census\_MN & 0.912 & 0.916 & 0.934 & 0.973 & 0.952 & 0.559 & 7 & 7.37 & 0.00 & 4.29 \\
    US\_census\_MI & 0.902 & 0.903 & 0.956 & 0.890 & 0.878 & 0.959 & 7 & 5.89 & 0.00 & 3.71 \\
    US\_census\_MS & 0.897 & 0.904 & 0.894 & 0.930 & 0.968 & 0.775 & 7 & 5.95 & 0.00 & 4.73 \\
    US\_census\_OK & 0.910 & 0.911 & 0.886 & 0.902 & 0.948 & 0.314 & 7 & 6.60 & 0.00 & 4.55 \\
    US\_census\_AR & 0.873 & 0.881 & 0.921 & 0.958 & 0.963 & 0.839 & 7 & 6.63 & 0.00 & 4.85 \\
    US\_census\_WI & 0.913 & 0.909 & 0.902 & 0.877 & 0.920 & 0.442 & 7 & 7.10 & 0.00 & 4.29 \\
    US\_census\_AL & 0.950 & 0.954 & 0.941 & 0.966 & 0.976 & 0.249 & 7 & 5.82 & 0.00 & 5.06 \\ \hline
    Average & 0.873 & 0.876 & 0.877 & 0.877 & 0.900 & 0.712 & 6.91 & 6.51 & 0.00 & 4.34 \\ \bottomrule
    \end{tabular}
    }    
  \scalebox{0.8}{
  \begin{tabular}{p{15.5cm}}
\begin{scriptsize}$\dagger$ IG, IL, FS, BG, MG, and GW stands for IGWR-G, IGWR-L, FS, BGWR, MGWR, and GWL, respectively. Note that IG, IL, and FS are global subset approaches giving the same subset for all local models, GW is a local subset approach giving different subsets for local models, and MG and GW do not select subsets. \end{scriptsize}
\end{tabular}
}
\end{table}  

Next, we evaluate the extent of spatially varying patterns using the range-to-mean ratio. For each IV $j$, the range-to-mean ratio is defined as 
\[
(\max_{o \in O}\{\beta_{oj}\} - \min_{o \in O}\{\beta_{oj}\}) / \operatorname{mean}_{o \in O}\{\beta_{oj}\}.
\]
While each model may operate on different scales, the range-to-mean ratio provides a robust metric because the range is normalized by the mean value. Since each model selects different variables, we compare the coefficients of variables common across all models. The complete results are provided in the online supplementary file.

\begin{figure}[h!]
\begin{center}
    \includegraphics[scale=0.5]{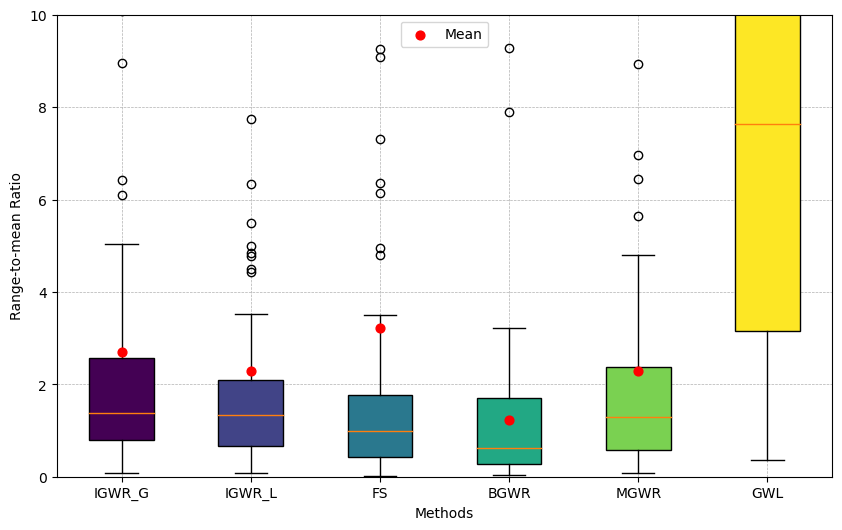} 
\end{center}
\vspace{-0.5cm}
    \caption{Box Plots Comparing Range-to-mean Ratio Across Models}
\label{boxplots_coeff}
\end{figure}

Given the goal of GWR to capture spatially varying patterns, these patterns should neither be extreme nor obscure. Instead, they should be distinct and interpretable. Figure \ref{boxplots_coeff} presents the range-to-mean ratio for all models. Based on the median and mean values, the models can be divided into three groups. The first group, GWL, exhibits the largest ratio among all models, reflecting the extreme variability of coefficients within the study region due to its local subset selection approach. The second group includes IGWR-G, IGWR-L, FS, and MGWR, which demonstrate moderate variability. Finally, the third group, BGWR, shows the smallest ratio, indicating that the coefficient patterns may be less distinct compared to the second group.

It is important to note that Figure \ref{boxplots_coeff} serves as a reference for understanding the extent of spatially varying patterns. Since the true underlying pattern is unknown, we cannot definitively determine which pattern is correct. However, by plotting the patterns for all models, we can identify those that are neither excessively unstable nor overly obscure. For instance, in Section \ref{section4.3}, we present detailed plots for all models and discuss which patterns are most desirable. IGWR and FS produce distinct and smooth spatial patterns, while BGWR exhibits a similar but somewhat obscure trend. MGWR tends to exaggerate, and GWL produces extremely unstable patterns. These patterns can vary from dataset to dataset; therefore, actual plotting provides the best insight.

Based on Table \ref{tab:adj_R_comparison} and Figures \ref{boxplots_adj_R2} and \ref{boxplots_coeff}, each model's characteristics are summarized in Table \ref{tab:overview}. IGWR and FS achieve good explanatory power comparable to the BGWR while maintaining distinct spatial patterns. GWL clearly demonstrates relatively poor performance in terms of explanatory power, coupled with extremely varying patterns. MGWR shows the best performance in explanatory power due to its use of different bandwidths for each IV, along with distinct spatial patterns. However, it tends to produce patterns that differ from those of IGWR, FS, and BGWR, requiring careful implementation. 

\begin{table}[h!]
  \centering
    \small
  \setlength{\tabcolsep}{3pt}
  \caption{Overview of Model and Output Characteristics}
  \scalebox{0.8}{
    \begin{tabular}{c|c|c|c|c|c}
    \toprule
          & IGWR & FS & BGWR   & MGWR  & GWL \\ \midrule
    Explanatory power & Good  & Good & Good & Excellent & Moderate \\
    Selected subset in local models & Homogeneous & Homogeneous & NA    & NA    & Heterogeneous \\ 
    Spatially varying pattern & Distinct & Distinct & Less distinct & Distinct & Extremely Varying \\ \bottomrule
    \end{tabular}%
  }
  \label{tab:overview}
\end{table}%

In summary, the IGWR algorithm is a competitive method that excels in subset selection tasks, maintaining strong explanatory power while achieving distinct and smooth spatial patterns. As demonstrated in Section \ref{section4}, the global subset selection methods offer the most desirable properties compared to others. The FS approach, which is a heuristic stepwise procedure, can be less precise for subset selection, as shown in Section \ref{section4.4}. Therefore, we recommend our proposed IGWR method as the preferred approach overall.

\section{Conclusion}\label{section5}

The algorithm proposed herein provides a unified framework for estimating all parameters of the GWR model simultaneously. By using a single objective function that incorporates all GWR parameters for every focal point and integrating a variable subset selection process into the framework, our approach offers a compelling alternative to existing methods. We formulate an MIQP model that optimizes this new objective function, ensuring a consistent subset across all focal points. Our proposed MIQP model is highly flexible and can accommodate additional constraints on selected subsets. To solve this non-convex MIQP model, we propose an ADM algorithm, IGWR, that exploits the partitioned subproblems and the block structures of variables, and guarantees convergence to a partial minimum. Computational experiments demonstrate that the proposed algorithm effectively selects variable subsets, estimates all GWR parameters simultaneously, and converges within a few iterations.

Compared with existing models such as the FS approach and standard GWR, our approach offers comparable explanatory power (in terms of $R^2_{\text{adj}}$) while providing a superior subset selection method. Furthermore, it generates homogeneous variable subsets and stable spatially varying patterns, setting itself apart from MGWR and GWL. Comprehensive computational results, utilizing multiple datasets, illustrate these advantages. We believe that our model offers an integrated framework with multiple advantages and added flexibility in model specifications.

It is important to note that the derivation of test statistics for our model is more challenging than for standard GWR approaches due to different variance assumptions. This area remains a topic for future research. Additionally, we suggest that our techniques can be extended to general local regression models or weighted regression models with similar structures. Overall, our proposed framework provides a significant contribution to the field of spatial data analysis.

\section*{Acknowledgement} The authors thank Southern Methodist University DataArts for allowing the authors to use the data.








\bibliographystyle{elsarticle-num} 
\bibliography{gwr.bib}

\setcounter{definition}{0}
\renewcommand\thedefinition{A\arabic{definition}}
\setcounter{assumption}{0}
\renewcommand\theassumption{A\arabic{assumption}}
\setcounter{theorem}{0}
\renewcommand\thetheorem{A\arabic{theorem}}
\setcounter{corollary}{0}
\renewcommand\thecorollary{A\arabic{corollary}}



\appendix

\setcounter{table}{0}
\renewcommand{\thetable}{B\arabic{table}}

\end{document}